\documentclass[conference]{IEEEtran}

\usepackage{times}
\usepackage{amsmath}
\usepackage{bm}
\usepackage[normalem]{ulem}

\usepackage{booktabs}
\usepackage{graphicx}
\usepackage{array}
\usepackage{longtable}
\usepackage{tabularx}
\usepackage{ragged2e}
\usepackage{wrapfig}
\usepackage{colortbl}
\usepackage{changepage}
\usepackage{boxedminipage}
\usepackage{adjustbox}

\usepackage{needspace}

\usepackage{float}
\usepackage{rotating}   

\PassOptionsToPackage{table}{xcolor}
\usepackage{xcolor}

\definecolor{sigpos}{RGB}{198,234,198}
\definecolor{signeg}{RGB}{234,198,198}

\newcolumntype{C}[1]{>{\centering\arraybackslash}p{#1}}

\usepackage{enumitem}
\usepackage{comment}
\usepackage{soul}

\definecolor{neg}{RGB}{245,220,220}  
\definecolor{pos}{RGB}{220,245,220}  
\definecolor{zer}{RGB}{235,235,235}  

\usepackage[hidelinks]{hyperref}

\usepackage{tikz}
\usepackage{fontawesome5} 
\usetikzlibrary{shapes.geometric, arrows.meta, positioning, calc, backgrounds, shadows, fit}
\usepackage{pgfplots}
\usepackage{pgfplotstable}
\usetikzlibrary{patterns}
\pgfplotsset{compat=newest}
\usepgfplotslibrary{statistics,groupplots}
\usepackage[most]{tcolorbox}

\colorlet{rowgray}{gray!12}
\definecolor{siguncorr}{RGB}{255, 230, 200}  
\definecolor{phaseonecolor}{HTML}{306fbb}
\definecolor{phasetwocolor}{HTML}{b02f1b}
\definecolor{phasethreecolor}{HTML}{673a9b}

\newcommand{\cnumone}[1]{\tikz[baseline=(char.base)]{
    \node[shape=circle, fill=phaseonecolor, text=white, inner sep=0.5pt, minimum size=12pt] (char) {\textbf{#1}};}}
\newcommand{\cnumtwo}[1]{\tikz[baseline=(char.base)]{
    \node[shape=circle, fill=phasetwocolor, text=white, inner sep=0.5pt, minimum size=12pt] (char) {\textbf{#1}};}}
\newcommand{\cnumthree}[1]{\tikz[baseline=(char.base)]{
    \node[shape=circle, fill=phasethreecolor, text=white, inner sep=0.5pt, minimum size=12pt] (char) {\textbf{#1}};}}

\usepackage{subcaption}  

\renewcommand{\arraystretch}{1.2}

\IEEEoverridecommandlockouts
\makeatletter
\def\@IEEEpubidpullup{6.5\baselineskip}
\makeatother

\IEEEpubid{\parbox{\columnwidth}{
Network and Distributed System Security (NDSS) Symposium 2027\\
22--26 March 2027, Seoul, Republic of Korea\\
ISBN 978-1-970672-09-1\\
https://dx.doi.org/10.14722/ndss.2027.230409\\
www.ndss-symposium.org
}
\hspace{\columnsep}\makebox[\columnwidth]{}}

\begin{document}

\title{User Perceptions of Deceptive In-Vehicle Infotainment User Interfaces}




\author{
\IEEEauthorblockN{
Pedram MohajerAnsari$^{\dagger}$,
Iwinosa Aideyan$^{\dagger}$,
Anna Galeano$^{\dagger}$,
Arjun Arunasalam$^{*}$, \\
Bart Knijnenburg$^{\dagger}$,
Richard Brooks$^{\dagger}$,
Mert D. Pesé$^{\dagger}$
}

\IEEEauthorblockA{
$^{\dagger}$ Clemson University,
\{pmohaje, iaideya, agalean, bartk, rrb, mpese\}@clemson.edu\\
$^{*}$ Florida International University,
aarunasa@fiu.edu
}
}

\maketitle

\begin{abstract}
Modern vehicles collect extensive data about drivers, passengers, and vehicle use, with privacy and consent decisions increasingly mediated through infotainment software. Although dark patterns have been widely studied in web, mobile, and e-commerce interfaces, little is known about how they appear in automotive infotainment systems or how users perceive them. We study this problem through Android Automotive OS (AAOS), conducting a three-phase empirical analysis. First, 26 HCI and privacy annotators audit four OEM AAOS emulators (Honda, Volvo, GM, and Polestar), documenting 17 dark-pattern types across 65 observations. Second, a user study with 100 participants compares OEM-specific implementations of five dark patterns across GMC, Honda, and Volvo, finding that implementation choices affect user perceptions. Third, an independent study with 49 participants compares automotive and e-commerce dark-pattern clips before and after a brief tutorial, showing that automotive's initial rating advantage largely collapses after instruction. Together, the studies identify a sophistication paradox: subtler dark patterns are less likely to trigger user suspicion and may therefore be harder to identify. These findings indicate that oversight should evaluate not just the presence but also the specific implementation of dark patterns.
\end{abstract}


\section{Introduction}
\label{sec:introduction}

Modern vehicles have become rolling data centers, continuously collecting and transmitting sensitive information about drivers, passengers, and their environments~\cite{mozilla2023carsprivacy,mozilla2023carsprivacyconcerns}. At the same time, the integration of full-featured operating systems such as Android Automotive OS (AAOS) has moved many privacy decisions from physical controls to software interfaces. As a result, the scope and granularity of in-vehicle data flows have dramatically expanded~\cite{androidautomotive_elevating2025,trijardcaddy2024enhancing}.

Meanwhile, research on manipulative interface design, commonly known as \textit{dark patterns}, has matured in e-commerce, web, and mobile contexts. Existing taxonomies by Gray \textit{et al}.~\cite{gray2018dark} and Brignull \emph{et al.}~\cite{brignull2023deceptive} document how interfaces can be designed to steer user choices and obtain consent that is neither fully informed nor freely given. In modern vehicles, this issue becomes important because extensive data collection is increasingly mediated through infotainment interfaces, a combination that creates a new and underexplored risk surface that existing dark-pattern research has not yet addressed~\cite{harrison2024privacy}.

\begin{figure}[t]
    \centering
    \begin{subfigure}[t]{\columnwidth}
        \centering
        \includegraphics[width=0.70\linewidth,height=3.2cm]{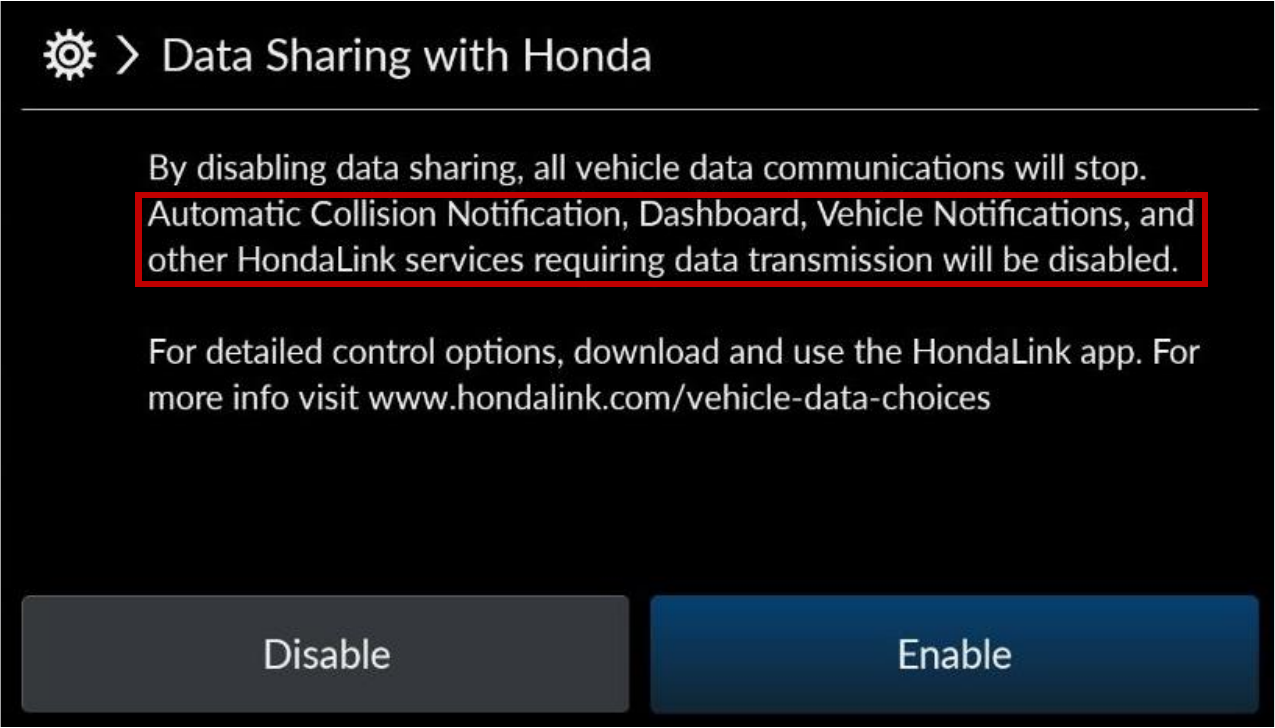}
        \caption{Bundling: opting out of data sharing also disables safety services such as Automatic Collision Notification.}
    \end{subfigure}
    \\[4pt]
    \begin{subfigure}[t]{\columnwidth}
        \centering
        \includegraphics[width=0.70\linewidth,height=3.2cm]{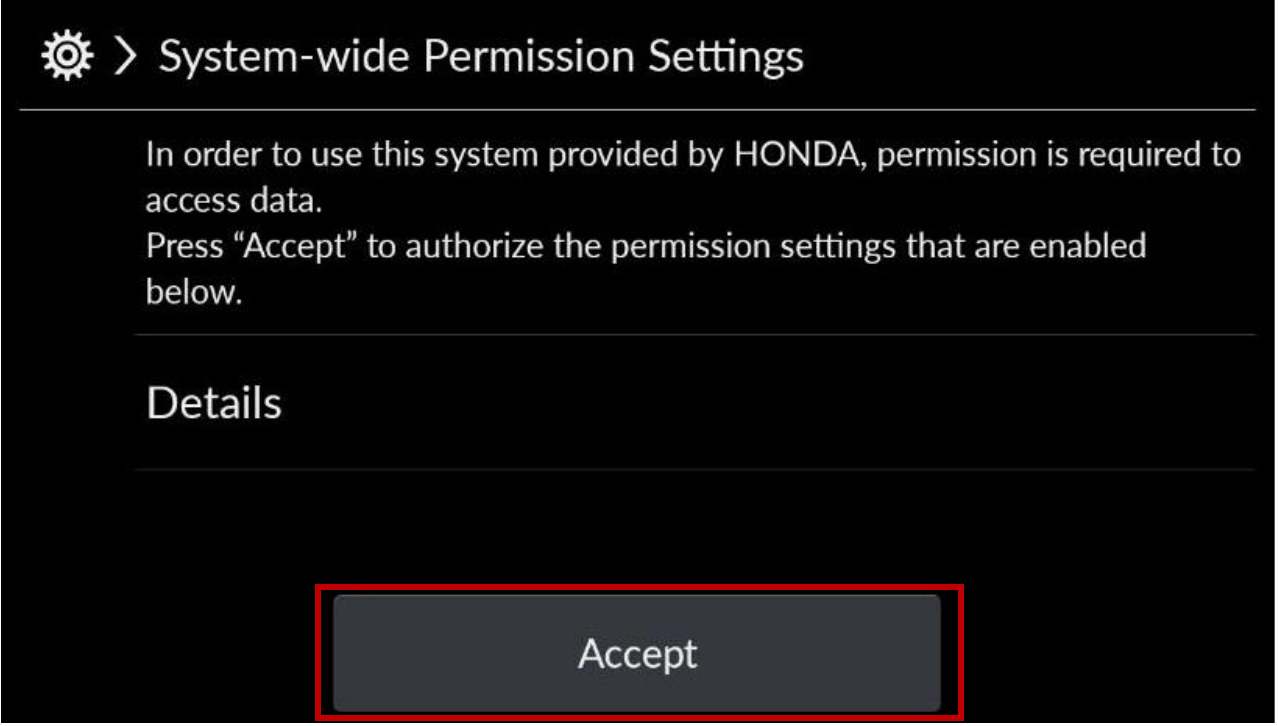}
        \caption{Forced Consent: the only available action is \emph{Accept}, with no decline or skip option.}
    \end{subfigure}
    \caption{Examples of dark patterns found in AAOS interfaces, captured from real manufacturer emulators. Red boxes highlight the specific UI elements that constitute each pattern.}
    \label{dp_car_pattern}
    \vspace{-15pt}
\end{figure}

Despite this emerging risk, the presence, implementation, and user perception of dark patterns in vehicle infotainment systems remain poorly understood. First, it is unclear which dark-pattern categories from these web and mobile taxonomies are actually present in AAOS and how consistently they do so across OEM builds. This gap is already visible in practice. For example, \autoref{dp_car_pattern} shows examples captured from manufacturer emulators: Bundling, where data permissions are tied to core vehicle features, and Forced Consent, where the interface offers no meaningful decline option.

While prior work has examined automotive privacy policies~\cite{gozubuyuk2024analyzing}, no prior study has systematically audited AAOS, or any production in-vehicle infotainment platform, for dark patterns at the UI level. Second, it is unknown whether manufacturers' versions of the same dark pattern lead users to perceive and respond to it differently. Third, no work has examined whether users can detect these patterns without instruction, nor how a brief educational intervention changes their perceptions or cross-domain comparisons with e-commerce. These three gaps motivate the following research questions:
\Needspace{6\baselineskip}
\begin{itemize}
    \item \emph{\textbf{RQ1:}} Which dark-pattern categories from existing web and mobile taxonomies manifest in AAOS-based vehicle infotainment interfaces, and how do their prevalence and distribution vary across OEM builds?
    \item \emph{\textbf{RQ2:}} How do different dark patterns and their OEM-specific implementations influence users' perceptions and intended behavioral responses?
    \item \emph{\textbf{RQ3:}} To what extent can users recognize dark patterns in automotive infotainment interfaces, and how does instruction alter their perceptions across automotive and e-commerce domains?
\end{itemize}

We address these gaps through a three-phase study that applies the taxonomies of Gray \textit{et~al.}~\cite{gray2018dark} and Brignull \emph{et al.}~\cite{brignull2023deceptive} to AAOS. In \textbf{Phase~1}, 26 HCI and privacy annotators inspect AAOS emulators from four manufacturers (Honda, Volvo, GM, and Polestar), producing the first systematic audit of dark patterns in AAOS. The audit documents 17 dark-pattern categories across 65 observed instances. These patterns are not limited to a single manufacturer. Instead, similar dark-pattern categories appear across multiple OEM builds, suggesting that manipulative interface design may be a broader issue across vehicle infotainment systems rather than a manufacturer-specific problem.

\textbf{Phase~2} includes 100 participants and uses the Phase~1 audit to select five dark patterns that appear across GMC, Honda, and Volvo AAOS builds. Each OEM implementation is recorded as a short screen recording, and each participant views one OEM version of each pattern, allowing cross-OEM differences to reflect implementation choices rather than pattern type. The results show that the same dark-pattern category can be implemented differently across OEMs, and that these implementation differences can affect user perceptions.

In \textbf{Phase~3}, 49 independent participants rate 14 clips, seven automotive and seven e-commerce, before and after a brief dark-pattern tutorial. This design examines both pattern recognition and how instruction changes perceptions across domains. Before instruction, participants rate automotive interfaces higher than e-commerce on ten of eleven attitudes, giving automotive interfaces a small but reliable benefit of the doubt. Afterward, this gap narrows substantially.

This paper makes the following main contributions:

\begin{itemize}
\item 
This paper presents the first systematic audit of AAOS interfaces for dark patterns, applying existing taxonomies across four OEM emulators (Honda, Volvo, GM, and Polestar). With 26 annotators, the audit documents 17 dark-pattern types across 65 observed instances.

\item 
We conduct the first empirical comparison of how users perceive OEM-specific implementations of the same dark patterns. In a study with 100 participants, Honda's Bundling implementation rates worse than GMC's on four attitudes, while uniformly coercive or hidden patterns such as Forced Consent and Sneaking show little cross-OEM variation.

\item 
We conduct a pre/post-tutorial study with an independent sample of 49 participants, comparing automotive and e-commerce dark-pattern clips. The results show that automotive's rating advantage over e-commerce largely collapses after a brief tutorial and reverses on avoidance of unintentional sharing, where automotive is rated worse than e-commerce.

\item 
Together, the studies identify a sophistication paradox in which subtler dark patterns are less likely to trigger user suspicion. Thus, the patterns hardest to identify may also be the least likely to provoke resistance.

\end{itemize}




These results carry three practical takeaways. First, dark patterns in AAOS are not monolithic: the same pattern can produce different user perceptions depending on how an OEM implements it, suggesting that oversight should evaluate \textit{how} manipulative patterns are designed, not only whether they are present. Second, users' initial trust in automotive interfaces is fragile: a brief educational intervention largely closes the rating advantage automotive interfaces hold over e-commerce, motivating clearer disclosures and user-facing education. Third, subtle patterns are especially difficult for users to resist because they do not reliably trigger suspicion, suggesting that disclosure alone is insufficient.

\section{Background}
\label{sec:background}

\subsection{Dark Patterns}

Dark patterns are interface designs that steer, pressure, or mislead users into choices that may not align with their interests. These designs often benefit the service provider by making certain actions easier, more salient, or more difficult to avoid. Brignull first popularized the term to describe deceptive interface tactics in online services, including patterns such as ``Roach Motels,'' where users can easily enter a service or agreement but face substantial friction when trying to leave~\cite{brignull2023deceptive}.

Our analysis builds on two existing dark-pattern taxonomies rather than proposing a new one. Gray \textit{et al.}~\cite{gray2018dark} organize dark patterns into five higher-level strategies:

\begin{itemize}
    \item \textbf{Nagging}: repeatedly interrupting users with prompts or requests in order to wear down resistance.
    \item \textbf{Obstruction}: making user-beneficial actions, such as opting out of data collection, unnecessarily difficult.
    \item \textbf{Sneaking}: hiding, delaying, or disguising information so that users cannot make fully informed decisions.
    \item \textbf{Interface Interference}: manipulating the visual or structural presentation of choices, such as through preselected defaults or misleading button emphasis.
    \item \textbf{Forced Action}: requiring users to complete an unwanted action, such as account creation or data sharing, before they can proceed.
\end{itemize}

Brignull's catalog~\cite{brignull2023deceptive} serves as a complementary practitioner-oriented reference for established dark-pattern terminology. In this paper, Gray \textit{et al.}'s taxonomy~\cite{gray2018dark} provides the strategy-level organization used in our analysis, while Brignull's catalog serves as a complementary practitioner-oriented reference for established dark-pattern terminology. The two frameworks are partly overlapping: Gray \textit{et al.} explicitly build on and reorganize several earlier Brignull categories within broader strategies. We therefore use them together as complementary references rather than as separate abstract- and interface-level taxonomies. Prior dark-pattern research has focused predominantly on web-based and online-service interfaces, with comparatively limited attention to embedded or operating-system interfaces such as automotive infotainment systems. We therefore used the two partially overlapping taxonomies as complementary references during our exploratory audit of the automotive interface context (see Section~\ref{sec:phase1}).

Although dark patterns have been widely studied in web, mobile, and e-commerce settings, vehicle infotainment systems create a different interaction context. Drivers often cannot abandon the interface, switch platforms, or avoid consent flows without affecting access to vehicle features. As a result, dark patterns in vehicles may carry distinct privacy and autonomy risks, when consent decisions are tied to data collection, account setup, navigation, telematics, or connected services.

\subsection{Android Automotive OS}

Android Automotive OS (AAOS) is a standalone operating system that runs directly on a vehicle's head unit. Unlike Android Auto, which projects a smartphone interface onto the vehicle display, AAOS is embedded in the vehicle itself and can support core infotainment functions such as navigation, media, climate controls, and third-party applications. This distinction is important because AAOS is not merely a display layer; it is part of the vehicle's software environment.

From a privacy perspective, AAOS matters because it can interact with vehicle-level data through the Vehicle Hardware Abstraction Layer (VHAL). This integration allows the operating system and applications to access vehicle properties such as gear position, battery status, speed-related signals, and other telemetry, depending on system configuration and permissions. Prior work shows that such data flows can raise privacy risks when they are collected at high frequency, insufficiently disclosed, or combined with other information to infer sensitive user behavior~\cite{pese2023first}.

AAOS has also become increasingly important as manufacturers adopt software-defined vehicle architectures and integrate full-featured infotainment platforms into production vehicles. Major OEMs, including Volvo, Polestar, General Motors, Honda, Ford, and Stellantis, have adopted or announced AAOS-based systems~\cite{luxoft2025gasaaos,wikipedia_androidautomotive}. As these systems become more common, privacy decisions that were once handled through physical controls or isolated settings are increasingly mediated through software interfaces.

This makes AAOS a useful and timely setting for studying dark patterns in vehicles. Its interfaces combine three characteristics that are central to this paper: they mediate consent and privacy choices, they are implemented differently across OEM builds, and they operate in a context where users may depend on the system for vehicle-related functionality. These properties make AAOS an important case for examining whether dark-pattern taxonomies developed in web and mobile settings also apply to in-vehicle infotainment interfaces.


\section{Approach Overview}
\label{sec:approach}

\begin{figure}[t]
    \centering
    \includegraphics[width=\columnwidth]{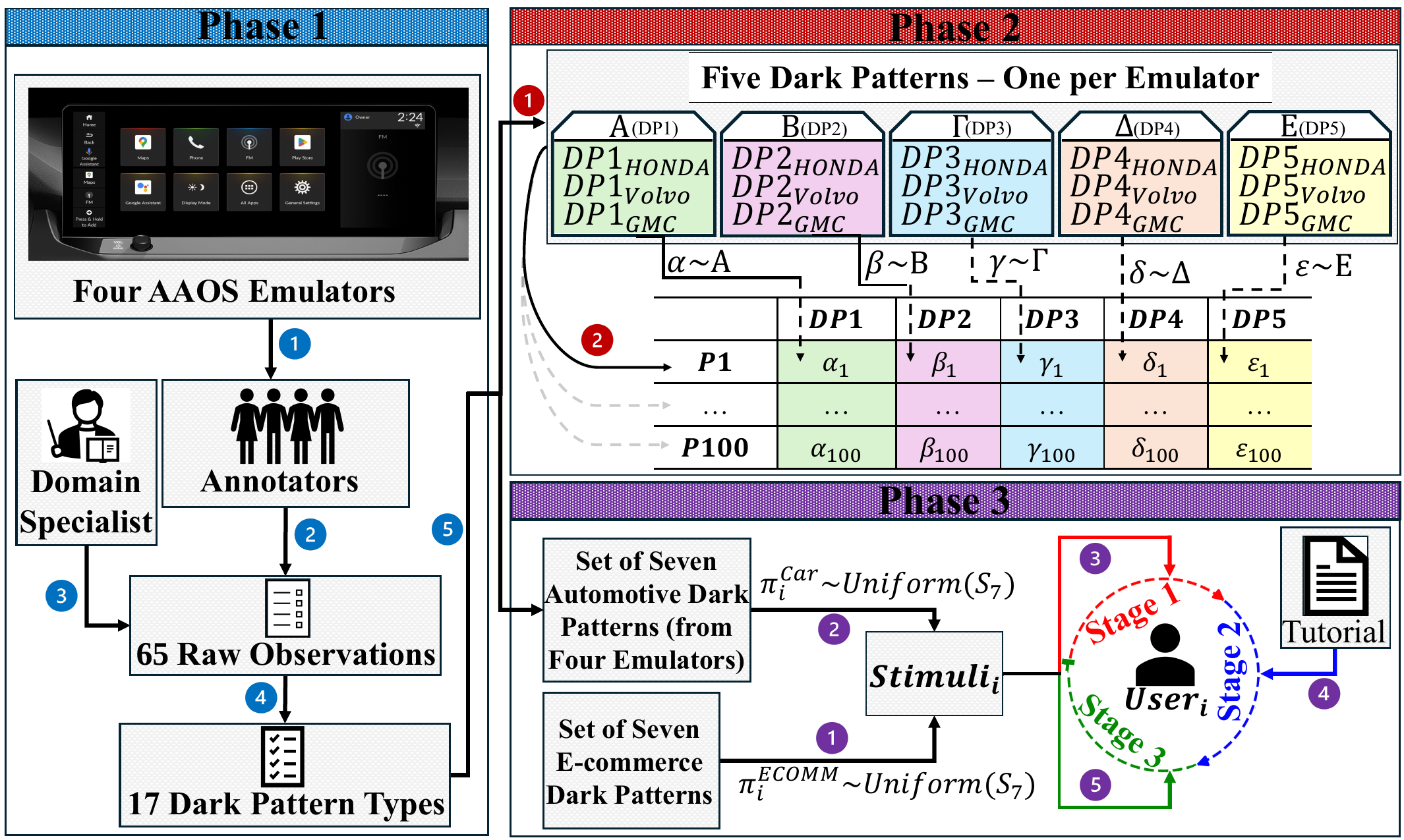}
    \caption{Overview of the three-phase methodology.
             Phase~1 documents which dark-pattern categories from existing taxonomies manifest in AAOS. Phase~2 uses the Phase~1 pattern catalog to select stimuli for a cross-OEM perception study comparing GMC, Honda, and Volvo implementations of five dark patterns. Phase~3 applies the same attitude scales in a pre/post-instruction study comparing automotive and e-commerce dark-pattern clips.}
    \label{fig:method}
    \vspace{-10pt}
\end{figure}

\autoref{fig:method} summarizes our three-phase methodology. Phase~1 addresses the presence and distribution of dark patterns in AAOS. Phase~2 examines how OEM-specific implementations shape user perception. Phase~3 evaluates pattern recognition and the effect of instruction across automotive and e-commerce interfaces.

\subsection{Phase 1: Audit of Dark Patterns in AAOS}

\noindent\textbf{RQ1: Presence and distribution.}
We conduct a human-annotated audit of AAOS emulator builds from four manufacturers: Honda, Volvo, GM, and Polestar. Annotators inspect each build for manipulative interface behaviors and document candidate dark patterns using the taxonomies of Gray \emph{et~al.}~\cite{gray2018dark} and Brignull \emph{et al.}~\cite{brignull2023deceptive}. The audit produces the dark-pattern catalog used to construct the stimuli for Phases~2 and~3.

\subsection{Phase 2: Cross-OEM Dark-Pattern Perception Study}

\noindent\textbf{RQ2: OEM implementation and user perception.}
Using the Phase~1 catalog, we select five dark patterns that appear across GMC, Honda, and Volvo AAOS builds.\footnote{Polestar's AAOS build did not implement all five selected dark patterns; we exclude it from Phase~2 to preserve valid cross-OEM comparisons.} For each selected pattern, we record one implementation from each OEM, yielding 15 screen-recorded stimuli in total. Participants view one OEM version of each dark pattern and rate the interaction on attitude and behavioral-intention measures, allowing us to compare how OEM-specific implementations shape user perception.

\subsection{Phase 3: Cross-Domain Perceptions and Instruction Effects}

\noindent\textbf{RQ3: Recognition and instruction effects.}
We conduct a within-subjects study in which participants rate 14 screen-recorded stimuli: seven automotive clips drawn from the AAOS audit and seven e-commerce clips matched by dark-pattern type. Participants rate the clips before and after a brief dark-pattern tutorial, enabling direct comparison of how instruction changes perception and recognition across automotive and e-commerce contexts.

\section{Phase 1: Dark-Pattern Categories in AAOS}
\label{sec:phase1}

This phase investigates \emph{RQ1} by conducting a systematic audit of AAOS implementations to document which manipulative design behaviors from the grounding taxonomies of Gray \textit{et al.}~\cite{gray2018dark} and Brignull \emph{et al.}~\cite{brignull2023deceptive} manifest in production vehicle infotainment systems, and to produce a codebook grounded in UI-level evidence for the empirical phases that follow.

\subsection{Audit Environment and Annotator Recruitment}

\noindent \textbf{Audit Environment.}
We leverage AAOS emulator builds from four major OEMs (Step~\cnumone{1}): Honda~\cite{honda2025carsapps}, Volvo~\cite{volvo2025emulator}, General Motors (GM)~\cite{gm2025invehicleapps}, and Polestar~\cite{polestar2025developer}. These OEMs were selected because they represented the primary production-grade, native AAOS implementations accessible to developers through official OEM portals at the time of study start. The concrete x86 emulator builds were obtained in April~2025. Appendix~\ref{app:emulators} reports their exact build numbers and screenshots. We validated the GMC and Honda emulator interfaces against production infotainment systems in a 2022 GMC Yukon and a 2023 Honda Accord Hybrid; details are provided in Appendix~\ref{app:emulator-validation}. Each emulator was configured to its factory default state with no user customization, ensuring that observed interface behaviors reflect manufacturer design choices rather than user preferences. GM required a free developer account; as of February~2026, its developer portal was no longer available.

\noindent \textbf{Apparatus \& Setup.} 
Annotators evaluated the four aforementioned emulator builds via Android Studio and were provided with URLs of the four builds and explicit instructions on how to download and run each emulator on their personal computers or workstations. Regardless of the hardware used, all annotators accessed the same emulator builds. This setup allowed annotators to perform unconstrained, real-time navigation of the operating system and capture documentary evidence of manipulative UI behaviors using desktop screen-recording and screenshot software. Annotators submitted 10--30-second video clips and high-resolution screenshots.

\noindent \textbf{Annotators.}
To improve coverage and minimize individual bias, we opted for an annotation methodology which involved recruiting 26 research-lab members with specialized backgrounds in Human-Computer Interaction (HCI), usability inspection, privacy, and automotive security. Participation was optional, and no monetary compensation was provided. We note that individual annotations were later followed by post-hoc verification by two authors and an automotive-privacy domain specialist to ensure consistency and resolve ambiguities (detailed below), with the reviewers serving as a consensus and reconciliation committee for automotive/privacy-specific implementations and boundary cases.

They did not generate observations, guide annotator exploration, or discard data. Before the audit, annotators received foundational training distinguishing dark patterns from anti-patterns~\cite{widdicks2020backfiring} (unintentional poor design). We did not provide annotators with a fixed category codebook. Instead, they received three instructional resources: one primary training deck and two supplementary further-reading resources. Prior to the audit, annotators completed a training module covering cybersecurity and privacy principles, culminating in a presentation on UX privacy risks and dark-pattern taxonomies. As reference material, annotators were provided with two resources: (1) \emph{Dark Patterns--Unfriendly UX Practices} by Zuleima Cota (UArizona)~\cite{cota_darkpatterns}, and (2) \emph{Dark Patterns--Manipulative UX Design and the Role of Regulation} by FPF~\cite{fpf_darkpatterns}. 

The primary deck introduced seven illustrative dark-pattern examples, while the supplementary materials provided additional examples, definitions, and characteristics of deceptive interfaces. These examples were intended as a non-exhaustive foundation rather than a closed set of permissible labels. Annotators could also report dark patterns not listed in the provided materials, and some submitted observations were ultimately mapped to categories that were not explicitly listed in those materials, including Preselection, Permission Creep, Illusion of Control, Illusion of Choice, Interface Interference, Identical Buttons, Obstruction, False Promises, and Upgrade Nagging.

\begin{table}[t]
\centering
\scriptsize
\setlength{\tabcolsep}{3pt}
\renewcommand{\arraystretch}{1.05}
\caption{Dark Pattern Distribution and Prevalence by OEM}
\label{tab:full_pattern_by_oem}

\begin{tabular}{@{} l c c c c c @{}}
\toprule
\rowcolor{white}
\textbf{Pattern Type} &
\textbf{\shortstack{Honda\\($n=34$)}} &
\textbf{\shortstack{Volvo\\($n=16$)}} &
\textbf{\shortstack{GMC\\($n=13$)}} &
\textbf{\shortstack{Polestar\\($n=2$)}} &
\textbf{Total} \\
\midrule

\rowcolor{rowgray}
Preselection
& 5 (14.7\%) & 5 (31.3\%) & 3 (23.1\%) & 0 (0.0\%) & 13 \\

\rowcolor{white}
Forced Consent
& 9 (26.5\%) & 1 (6.3\%) & 1 (7.7\%) & 1 (50.0\%) & 12 \\

\rowcolor{rowgray}
Privacy Zuckering
& 3 (8.8\%) & 1 (6.3\%) & 1 (7.7\%) & 1 (50.0\%) & 6 \\

\rowcolor{white}
Confirmshaming
& 3 (8.8\%) & 2 (12.5\%) & 0 (0.0\%) & 0 (0.0\%) & 5 \\

\rowcolor{rowgray}
Permission Creep
& 2 (5.9\%) & 1 (6.3\%) & 2 (15.4\%) & 0 (0.0\%) & 5 \\

\rowcolor{white}
Bundling Consent
& 2 (5.9\%) & 1 (6.3\%) & 1 (7.7\%) & 0 (0.0\%) & 4 \\

\rowcolor{rowgray}
Roach Motel
& 1 (2.9\%) & 1 (6.3\%) & 2 (15.4\%) & 0 (0.0\%) & 4 \\

\rowcolor{white}
Obstruction
& 3 (8.8\%) & 0 (0.0\%) & 0 (0.0\%) & 0 (0.0\%) & 3 \\

\rowcolor{rowgray}
Hidden Costs
& 1 (2.9\%) & 1 (6.3\%) & 1 (7.7\%) & 0 (0.0\%) & 3 \\

\rowcolor{white}
Illusion of Control
& 1 (2.9\%) & 1 (6.3\%) & 1 (7.7\%) & 0 (0.0\%) & 3 \\

\rowcolor{rowgray}
Sneaking
& 1 (2.9\%) & 0 (0.0\%) & 0 (0.0\%) & 0 (0.0\%) & 1 \\

\rowcolor{white}
Misdirection
& 1 (2.9\%) & 0 (0.0\%) & 0 (0.0\%) & 0 (0.0\%) & 1 \\

\rowcolor{rowgray}
Illusion of Choice
& 1 (2.9\%) & 0 (0.0\%) & 0 (0.0\%) & 0 (0.0\%) & 1 \\

\rowcolor{white}
Interface Interference
& 1 (2.9\%) & 0 (0.0\%) & 0 (0.0\%) & 0 (0.0\%) & 1 \\

\rowcolor{rowgray}
False Promises
& 0 (0.0\%) & 1 (6.3\%) & 0 (0.0\%) & 0 (0.0\%) & 1 \\

\rowcolor{white}
Identical Buttons
& 0 (0.0\%) & 1 (6.3\%) & 0 (0.0\%) & 0 (0.0\%) & 1 \\

\rowcolor{rowgray}
Upgrade Nagging
& 0 (0.0\%) & 0 (0.0\%) & 1 (7.7\%) & 0 (0.0\%) & 1 \\

\bottomrule
\end{tabular}
\vspace{-10pt}
\end{table}

\subsection{Annotation Protocol}

The transition from raw observations to a validated codebook followed an unconstrained, exploratory protocol designed to capture the diverse ways manipulative design manifests in a vehicle's software ecosystem. Two authors brainstormed a preliminary protocol that was refined after piloting. The final protocol is detailed as follows.

\noindent \textbf{Exploratory Audit Protocol.} Annotators independently labeled observable instances of potentially manipulative interface behavior (Step~\cnumone{2}). During post-hoc review, the two authors and the specialist evaluated the submitted observations based on asymmetrical benefit. If an interface element introduced friction to privacy-protective actions while smoothing data extraction or service retention for the OEM, they retained the observation as a dark pattern. If an interface was clunky or confusing without benefiting the service provider over the user, they categorized the observation as unintentional poor usability (an anti-pattern) and excluded it from our analysis. Annotators were not instructed in how to approach identification, so as to reduce bias; for example, no systematic walkthroughs through menus were given. No fixed time limits were given to annotators and they were instructed to spend as much time as they think is appropriate with the emulators.

They could self-select which emulators to examine and were not required to examine every build. 19 annotators audited one emulator, three audited two, and four audited three; in total, Honda, Volvo, GMC, and Polestar were examined by 18, 11, 6, and 2 annotators, respectively. Table~\ref{tab:annotator-emulator-pattern} (Appendix~\ref{app:annotator-coverage}) reports the examined emulators and the patterns identified by each annotator, showing overlap most notably for Forced Consent in Honda and Preselection across Honda, Volvo, and GMC.

\noindent \textbf{Documentation Requirements.} Each reported observation had to include visual evidence and a written explanation of why the behavior was potentially manipulative or deceptive. After the annotations were collected, the automotive-privacy domain specialist reviewed the annotators' labels, descriptions, and supporting evidence against the categories and examples represented in the provided materials and the taxonomies of Gray \textit{et al.}~\cite{gray2018dark} and Brignull \emph{et al.}~\cite{brignull2023deceptive}. Some observations mapped directly to an existing category, whereas others were described in plain language or represented an automotive/privacy-specific implementation. In those cases, the specialist assigned an operational label based on the observable interface behavior and linked it to the closest Gray strategy or Brignull pattern (Step~\cnumone{3}).

\noindent \textbf{Verification.}
The combined dataset comprised 65 unique observations spanning 17 dark-pattern types (Step~\cnumone{4}). Because annotators inspected the emulators independently and reported only the instances they identified, rather than labeling a shared fixed set of UI events, a single inter-rater agreement coefficient over the full corpus is not well defined. Instead, all candidate observations were jointly reviewed by the aforementioned three authors on the research team (researchers in the security and privacy of automotive vehicles). Researchers jointly reviewed the documentation for each dark pattern, and any disagreements were reconciled.

\begin{table}[tbp]
\centering
\scriptsize
\setlength{\tabcolsep}{4pt}
\renewcommand{\arraystretch}{1.05}
\caption{Pattern co-occurrence matrix (number of annotator--emulator audits identifying both patterns).}
\label{tab:cooccurrence}

\begin{tabularx}{\columnwidth}{@{} >{\raggedright\arraybackslash}X c c c c c c c @{}}
\toprule
\rowcolor{white}
\textbf{Pattern} & \textbf{Pre} & \textbf{FC} &
\textbf{PZ} & \textbf{CS} & \textbf{PC} &
\textbf{BC} & \textbf{RM} \\
\midrule
\rowcolor{rowgray}
Preselection (Pre)     & 13 & 3  & 0 & 3 & 5 & 0 & 0 \\
\rowcolor{white}
Forced Consent (FC)    & 3  & 12 & 3 & 3 & 1 & 0 & 1 \\
\rowcolor{rowgray}
Privacy Zuckering (PZ) & 0  & 3  & 6 & 0 & 0 & 0 & 0 \\
\rowcolor{white}
Confirmshaming (CS)    & 3  & 3  & 0 & 5 & 0 & 1 & 0 \\
\rowcolor{rowgray}
Permission Creep (PC)  & 5  & 1  & 0 & 0 & 5 & 0 & 0 \\
\rowcolor{white}
Bundling Consent (BC)  & 0  & 0  & 0 & 1 & 0 & 4 & 3 \\
\rowcolor{rowgray}
Roach Motel (RM)       & 0  & 1  & 0 & 0 & 0 & 3 & 4 \\
\bottomrule
\end{tabularx}
\vspace{-15pt}
\end{table}



\begin{figure}[t]
    \centering
    \begin{subfigure}[t]{\columnwidth}
        \centering
        \includegraphics[width=0.70\linewidth,height=3.2cm]{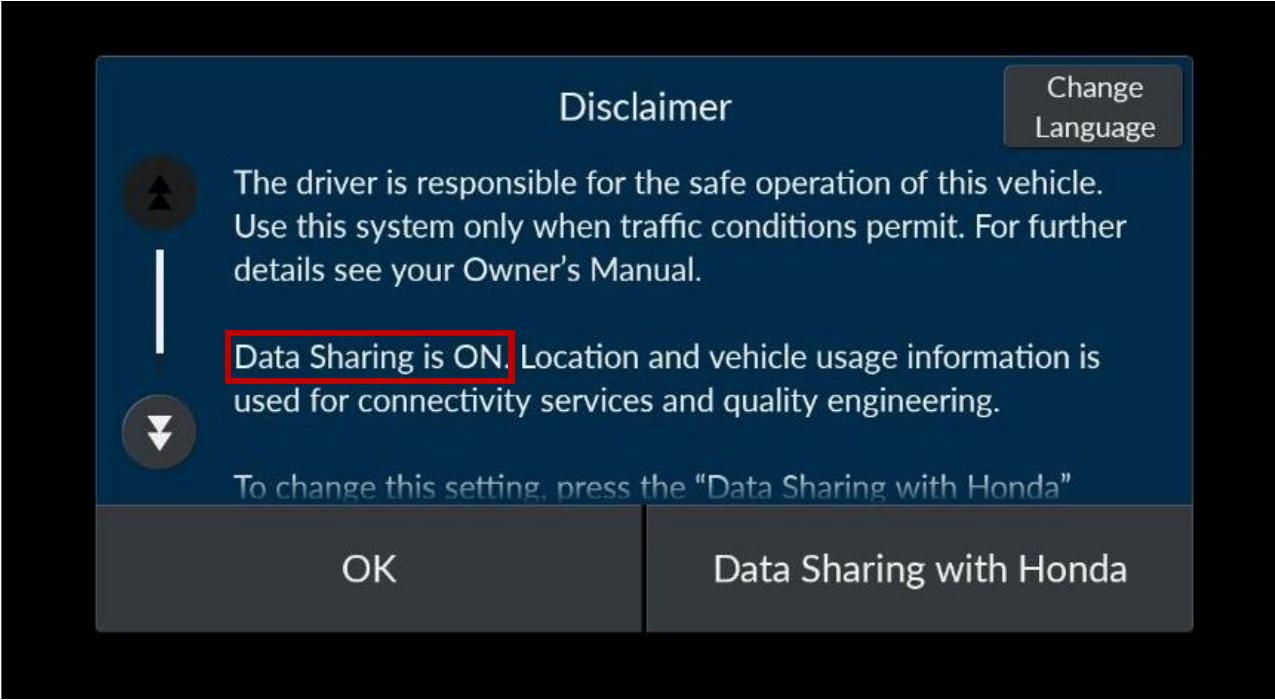}
        \caption{Sneaking: a disclaimer informs the user that ``Data Sharing is ON'' by default, without prior explicit consent.}
    \end{subfigure}
    \caption{An additional example of a dark pattern found in AAOS interfaces, captured from a real manufacturer emulator. Red boxes highlight the specific UI elements that constitute the pattern.}
    \label{dp_car_pattern_extra}
    \vspace{-15pt}
\end{figure}

\subsection{Data Analysis}

\autoref{tab:full_pattern_by_oem} presents the full distribution of all 17 identified pattern types, showing both absolute counts and relative prevalence based on the total observations per emulator ($n$). We characterize the distribution of these patterns across the corpus and across the evaluated OEMs, and identify common pattern co-occurrences that suggest layered manipulative strategies.
Illustrative examples of three dark patterns are depicted in Figures~\ref{dp_car_pattern} and \ref{dp_car_pattern_extra}.

\noindent\textbf{Annotator Discovery Convergence.}
We report descriptive discovery convergence for pattern--OEM combinations in which the pattern was observed at least once, as the proportion of annotators who identified the pattern among those who audited that emulator. Preselection showed comparatively broad overlap across Honda, Volvo, and GMC, while Forced Consent showed high overlap for Honda. Privacy Zuckering and Sneaking showed lower overlap across the more extensively audited OEMs. Discovery-convergence percentages are visualized in Appendix~\ref{app:annotator-discovery-convergence}, Figure~\ref{fig:annotator-discovery-convergence}. We distinguish pattern--OEM combinations with no observed occurrence from those with at least one observed occurrence; combinations with no observed occurrence are shown as \emph{N/O} (not observed).

\noindent \textbf{Pattern Prevalence.}
We summarize how often each pattern type appeared in the 65 observations and, as a coarse descriptive check, compare these frequencies against a uniform baseline in which each of the 17 patterns would appear with equal probability ($1/17 \approx 5.9\%$). We note that this baseline provides a simple reference point for which tactics are the most prevalent within the corpus, as detailed below:

\begin{itemize}
    \item \textbf{Preselection}: identified in 20.0\% of observations, the most frequently observed pattern (exact binomial $p < 0.001$ against the uniform reference).
    \item \textbf{Forced Consent}: appeared in 18.5\% of instances, the second most frequent pattern ($p < 0.001$).
    \item \textbf{Privacy Zuckering}: observed in 9.2\% of cases, above the uniform reference but not significantly so ($p = 0.183$).
\end{itemize}

\noindent \textbf{Cross-OEM Distribution.}
While patterns such as \textit{Preselection} and \textit{Forced Consent} appear across multiple OEMs, Honda's build surfaces the broadest range of pattern types (fourteen categories across $n = 34$ observations), reflecting the richer set of consent flows, account-provisioning steps, and permission dialogs exposed in its emulator. Polestar, in contrast, exhibits a ``minimalist but restrictive'' pattern mix in which the few identified patterns ($n = 2$) fall entirely into categories such as \textit{Forced Consent} and \textit{Privacy Zuckering}. A chi-square test grouping patterns into Preselection, Forced Consent, and Other ($\chi^{2}(6) = 6.53$, $p = 0.367$) did not detect significant brand-specific variation in the distribution of pattern groups. The data are therefore consistent with an industry-wide pattern of manipulative choice architecture rather than a manufacturer-specific phenomenon. Interestingly, 5 of 26 annotators could not identify any dark patterns at all which indicates that even individuals with specialized training in HCI or privacy and knowledge about privacy dark patterns may have difficulties detecting such patterns in interfaces.

\noindent \textbf{Pattern Co-Occurrence.} Manipulative interface tactics rarely appeared as isolated events; instead, they were often strategically layered to amplify user harm and eliminate opt-out opportunities. To analyze these relationships, we constructed a co-occurrence matrix shown in \autoref{tab:cooccurrence}, uncovering several noteworthy pairings:

\begin{itemize}[noitemsep, topsep=0pt]
    \item \textbf{Preselection + Forced Consent} ($n = 3$): a ``double-tap'' strategy, where the system first nudges the user via a pre-selected default and then effectively mandates that choice by making progression contingent on consent.
    \item \textbf{Privacy Zuckering + Forced Consent} ($n = 3$): notably present in the Polestar and Honda emulators, where users were tricked into sharing more data than intended while simultaneously being told that refusal would disable core features.
    \item \textbf{Bundling Consent + Roach Motel} ($n = 3$): often found in account-setup flows, where these patterns force users to accept multiple data-sharing practices at once and then make it intentionally difficult to unbundle or revoke those permissions later in the settings menu.
\end{itemize}

While Bundling Consent and Roach Motel co-occur only three times, they share a Jaccard similarity coefficient higher than their absolute frequency suggests, as they frequently appear together in the same specific sub-menus (e.g., account settings). This indicates that while they are not the most frequent pairing across the entire OS, they form a specialized ``cluster'' used for data extraction and retention.

\begin{tcolorbox}[colback=phaseonecolor!6,colframe=phaseonecolor!70!black,
  boxrule=0.8pt,arc=2pt,left=6pt,right=6pt,top=4pt,
  bottom=4pt]
\textbf{Finding 1.}
The audit confirmed that 17 distinct dark-pattern categories from existing taxonomies manifest in AAOS, with \textit{Preselection} and \textit{Forced Consent} as the two most frequently observed patterns. The cross-OEM distribution showed no evidence of brand-specific design philosophies, consistent with an industry-wide rather than manufacturer-specific phenomenon.
\end{tcolorbox}


\begin{table*}[t]
\centering
\caption{Mean attitude scores (1--7) for GMC (G), Honda (H),
         and Volvo (V) within each dark pattern ($N{=}100$,
         32--35 per OEM group).
         \colorbox{signeg}{Red} = Bonferroni-significant lower than GMC
         (linear regression per dark pattern, Bonferroni $\alpha = 0.025$
         for the two GMC contrasts).
         \colorbox{siguncorr}{Orange} = significant at uncorrected
         $\alpha = 0.05$ but not surviving Bonferroni.
         $^{*}p{<}0.05$, $^{**}p{<}0.01$, $^{***}p{<}0.001$.
         White cells are non-significant.}
\label{tab:phase2_means}
\scriptsize
\renewcommand{\arraystretch}{1.05}
\resizebox{\textwidth}{!}{%
\begin{tabular}{@{}
  p{2.4cm}
  ccc ccc ccc ccc ccc
@{}}
\toprule
& \multicolumn{3}{c}{\textbf{Bundling}}
& \multicolumn{3}{c}{\textbf{Confirmshaming}}
& \multicolumn{3}{c}{\textbf{Forced Consent}}
& \multicolumn{3}{c}{\textbf{Privacy Zuckering}}
& \multicolumn{3}{c}{\textbf{Sneaking}} \\
\cmidrule(lr){2-4}\cmidrule(lr){5-7}
\cmidrule(lr){8-10}\cmidrule(lr){11-13}
\cmidrule(lr){14-16}
\textbf{Attitude}
& \textbf{G} & \textbf{H} & \textbf{V}
& \textbf{G} & \textbf{H} & \textbf{V}
& \textbf{G} & \textbf{H} & \textbf{V}
& \textbf{G} & \textbf{H} & \textbf{V}
& \textbf{G} & \textbf{H} & \textbf{V} \\
\midrule
\rowcolor{rowgray} Fairness
  & 4.06 & 3.15 & 3.97
  & 4.24 & 4.42 & 3.64
  & 3.29 & 3.24 & 3.47
  & 3.30 & 3.85 & 3.44
  & 3.70 & 3.88 & 3.65 \\
\rowcolor{white} Honesty
  & 3.94 & 3.33 & 4.03
  & 4.06 & 4.39 & 3.67
  & 3.34 & 3.58 & 3.69
  & 3.12 & 3.76 & 3.35
  & 3.70 & 3.67 & 3.59 \\
\rowcolor{rowgray} Control
  & 4.97
  & \cellcolor{signeg}3.33$^{***}$
  & \cellcolor{siguncorr}4.00$^{*}$
  & 4.38 & 5.15 & 4.18
  & 3.37 & 3.18 & 3.38
  & 3.76 & 4.27 & 4.12
  & 4.58 & 3.91 & 4.18 \\
\rowcolor{white} Privacy
  & 4.18
  & \cellcolor{siguncorr}3.18$^{*}$
  & 3.82
  & 4.32 & 4.30 & 3.67
  & 3.40 & 3.21 & 3.25
  & 3.24 & 3.61 & 3.29
  & 3.55 & 3.85 & 3.59 \\
\rowcolor{rowgray} Intuitiveness
  & 4.19 & 3.73 & 3.73
  & 4.45 & 4.56
  & \cellcolor{signeg}3.58$^{*}$
  & 4.18 & 3.53 & 3.65
  & 3.13 & 3.77 & 3.59
  & 4.11 & 3.53 & 3.53 \\
\rowcolor{white} Transparency
  & 4.19
  & \cellcolor{signeg}3.17$^{*}$
  & 3.68
  & 4.38 & 4.35 & 3.72
  & 3.88 & 3.37 & 3.75
  & 3.66 & 3.74 & 3.28
  & 3.97 & 4.06 & 3.61 \\
\rowcolor{rowgray} Avoids Unint.\ Shar.
  & 3.45 & 3.15 & 3.47
  & 3.62 & 3.27 & 3.73
  & 2.94 & 3.03 & 3.31
  & 2.82 & 2.94 & 2.97
  & 2.73 & 2.82 & 2.82 \\
\rowcolor{white} Trust
  & 3.88
  & \cellcolor{siguncorr}2.94$^{*}$
  & 3.56
  & 3.94 & 3.82 & 3.30
  & 3.09 & 2.94 & 3.16
  & 2.97 & 3.45 & 3.09
  & 3.15 & 3.52 & 3.24 \\
\rowcolor{rowgray} Competence
  & 4.91
  & \cellcolor{signeg}3.97$^{*}$
  & \cellcolor{siguncorr}4.09$^{*}$
  & 4.50 & 4.70 & 4.24
  & 4.20 & 3.85 & 4.41
  & 4.24 & 4.36 & 4.26
  & 4.36 & 4.12 & 4.24 \\
\rowcolor{white} Emotion
  & 4.12 & 3.39 & 3.59
  & 3.79 & 3.91 & 3.21
  & 3.26 & 3.06 & 3.06
  & 3.15 & 3.76 & 3.29
  & 3.52 & 3.52 & 3.41 \\
\rowcolor{rowgray} Confidence
  & 4.36
  & \cellcolor{signeg}3.33$^{*}$
  & 3.74
  & 4.29 & 4.58 & 3.64
  & 3.40 & 3.42 & 3.62
  & 3.06
  & \cellcolor{siguncorr}3.88$^{*}$
  & 3.35
  & 4.03 & 3.61 & 3.38 \\
\bottomrule
\multicolumn{16}{l}{%
  \footnotesize G\,=\,GMC (reference), H\,=\,Honda,
  V\,=\,Volvo. Higher scores = more positive/user-favourable.}\\
\end{tabular}}
\vspace{-10pt}
\end{table*}

\begin{table}[tbp]
\centering
\scriptsize
\setlength{\tabcolsep}{2pt}
\renewcommand{\arraystretch}{0.88}
\renewcommand{\tabularxcolumn}[1]{m{#1}}

\caption{Definitions of the eleven attitude constructs.}
\label{tab:attitude-constructs}

\begin{tabularx}{\columnwidth}{@{}
>{\raggedright\arraybackslash}m{0.20\columnwidth}   
>{\raggedright\arraybackslash}m{0.50\columnwidth}   
>{\centering\arraybackslash}m{0.23\columnwidth}     
@{}}
\toprule

\textbf{Construct} &
\textbf{What It Measures} &
\textbf{Scale Endpoints (1--7)} \\
\midrule

\rowcolor{rowgray}
Fairness &
Whether the interface treats the user fairly or unfairly. &
Unfair $\ldots$ Fair \\

Honesty &
Whether the interface comes across as deceptive or honest. &
Deceptive $\ldots$ Honest \\

\rowcolor{rowgray}
Control &
How much control the user feels they have over the interaction. &
No Control $\ldots$ Full Control \\

Privacy Respect &
Whether the interface respects or invades the user's privacy. &
Invasive $\ldots$ Respectful \\

\rowcolor{rowgray}
Ease of Use &
Whether the interaction feels confusing or intuitive. &
Confusing $\ldots$ Intuitive \\

Transparency &
How clearly the design reveals what it is doing. &
Opaque $\ldots$ Transparent \\

\rowcolor{rowgray}
Avoidance of Unintentional Sharing &
How likely the user is to share private information without meaning to. &
Very Likely $\ldots$ Very Unlikely \\

Trust &
Whether the interaction increases or decreases trust in the manufacturer. &
Much Less Trust $\ldots$ Much More Trust \\

\rowcolor{rowgray}
Competence &
How competent the manufacturer appears based on the interface. &
Not at All Competent $\ldots$ Very Competent \\

Emotional Response &
Whether the interaction feels frustrating or supportive. &
Very Frustrating $\ldots$ Very Supportive \\

\rowcolor{rowgray}
Confidence &
How doubtful or confident the user feels during the interaction. &
Very Doubtful $\ldots$ Very Confident \\

\bottomrule
\end{tabularx}

\vspace{1pt}
{\scriptsize
\raggedright
All eleven constructs were measured using seven-point semantic-differential scales developed for this study.\par
}
\vspace{-15pt}
\end{table}

\section{Phase 2: Cross-OEM Perceptions of Dark-Pattern Implementations}
\label{sec:phase2}

Phase~2 examines whether OEM-specific implementations of the same dark-pattern categories influence users' perceptions and intended behavioral responses, addressing \emph{RQ2}. Participants evaluated each interaction using eleven attitude constructs, which are defined in Table~\ref{tab:attitude-constructs}. We decompose this question into three sub-research questions:

\noindent\textbf{RQ2.1 (OEM-Specific Implementation Effects):}
Within each dark-pattern type, do GMC, Honda, and Volvo implementations differ significantly in how users perceive them across the eleven attitude constructs?\\
\noindent\textbf{RQ2.2 (Perceived Dark-Pattern Severity):}
Across all manufacturers, which dark-pattern types produce the most
negative perceptions across the eleven attitude constructs?\\
\textbf{RQ2.3 (Reasoning and Behavioral Intention):}
Do participants' explanations of their design preferences and trust changes align with their stated intentions to engage with consent screens and privacy controls?

\subsection{Study Design}

\noindent\textbf{OEM scope.}
Following the Phase~1 audit, Polestar's AAOS build did not exhibit the full range of dark-pattern types identified in the other three OEMs. Because valid within-pattern cross-OEM comparisons require each participating OEM to implement all five selected patterns, we exclude Polestar from Phase~2 and retain GMC, Honda, and Volvo.

\noindent\textbf{Design.}
Phase~2 manipulates dark-pattern type within subjects, with OEM varied between subjects within each dark-pattern type. From the Phase~1 audit, we select five dark patterns that appear across GMC, Honda, and Volvo: Bundling, Confirmshaming, Forced Consent, Privacy Zuckering, and Sneaking~(Step~\cnumtwo{1}). We denote these patterns as $DP_1, \ldots, DP_5$. For each dark pattern $DP_k$, we produce one OEM-specific implementation from GMC, Honda, and Volvo, yielding the variant set $\{DP_{k,\text{GMC}}, DP_{k,\text{Honda}}, DP_{k,\text{Volvo}}\}$.

For every participant $P_i$ and every dark pattern $DP_k$, one OEM variant is drawn uniformly at random from the corresponding variant set~(Step~\cnumtwo{2}). This design ensures that each participant views exactly one OEM implementation of each dark-pattern type and never sees the same dark-pattern type more than once. As a result, within-pattern cross-OEM differences can be attributed to implementation choices rather than repeated exposure to the same pattern category. The single-OEM-per-pattern design was adopted after a pilot version that asked participants to compare the three OEM implementations side by side proved cognitively demanding and prone to comparative anchoring. Details of the pilot study are presented in Appendix~\ref{app:pilot}.
The presentation order of the five dark patterns is independently sampled for each participant from the uniform distribution over permutations, $\sigma_i \sim \mathrm{Uniform}(S_5)$, to reduce order effects.

\noindent\textbf{Stimuli.}
Each stimulus is a short, silent screen-recorded video of a real AAOS interface interaction demonstrating one dark pattern from one OEM. Videos are similar in length, averaging approximately 30 seconds, and follow a consistent presentation style so that pacing does not affect responses. We produce 15 videos in total (5 dark patterns $\times$ 3 OEMs). Each video is reviewed by two members of the research team to confirm that the target pattern is clearly visible and that no additional dark patterns are inadvertently introduced.

\noindent\textbf{Measures.} After each video, participants completed the eleven semantic-differential ratings summarized in Table~\ref{tab:attitude-constructs}. After the five rating tasks, participants completed a post-survey with three open-ended prompts about manufacturer-side design choices, trust change, and consent-screen design, as well as two behavioral-intention items about intended consent-screen reading and per-data-type privacy toggles.

\noindent\textbf{Ethics.}
The study was reviewed by our institutional review board (IRB) and deemed exempt. We did not collect personally identifiable information beyond basic demographics; any incidental identifiers were removed, and all responses were stored in encrypted online storage.

\subsection{Recruitment}
We recruited 100 adult participants via Prolific~\cite{prolific}, and administered the study through Qualtrics~\cite{qualtrics_website}. Participants were compensated at a rate of \$12.00 per hour. Each participant completed five video-rating tasks, yielding 500 observations in total. OEM assignment was balanced across participants, with 32 to 35 participants per OEM per dark pattern (mean 33.3). Of the 100 participants, 98 provided valid post-survey responses and form the base sample for RQ2.3. Participants spanned a range of age groups (26--35: 31\%, 36--45: 26\%, 46 and above: 29\%) and were 51\% female. In addition, 65\% reported some prior awareness of AAOS, although only 17\% reported current use. Detailed demographics appear in Appendix~\ref{app:demographics}.

\subsection{Data Analysis}
We analyze the Phase~2 data using quantitative models for RQ2.1 and RQ2.2 and qualitative thematic coding for RQ2.3. For RQ2.1, we fit a separate linear regression for each attitude construct $j = 1, \ldots, 11$ and each dark-pattern type, with OEM as a fixed effect:
\begin{equation*}
Y_{j} \sim C(\mathrm{OEM}).
\end{equation*}
Because each model contains one observation per participant, no random intercept is required. GMC serves as the reference OEM, yielding $5 \times 11 = 55$ primary models. Each model produces two contrasts of interest, Honda vs.\ GMC and Volvo vs.\ GMC. We apply a Bonferroni correction within each attitude at $\alpha = 0.05/2 = 0.025$, with no correction across the eleven attitudes. We report bootstrapped 95\% confidence intervals ($B = 1000$) for OEM means.

The raw data for RQ2.2 consisted of 500 participant--stimulus observations from 100 participants, each rating five dark-pattern stimuli on eleven seven-point semantic-differential scales. For RQ2.2, we fit a linear mixed-effects model for each attitude construct, with OEM, dark-pattern type, and their interaction as fixed effects and participant as a random intercept:
\begin{equation*}
Y_{j} \sim C(\mathrm{OEM}) * C(DP) + (1 \mid P_i).
\end{equation*}
Bundling ($DP_1$) and GMC are the reference levels. Effects of dark-pattern type estimate how each pattern shifts perceptions relative to Bundling for GMC, while interaction terms test whether the OEM contrasts, Honda vs.\ GMC and Volvo vs.\ GMC, differ across dark-pattern types. For attitudes where the interaction model fails to converge because of a singular matrix, we report main effects from the additive model.

For RQ2.3, we analyze the open-ended post-survey responses using inductive thematic coding. Two researchers independently code responses, with disagreements resolved through discussion. Inter-rater reliability reaches almost-perfect agreement on the manufacturer-design ($\kappa = 0.871$) and trust-change ($\kappa = 0.878$) prompts, and substantial agreement on the consent-screen-design prompt ($\kappa = 0.761$), with a mean $\kappa = 0.837$~\cite{landis1977measurement}. Codes are grouped into four design-reasoning themes (\emph{Transparency \& Clarity}, \emph{Control \& Fair Design}, \emph{Contextual \& Purposeful Consent}, \emph{Fatalism / No Change}) and three trust-change themes (\emph{Trust Decreased}, \emph{Trust Unchanged}, \emph{Awareness Raised}). Because the two design-reasoning prompts produce overlapping themes, we use only the manufacturer-perspective prompt. For RQ2.3, we supplemented the descriptive percentages with the inferential statistics. The relevant contingency-table counts, test statistics, corrected \(p\)-values, and effect sizes are reported in Appendix~\ref{app:rq2-3-contingency}.


\subsection{RQ2.1: OEM-Specific Implementation Effects}

\autoref{tab:phase2_means} shows construct means by OEM and dark pattern. Cross-OEM differences are not evenly distributed across dark patterns. They concentrate in patterns where the OEM has room to shape the consent flow, and disappear when the manipulation is structurally coercive or hidden from view. Bundling shows the clearest evidence that OEM implementation choices affect user perception. Honda's version performed worse than GMC's on four Bonferroni-significant attitudes, with the largest drop on \emph{Control} ($\beta = -1.636$, $p < 0.001$). This pattern suggests that participants were not only reacting to the presence of Bundling, but also to how the bundled consent request was presented. Notably, Honda's implementation appears to make the consent flow feel more difficult to control, less transparent, and less confidence-inducing than GMC's. Volvo followed the same direction on some attitudes, but with smaller and mostly uncorrected effects. 

The remaining patterns show much weaker OEM differentiation. Confirmshaming produced only one Bonferroni-significant contrast, with Volvo rated lower than GMC on \emph{Intuitiveness}. Privacy Zuckering produced no Bonferroni-significant OEM differences, despite one uncorrected increase in \emph{Confidence} for Honda relative to GMC. Forced Consent and Sneaking produced no significant OEM contrasts at all. These null effects are informative. 

Forced Consent offers no meaningful choice, and Sneaking conceals the relevant action; in both cases, the pattern itself dominates the experience, leaving little room for manufacturer-specific presentation to change user perception. Privacy Zuckering, by contrast, may produce little OEM variation because its manipulation is harder for users to notice, reflecting the \emph{sophistication paradox}, whereby subtler dark patterns are less likely to trigger user suspicion.

\begin{tcolorbox}[colback=phasetwocolor!8,colframe=phasetwocolor!70!black,
  boxrule=0.8pt,arc=2pt,left=6pt,right=6pt,top=4pt,
  bottom=4pt]
\textbf{Finding 2.}
OEM differences emerge when dark patterns leave room for interface-level variation. When the manipulation is either obvious or hidden from view, users respond similarly across manufacturers.
\end{tcolorbox}

\subsection{RQ2.2: Perceived Dark-Pattern Severity Across Manufacturers}

To assess relative perceived severity across manufacturers, the analysis compares how each dark-pattern type affects user perceptions across the eleven attitude constructs. Across manufacturers, Forced Consent produces the strongest negative reactions. Relative to Bundling, it shows the largest and most consistent declines, especially on \emph{Confidence}, \emph{Trust}, and \emph{Privacy}. This result matches the structure of the pattern: when an interface offers no meaningful decline option, participants respond with lower confidence, lower trust, and greater privacy concern regardless of manufacturer.

Sneaking and Privacy Zuckering show a different form of perceived severity. Rather than producing the strongest negative attitudes, they produced the largest reductions in ratings on \emph{Avoidance of Unintentional Sharing} relative to Bundling (Sneaking: $\beta=-0.570$, $p=0.003$, $d=0.33$; Privacy Zuckering: $\beta=-0.450$, $p=0.017$, $d=0.26$), indicating a greater perceived likelihood of inadvertent disclosure. This suggests that hidden or subtle patterns can be consequential even when they do not provoke the same overtly negative reaction as Forced Consent.

Cross-OEM differences in perceived severity concentrate clearly on \emph{Confidence}. This pattern is consistent with RQ2.1: OEM differences are largest for Bundling, where manufacturers have more room to shape the consent flow, and smaller for patterns whose coercive or hidden structure dominates the experience. Means and within-pattern contrasts for all attitude constructs appear in \autoref{tab:phase2_means}; interaction terms are interpreted only for constructs whose mixed-effects models converged.

\begin{tcolorbox}[colback=phasetwocolor!8,colframe=phasetwocolor!70!black,
  boxrule=0.8pt,arc=2pt,left=6pt,right=6pt,top=4pt,
  bottom=4pt]
\textbf{Finding 3.}
Forced Consent produces the strongest negative reactions, while Sneaking and Privacy Zuckering produce the largest reductions in ratings on \emph{Avoidance of Unintentional Sharing}, indicating a greater perceived likelihood of inadvertent disclosure.
\end{tcolorbox}


\subsection{RQ2.3: Reasoning and Behavioral Intention}

Most participants reported stronger privacy intentions after viewing the dark-pattern stimuli. Of 98 valid post-survey respondents, 81.6\% intended to \emph{Read More Carefully} when shown a consent screen and 77.6\% intended to go \emph{Through Each Toggle Carefully} when reviewing privacy controls. Smaller groups reported \emph{No Difference}, \emph{No Real Choice Anyway}, would \emph{Accept Defaults}, or would \emph{Turn Everything Off}. Because these groups are small ($n = 2$ to $12$), results involving them are interpreted as directional rather than confirmatory.

\begin{figure}[t]
    \centering
    \includegraphics[width=\columnwidth]
      {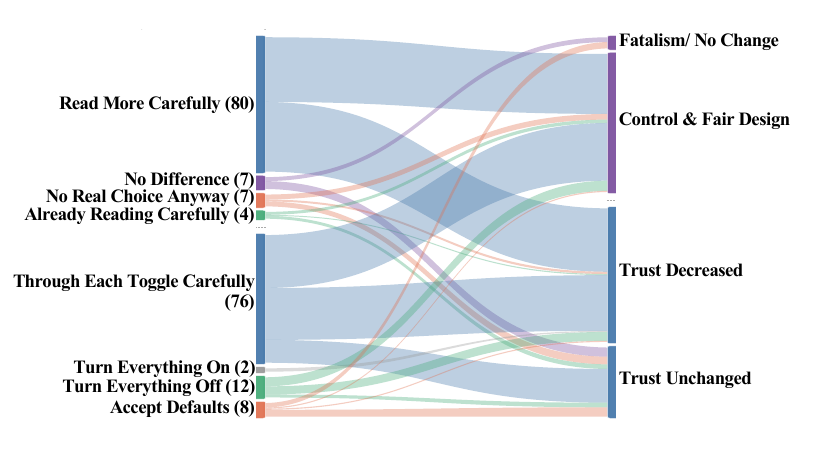}
    \caption{Mapping of behavioral-intention groups to design-reasoning and trust-change themes ($N=98$). Line weight is proportional to the percentage of each group expressing a theme; flows below 10\% are omitted.}
    \label{fig:mapping}
    \vspace{-15pt}
\end{figure}

We next examined how these behavioral-intention groups related to participants' qualitative reasoning. \autoref{fig:mapping} maps these behavioral groups to two design-reasoning themes and two trust-change themes. \emph{Control \& Fair Design} captures demands for clearer choices, less bundling, and fairer opt-in and opt-out paths. \emph{Fatalism / No Change} captures participants who recognized problematic design but did not expect to change their behavior. The trust-change themes capture whether participants reported decreased trust in automotive manufacturers (\emph{Trust Decreased}) or no change in trust (\emph{Trust Unchanged}).

Trust change was most clearly associated with consent-screen reading intentions. Among participants who intended to \emph{Read More Carefully}, 71.2\% reported \emph{Trust Decreased} in automotive manufacturers. By contrast, the smaller \emph{No Difference} and \emph{Accept Defaults} groups were dominated by \emph{Trust Unchanged}. The toggle scenario instead showed a significant association with \emph{Fatalism / No Change}, which had the strongest observed association (\(V=0.62\)). Overall, trust-change themes were associated with consent-screen reading intentions, whereas \emph{Fatalism / No Change} was associated with privacy-toggle intentions.

\emph{Fatalism / No Change} was most common among participants reporting \emph{No Difference} and those who would \emph{Accept Defaults}. Representative responses capture this resignation directly: \textit{``Nothing. It's a car,''} and \textit{``I would still do it the way other manufacturers do it.''} Importantly, fatalism does not mean participants failed to notice manipulation. Many fatalistic participants also expressed demands for \emph{Control \& Fair Design}, including the view that manufacturers \textit{``always get their way to get personal data.''} This suggests that recognizing manipulation does not necessarily produce protective action when users feel they have no practical alternative.

\begin{tcolorbox}[colback=phasetwocolor!8,colframe=phasetwocolor!70!black,
  boxrule=0.8pt,arc=2pt,left=6pt,right=6pt,top=4pt,
  bottom=4pt]
\textbf{Finding 4.}
Trust change was associated with consent-screen reading intention, while \emph{Fatalism / No Change} was associated with privacy-toggle intention: recognizing manipulation does not, on its own, translate into protective action.
\end{tcolorbox}

\section{Phase 3: User Perceptions Before and After Dark-Pattern Instruction}
\label{sec:phase3}

Phase~3 examines whether a brief educational tutorial shifts users' perceptions of dark patterns in automotive interfaces and whether users can identify dark-pattern types after instruction. This phase addresses \emph{RQ3} through three questions:

\noindent\textbf{RQ3.1 (Instructional Shift and Domain Gap):}
How does a brief tutorial reshape ratings and the cross-domain gap between automotive ($\mathrm{CAR}$) and e-commerce ($\mathrm{ECOMM}$) interfaces?\\
\textbf{RQ3.2 (Post-Instruction Pattern Recognition):}
How accurately do participants label dark-pattern types across automotive and e-commerce interfaces after instruction?\\
\textbf{RQ3.3 (Retrospective Domain Judgment):}
How do participants' attitudes toward resisting dark patterns align with their self-reported recognition and perceived avoidability?

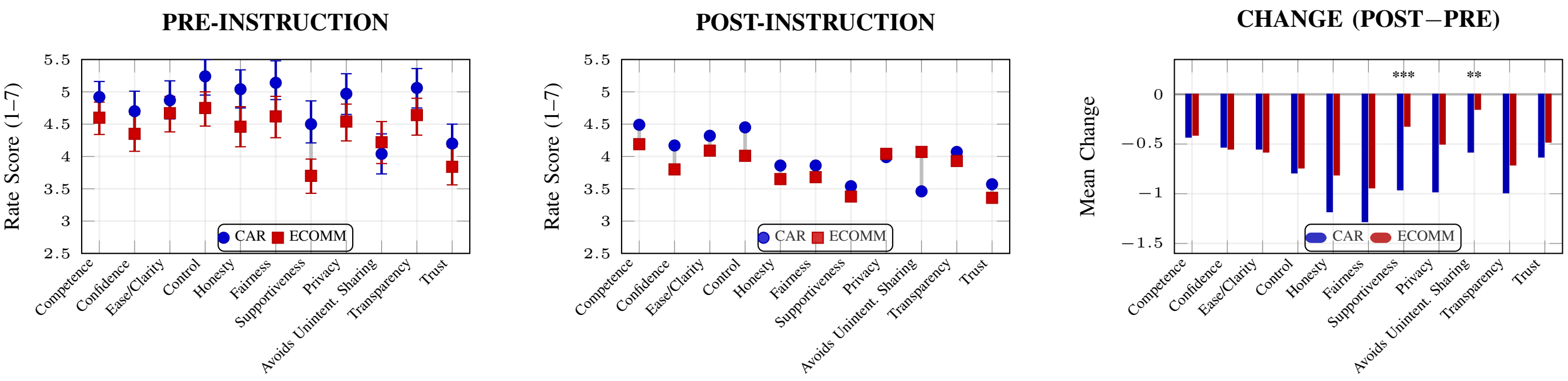
\begin{figure*}[t]
\centering

\begin{subfigure}[t]{0.31\textwidth}
\centering
\begin{tikzpicture}
\pgfplotstableread[col sep=comma]{
construct,y,car,car_lo,car_hi,ecomm,ecomm_lo,ecomm_hi
competence,1,4.92,4.68,5.16,4.60,4.34,4.84
confidence,2,4.70,4.34,5.01,4.35,4.08,4.65
ease_clarity,3,4.87,4.58,5.17,4.67,4.38,4.93
control,4,5.24,4.95,5.50,4.75,4.47,5.00
honesty,5,5.04,4.75,5.34,4.46,4.15,4.77
fairness,6,5.14,4.88,5.48,4.62,4.29,4.93
supportiveness,7,4.50,4.21,4.86,3.70,3.43,3.96
privacy,8,4.97,4.65,5.28,4.54,4.24,4.81
unintentional,9,4.04,3.73,4.35,4.22,3.89,4.54
transparency,10,5.06,4.75,5.36,4.64,4.33,4.90
trust,11,4.20,3.89,4.50,3.84,3.56,4.15
}\datatablePRE
\begin{axis}[
    scale only axis,
    width=0.8\columnwidth,
     height=0.4\columnwidth,
    title={PRE-INSTRUCTION},
    title style={font=\footnotesize\bfseries},
    ylabel={Rate Score (1--7)},
    ylabel style={font=\scriptsize},
    ymin=2.5, ymax=5.5,
    ytick={2.5,3,3.5,4,4.5,5,5.5},
    xmin=0.5, xmax=11.5,
    xtick={1,...,11},
    xticklabels={%
        Competence, Confidence, Ease/Clarity, Control, Honesty,
        Fairness, Supportiveness, Privacy,
        Avoids Unintent.\ Sharing, Transparency, Trust%
    },
    xticklabel style={rotate=45, anchor=east, font=\tiny},
    ticklabel style={font=\tiny},
    grid=major,
    grid style={opacity=0.25, line width=0.5pt},
legend style={
    font=\tiny,
    at={(0.7,0.008)}, anchor=south east,
    draw=black, fill=none,   
    cells={anchor=west}, legend columns=2,
    rounded corners=2pt,
    inner sep=0pt,   
    column sep=0pt,  
    row sep=0pt      
},
    enlarge x limits=false,
]
\addplot[
    gray!50, line width=1.2pt, forget plot,
    quiver={u=0, v=\thisrow{ecomm}-\thisrow{car},
            every arrow/.style={-}},
] table[x=y, y=car] {\datatablePRE};
\addplot+[
    only marks, mark=*, mark size=1.8pt, color=blue!70!black,
    error bars/y dir=both, error bars/y explicit,
    error bars/error mark=-,
    error bars/error bar style={line width=0.8pt}
] table[
    x=y, y=car,
    y error plus expr=\thisrow{car_hi}-\thisrow{car},
    y error minus expr=\thisrow{car}-\thisrow{car_lo}
] {\datatablePRE};
\addlegendentry{CAR}
\addplot+[
    only marks, mark=square*, mark size=1.8pt, color=red!70!black,
    error bars/y dir=both, error bars/y explicit,
    error bars/error mark=-,
    error bars/error bar style={line width=0.8pt}
] table[
    x=y, y=ecomm,
    y error plus expr=\thisrow{ecomm_hi}-\thisrow{ecomm},
    y error minus expr=\thisrow{ecomm}-\thisrow{ecomm_lo}
] {\datatablePRE};
\addlegendentry{ECOMM}
\end{axis}
\end{tikzpicture}
\subcaption{Pre-instruction ratings.}
\label{fig:results_pre}
\end{subfigure}%
\hfill
\begin{subfigure}[t]{0.31\textwidth}
\centering
\begin{tikzpicture}
\pgfplotstableread[col sep=comma]{
construct,y,car,ecomm
competence,1,4.49,4.19
confidence,2,4.17,3.80
ease_clarity,3,4.32,4.09
control,4,4.45,4.01
honesty,5,3.86,3.65
fairness,6,3.86,3.68
supportiveness,7,3.54,3.38
privacy,8,3.99,4.04
unintentional,9,3.46,4.07
transparency,10,4.07,3.93
trust,11,3.57,3.36
}\datatablePOST
\begin{axis}[
    scale only axis,
    width=0.8\columnwidth,
     height=0.4\columnwidth,
    title={POST-INSTRUCTION},
    title style={font=\footnotesize\bfseries},
    ylabel={Rate Score (1--7)},
    ylabel style={font=\scriptsize},
    ymin=2.5, ymax=5.5,
    ytick={2.5,3,3.5,4,4.5,5,5.5},
    xmin=0.5, xmax=11.5,
    xtick={1,...,11},
    xticklabels={%
        Competence, Confidence, Ease/Clarity, Control, Honesty,
        Fairness, Supportiveness, Privacy,
        Avoids Unintent.\ Sharing, Transparency, Trust%
    },
    xticklabel style={rotate=45, anchor=east, font=\tiny},
    ticklabel style={font=\tiny},
    grid=major,
    grid style={opacity=0.25, line width=0.5pt},
    legend style={
        font=\tiny,
        at={(0.7,0.008)}, anchor=south east,
        draw=black, fill=none, fill opacity=0.8,
        cells={anchor=west}, legend columns=2,
        rounded corners=2pt,
        inner sep=0pt,
        column sep=0pt,
        row sep=0pt
    },
    enlarge x limits=false,
]
\addplot[
    gray!50, line width=1.2pt, forget plot,
    quiver={u=0, v=\thisrow{ecomm}-\thisrow{car},
            every arrow/.style={-}},
] table[x=y, y=car] {\datatablePOST};
\addplot+[
    only marks, mark=*, mark size=1.8pt, color=blue!70!black,
] table[x=y, y=car] {\datatablePOST};
\addlegendentry{CAR}
\addplot+[
    only marks, mark=square*, mark size=1.8pt, color=red!70!black,
] table[x=y, y=ecomm] {\datatablePOST};
\addlegendentry{ECOMM}
\end{axis}
\end{tikzpicture}
\subcaption{Post-instruction ratings.}
\label{fig:results_post}
\end{subfigure}%
\hfill
\begin{subfigure}[t]{0.31\textwidth}
\centering
\begin{tikzpicture}
\pgfplotstableread[col sep=comma]{
construct,y,car,ecomm
competence,1,-0.43,-0.41
confidence,2,-0.53,-0.55
ease_clarity,3,-0.55,-0.58
control,4,-0.79,-0.74
honesty,5,-1.18,-0.81
fairness,6,-1.28,-0.94
supportiveness,7,-0.96,-0.32
privacy,8,-0.98,-0.50
unintentional,9,-0.58,-0.15
transparency,10,-0.99,-0.71
trust,11,-0.63,-0.48
}\datatableCHANGE
\begin{axis}[
    scale only axis,
    width=0.8\columnwidth,
     height=0.4\columnwidth,
    title={CHANGE (POST$-$PRE)},
    title style={font=\footnotesize\bfseries},
    ylabel={Mean Change},
    ylabel style={font=\scriptsize},
    ymin=-1.6, ymax=0.35,
    ytick={-1.5,-1.0,-0.5,0},
    yticklabel style={font=\tiny},
    xmin=0.5, xmax=11.5,
    xtick={1,...,11},
    xticklabels={%
        Competence, Confidence, Ease/Clarity, Control, Honesty,
        Fairness, Supportiveness, Privacy,
        Avoids Unintent.\ Sharing, Transparency, Trust%
    },
    xticklabel style={rotate=45, anchor=east, font=\tiny},
    ticklabel style={font=\tiny},
    grid=major,
    grid style={opacity=0.25, line width=0.5pt},
    legend style={
        font=\tiny,
        at={(0.74,0.008)}, anchor=south east,
        draw=black, fill=none, fill opacity=0.8,
        cells={anchor=west}, legend columns=2,
        rounded corners=2pt,
        inner sep=0pt,
        column sep=0pt,
        row sep=0pt
    },
    legend image code/.code={%
        \draw[#1, draw=none] (0cm,-0.06cm)
            rectangle (0.25cm,0.06cm);
    },
    enlarge x limits=false,
    extra y ticks={0},
    extra y tick style={
        grid=major,
        grid style={black, line width=0.8pt}
    },
    extra y tick labels={},
]
\addplot[
    ybar, bar width=0.15,
    fill=blue!70!black, draw=blue!70!black,
    bar shift=-0.1,
] table[x=y, y=car] {\datatableCHANGE};
\addlegendentry{CAR}
\addplot[
    ybar, bar width=0.15,
    fill=red!70!black, draw=red!70!black,
    bar shift=0.1,
] table[x=y, y=ecomm] {\datatableCHANGE};
\addlegendentry{ECOMM}
\node[font=\tiny\bfseries, color=black, anchor=south]
    at (axis cs:7, 0.02) {***};
\node[font=\tiny\bfseries, color=black, anchor=south]
    at (axis cs:9, 0.02) {**};
\end{axis}
\end{tikzpicture}
\subcaption{Pre-to-post change.}
\label{fig:change_scores}
\end{subfigure}

\caption{CAR versus ECOMM ratings across the eleven attitude constructs ($N{=}49$). 
\textbf{(a)}~Baseline ratings show CAR higher than ECOMM on ten of eleven attitudes, with error bars showing 95\% bootstrap confidence intervals. 
\textbf{(b)}~After a brief tutorial, only \emph{Avoids Unintentional Sharing} ($p{=}0.0001$) and \emph{Competence} ($p{=}0.0032$) remain significantly different (Wilcoxon signed-rank, Bonferroni-corrected $\alpha{=}0.0045$). 
\textbf{(c)}~Mean pre-to-post change per domain. \emph{Supportiveness} (***~$p{=}0.0001$, $r_{rb}{=}0.804$) and \emph{Avoids Unintentional Sharing} (**~$p{=}0.0065$, $r_{rb}{=}0.714$) show significantly larger drops in CAR than ECOMM.}
\label{fig:prepost_combined}
\vspace{-10pt}
\end{figure*}

\subsection{Study Design}
Phase~3 uses a fully within-subjects design with two measurement rounds, pre-instruction and post-instruction, separated by a brief tutorial. The study spans two domains: automotive ($\mathrm{CAR}$) and e-commerce ($\mathrm{ECOMM}$). All participants view all clips in both rounds, enabling direct within-participant comparisons before and after instruction as well as cross-domain comparisons under identical conditions.

\noindent\textbf{Stimuli.} The stimulus set shown to each participant consists of 14 short, silent screen-recorded clips: seven $\mathrm{ECOMM}$ clips recorded from live online interfaces~(Step~\cnumthree{1}) and seven $\mathrm{CAR}$ clips drawn from the Phase~1 audit~(Step~\cnumthree{2}). The $\mathrm{ECOMM}$ clips include examples from Amazon Prime~\cite{amazonprime_website}, KAY Jewelers~\cite{kay_website}, etc.; all sources are shown in Table~\ref{tab:ecomm-stimuli} in Appendix~\ref{app:ecomm-stimuli}. The seven $\mathrm{ECOMM}$ stimuli were selected to include the same dark-pattern types as the seven $\mathrm{CAR}$ stimuli.

\noindent\textbf{Procedure.} Participants proceed through three stages. Stage~1, the pre-instruction round, asks participants to rate all 14 clips~(Step~\cnumthree{3}). The seven $\mathrm{CAR}$ clips are presented first, followed by the seven $\mathrm{ECOMM}$ clips, with the within-domain order independently sampled for each participant ($\pi_i^{\mathrm{CAR}} \sim \mathrm{Uniform}(S_7)$ and $\pi_i^{\mathrm{ECOMM}} \sim \mathrm{Uniform}(S_7)$). Stage~2 introduces a brief tutorial that names and illustrates each dark-pattern type from the Phase~1 catalog~(Step~\cnumthree{4}). Stage~3, the post-instruction round, asks participants to rate the same 14 clips again in the same domain order, with the automotive clips followed by the e-commerce clips, and additionally label each clip with the dark-pattern types they observe~(Step~\cnumthree{5}). The within-domain order is freshly sampled again in Stage~3, independently of the Stage~1 ordering.

\noindent\textbf{Measures.}
In Stage~1 and Stage~3, after each clip, participants answer the eleven seven-point semantic-differential attitude scales reused from Phase~2: \emph{Fairness}, \emph{Honesty}, \emph{Control}, \emph{Privacy Respect}, \emph{Ease of Use}, \emph{Transparency}, \emph{Avoidance of Unintentional Sharing}, \emph{Trust}, \emph{Competence}, \emph{Emotional Response}, and \emph{Confidence}. Participants may also provide an optional open-ended justification after each clip. In Stage~3, participants select all dark-pattern types they observe in each clip from a list of seven labels: \emph{Forced Consent}, \emph{Privacy Zuckering}, \emph{Confirm Shaming}, \emph{Roach Motel}, \emph{Bundling Consent}, \emph{Illusion of Control}, \emph{Hidden Cost}, or \emph{None}. They also rate their confidence in each label selection. After Stage~3, participants complete a post-survey containing five domain-comparison Likert items, a binary avoidability question, seven per-pattern confidence ratings, and four open-ended reflection prompts.

\noindent\textbf{Power analysis and eligibility.}
We conduct an a priori power analysis for paired Wilcoxon signed-rank tests using a Bonferroni-corrected threshold of $0.05/11 = 0.0045$ across the eleven attitude constructs, power $= 0.80$, and effect size $d = 0.5$, yielding a minimum sample size of $N = 45$. Eligibility criteria and incentive structure match Phase~2.

\noindent\textbf{Ethics.}
Phase~3 protocol was approved by our IRB.

\subsection{Recruitment}

The study is administered through Qualtrics~\cite{qualtrics_website}, with participants recruited through Prolific~\cite{prolific}. We recruited 49 participants via Prolific, all of whom completed the study with no mid-survey attrition. Participants received \$4.00 for the estimated 20-minute study, corresponding to a compensation rate of \$12.00 per hour. Participants were 49\% female with a balanced age distribution across groups (26--35: 35\%, 46 and above: 39\%); 78\% reported prior awareness of AAOS. Detailed demographics appear in Appendix~\ref{app:demographics}.

\subsection{Data Analysis}
We summarize Likert ratings with descriptive statistics and assess within-participant changes and cross-domain differences using paired Wilcoxon signed-rank tests across the eleven attitude constructs, with Bonferroni correction where reported. We report rank-biserial correlations ($r_{rb}$) as effect sizes, with bootstrapped 95\% confidence intervals ($B = 1000$). To complement the per-attitude Wilcoxon tests, we fit linear mixed-effects models for each attitude with the form Score $\sim$ Domain $\times$ Instruction + (1\,$|$\,Participant), where Domain ($\mathrm{CAR}$ vs.\ $\mathrm{ECOMM}$) and Instruction (pre vs.\ post) are fixed effects and Participant is a random intercept. The Domain $\times$ Instruction interaction tests whether the tutorial shifts one domain more than the other.

For post-instruction pattern labels, we compute contingency matrices and run Fisher--Freeman--Halton exact tests on the ground-truth $\times$ predicted-label tables, separately for each domain. For post-survey analyses, we run paired Wilcoxon tests on the five direct domain-comparison items with Bonferroni correction at $\alpha = 0.05/5 = 0.01$, a Friedman test across the seven per-pattern confidence ratings, and Fisher's exact tests relating qualitative profiles to the binary avoidability question and other self-report outcomes.

\subsection{RQ3.1: Instructional Shift and Domain Gap}

\autoref{fig:results_pre} shows the pre-instruction rating gap between automotive and e-commerce interfaces. Before receiving the tutorial, participants gave automotive interfaces more favorable ratings on ten of the eleven attitude constructs. The largest gap appeared on \emph{Supportiveness} ($\Delta = 0.80$), with clear advantages also on \emph{Honesty}, \emph{Fairness}, and \emph{Control}. The only exception was \emph{Avoidance of Unintentional Sharing}, where e-commerce received more favorable ratings. This suggests that participants initially gave automotive interfaces a broad benefit of the doubt, but that advantage was weakest when the question focused on unintended data sharing.

The tutorial substantially reduced this initial advantage. As shown in \autoref{fig:results_post}, only two attitudes remained significantly different after instruction. Automotive received more favorable ratings on \emph{Competence} ($p = 0.0032$), while e-commerce received more favorable ratings on \emph{Avoidance of Unintentional Sharing} ($p = 0.0001$). In other words, the pre-instruction advantage for automotive narrowed to one remaining strength and one privacy-specific weakness. \autoref{fig:change_scores} shows where this shift was largest. Automotive ratings declined across attitude constructs, with the steepest losses on \emph{Fairness} ($\Delta = -1.28$), followed by \emph{Honesty}, \emph{Transparency}, and \emph{Privacy Respect}. More procedural attitudes, such as \emph{Ease of Use} and \emph{Confidence}, declined less. This suggests that the tutorial weakened participants' belief that automotive interfaces were fair, honest, transparent, and privacy-respecting.

\begin{tcolorbox}[colback=phasethreecolor!8,colframe=phasethreecolor!70!black,
  boxrule=0.8pt,arc=2pt,left=6pt,right=6pt,top=4pt,
  bottom=4pt]
\textbf{Finding 5.}
A brief tutorial turns automotive's initial benefit of the doubt into targeted skepticism. Participants no longer broadly favor automotive interfaces after instruction, and their strongest remaining concern centers on unintended data sharing.
\end{tcolorbox}


\subsection{RQ3.2: Post-Instruction Pattern Recognition}

\autoref{fig:combined_confusion} shows that post-instruction recognition differs across domains. In e-commerce, participants correctly identify several dark-pattern types, particularly \emph{Hidden Cost}, \emph{Forced Consent}, \emph{Confirm Shaming}, and \emph{Roach Motel}. In automotive interfaces, correct recognition is generally weaker and more uneven across pattern types. This suggests that participants are better able to identify dark-pattern types in e-commerce interfaces than in vehicle infotainment interfaces.

The confusion patterns also indicate important differences in how participants distinguish among dark-pattern types. In the automotive domain, responses are distributed across several incorrect categories, suggesting difficulty in identifying the specific mechanism underlying a manipulative design. E-commerce responses show clearer category-level recognition, although confusion between related patterns remains. Across both domains, \emph{Privacy Zuckering} is among the most difficult patterns to recognize, indicating that subtle privacy-oriented manipulation may be harder for participants to identify even after instruction.

\begin{tcolorbox}[colback=phasethreecolor!8,colframe=phasethreecolor!70!black,
boxrule=0.8pt,arc=2pt,left=6pt,right=6pt,top=4pt,
bottom=4pt]
\textbf{Finding 6.}
Dark-pattern recognition differs across domains. Participants show stronger category-level recognition in e-commerce, whereas automotive responses exhibit greater confusion among dark-pattern types, suggesting that identifying the specific manipulation is more difficult in vehicle interfaces.
\end{tcolorbox}


\subsection{RQ3.3: Retrospective Domain Judgment}

\begin{figure}[t]
\centering
\begin{tikzpicture}
\begin{axis}[
    scale only axis,
    width=0.65\columnwidth,
    height=0.3\columnwidth,
    xmin=-1.6, xmax=1.6,
    xlabel={Mean difference: CAR $-$ ECOMM},
    xlabel style={font=\scriptsize},
    xtick={-1.5,-1.0,-0.5,0,0.5,1.0,1.5},
    xticklabel style={font=\scriptsize},
    ymin=0.5, ymax=5.5,
    ytick={1,2,3,4,5},
    yticklabels={%
        Self-Efficacy,Cognitive Effort,Privacy Trust,%
        Control Loss,Manipulation%
    },
    yticklabel style={font=\scriptsize},
    grid=major,
    grid style={opacity=0.25, line width=0.5pt},
    enlarge y limits=0.15,
    extra x ticks={0},
    extra x tick style={
        grid=major,
        grid style={black, line width=1.2pt}
    },
    extra x tick labels={},
    legend columns=1,
    legend style={
        font=\tiny,
        at={(0.98,0.55)}, anchor=south east,
        draw=black, fill=white, fill opacity=0.85,
        cells={anchor=west},
        rounded corners=2pt
    },
    legend image code/.code={%
        \fill[#1] (0cm,-0.04cm) rectangle (0.15cm,0.04cm);
    },
]
\fill[red!70!black]   (axis cs:0,4.85)  rectangle (axis cs:-1.25,5.15);
\fill[red!70!black]   (axis cs:0,3.85)  rectangle (axis cs:-0.96,4.15);
\fill[blue!70!black]  (axis cs:0,2.85)  rectangle (axis cs: 0.83,3.15);
\fill[black!30]       (axis cs:0,1.85)  rectangle (axis cs:-0.21,2.15);
\draw[black!50]       (axis cs:0,1.85)  rectangle (axis cs:-0.21,2.15);
\fill[black!30]       (axis cs:0,0.85)  rectangle (axis cs:-0.27,1.15);
\draw[black!50]       (axis cs:0,0.85)  rectangle (axis cs:-0.27,1.15);
\node[font=\tiny\bfseries, anchor=east] at (axis cs:-1.25,5)
    {***\;};
\node[font=\tiny\bfseries, anchor=east] at (axis cs:-0.96,4)
    {**\;};
\node[font=\tiny\bfseries, anchor=west] at (axis cs:0.83,3)
    {\;**};
\addplot[fill=red!70!black, draw=red!70!black] coordinates {(0,0) (0,0)};
\addlegendentry{ECOMM rated higher}
\addplot[fill=blue!70!black, draw=blue!70!black] coordinates {(0,0) (0,0)};
\addlegendentry{CAR rated higher}
\addplot[fill=black!30, draw=black!50] coordinates {(0,0) (0,0)};
\addlegendentry{Not significant}
\end{axis}
\end{tikzpicture}
\caption{Mean rating differences between CAR (automotive) and
         ECOMM (e-commerce) interfaces on five post-survey
         Likert items (Phase~3, $N{=}49$). Wilcoxon signed-rank
         with Bonferroni-corrected $\alpha{=}0.01$:
         Manipulation $^{***}p{<}0.001$,
         Control Loss $^{**}p{=}0.002$,
         Privacy Trust $^{**}p{=}0.001$;
         Cognitive Effort and Self-Efficacy non-significant.
         Negative values indicate ECOMM was rated higher on
         that item; positive values indicate CAR was rated
         higher.}
\label{fig:phase3_postsurvey_domain}
\vspace{-12pt}
\end{figure}
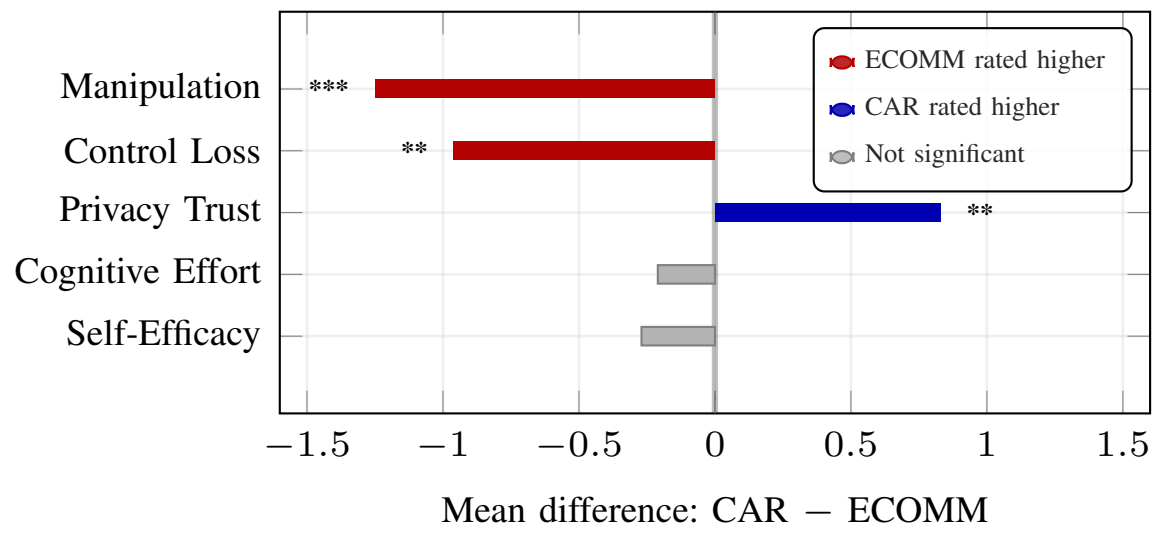

Thematic coding of the post-survey reflections surfaces four mental-model groups:
\begin{itemize}[leftmargin=*, itemsep=1pt, topsep=2pt]
    \item \emph{Harm-focused} participants emphasize concrete harms, such as privacy loss, unwanted sharing, or manipulation.
    \item \emph{Structurally fatalistic} participants recognize manipulation but view it as difficult to avoid because companies control the interface or users have limited alternatives.
    \item \emph{Self-efficacious} participants believe they can resist dark patterns by reading carefully, changing settings, or avoiding services.
    \item \emph{Change-oriented} participants focus on external remedies, such as clearer disclosures, better defaults, regulation, or manufacturer accountability.
\end{itemize}
Participants could belong to more than one group, but no participant was both \emph{structurally fatalistic} and \emph{self-efficacious}. Fatalistic participants saw little ability to avoid dark patterns, while self-efficacious participants believed users could resist them through careful action. Participants rated e-commerce higher on \emph{Manipulation} and \emph{Control Loss}, while rating automotive higher on \emph{Privacy Trust}. As shown in \autoref{fig:phase3_postsurvey_domain}, these differences remain significant after Bonferroni correction. On the direct autonomy-risk question, most participants named e-commerce as the greater risk, citing financial exposure and frequent interaction. A smaller group named automotive instead, emphasizing that vehicle interfaces are harder to avoid or replace when privacy choices are tied to vehicle features.

These mental-model groups align with perceived avoidability. All \emph{self-efficacious} participants report that dark patterns can be avoided, while nearly all \emph{structurally fatalistic} participants report that they cannot. Yet this difference is not about recognition confidence: \emph{self-efficacious} participants do not report greater confidence in identifying specific dark-pattern types. Instead, \emph{structurally fatalistic} participants describe resignation or withdrawal, while \emph{self-efficacious} participants describe practical workarounds such as reading carefully or changing settings.

\begin{tcolorbox}[colback=phasethreecolor!8,colframe=phasethreecolor!70!black,
  boxrule=0.8pt,arc=2pt,left=6pt,right=6pt,top=4pt,
  bottom=4pt]
\textbf{Finding 7.}
Some participants believe careful action can help them resist manipulation, while others recognize the problem but see few practical alternatives, especially when privacy choices are tied to vehicle features. What distinguishes these groups is perceived control, not recognition confidence, both recognize the manipulation, but only the former feel they can act on that recognition.
\end{tcolorbox}

\input{Figures/matrix_dp}

\section{Discussion}
\subsection{Takeaways and Recommendations}

Phase~1 confirms that in-vehicle dark patterns are not exotic: they reflect the same categories documented in web and mobile taxonomies, but appear in a safety-critical and lock-in-heavy setting. Because these patterns are distributed across OEMs, the results suggest a need for industry-specific accountability rather than manufacturer-specific blame. Bodies such as SAE International and ISO should develop anti-dark-pattern guidelines that distinguish safety-critical interactions from data-collection practices. Regulators should consider pre-release design reviews of consent flows and default configurations.

Phase~2's sophistication paradox carries an important policy implication. Honda's Confirmshaming and Privacy Zuckering implementations are rated more positively on attitudes such as \emph{Fairness} and \emph{Supportiveness} than GMC's implementations of the same patterns, and Phase~3 shows that Privacy Zuckering is also the hardest pattern for users to name. A dark pattern that does not feel manipulative may fail to trigger resistance, and standard disclosure requirements may not help when users do not realize they are being steered. Oversight should therefore evaluate \emph{how} a pattern is implemented, not only whether it is present. Manufacturers should adopt privacy-forward defaults, separate essential functionality from data-sharing requests, allow individual rather than bundled permissions, and provide equivalent opt-in and opt-out effort.

Phase~3 exposes a fragile automotive rating premium. Before instruction, participants rate automotive interfaces higher than e-commerce on ten attitude constructs, yet a brief tutorial within the same session narrows this gap to two constructs. Although most participants (69.4\%) judge e-commerce to be the greater autonomy risk, a smaller group (18.4\%) views automotive interfaces as more threatening because of structural lock-in. Unlike a website, an automotive OS is tied to the vehicle and cannot be easily abandoned or replaced. Policy should therefore guarantee a ``right to disconnect'' that preserves full functionality when users decline non-essential data sharing, make user data deletable and accounts unlinkable without loss of functionality, and mandate interoperability standards that allow interface migration.

\subsection{Limitations}

\noindent\textbf{Phase 1: Taxonomy and Annotation.} The examples provided may have focused annotators on the identified patterns, and broader training materials with greater coverage might have led to the identification of additional patterns. Because annotators self-selected which emulators to audit, the emulator--annotator assignment counts were uneven, and the order in which annotators examined multiple emulators was not controlled. Consequently, order effects may have contributed to some variation in the observed coverage across emulators. Finally, review and mapping by a single automotive-privacy specialist may have introduced bias. However, the annotation exercise was deliberately open-ended, and annotators could also report dark patterns not listed in the provided materials; some submitted observations were mapped to categories that were not explicitly listed in those materials. Each documented dark pattern was confirmed to be real.

\noindent\textbf{Phase 2: Perceived Severity and Stimulus Design.} The perceived-severity analysis captures how negatively participants perceived the identified patterns rather than their actual privacy, security, or safety impacts. Because each OEM was represented by a specific interface implementation, we cannot determine whether observed differences were due to the OEM itself or to design features such as button size, wording, and interaction flow. Our results therefore characterize differences among the specific OEM implementations tested rather than general brand effects.

Our stimuli were screen-recorded clips rather than hands-on interactive tasks. Participants could not navigate, make and correct mistakes, or observe the downstream consequences of their choices. 
Our results reflect perceptions of manipulation when \emph{observing} an interaction, not behavior during real vehicle or emulator use. Future work should complement video-based evaluation with interactive emulator tasks or in-vehicle studies that capture behavioral measures such as time to find opt-out controls, error recovery, and follow-up settings changes. 

We focus on UI flows visible in short clips, but privacy decision-making is also shaped by notices, explanations, and policy language that often appear in deeper settings screens. Because these elements are not foregrounded in our stimuli, participants may rely more heavily on salient UI cues than they would in contexts where detailed disclosures are prominent. Future stimuli should include text-based disclosures to test how disclosure quality interacts with UI patterns.

\noindent\textbf{Phase 3: Sample and Intervention Design.}
Phase~3's sample is modest for the passive subgroups within the post-survey qualitative profiles; results involving these subgroups are directional rather than confirmatory. The intervention is a single brief tutorial measured in the same session. Whether the perceptual shift persists over time or translates into real-world behavior remains an open question. For this study, a separate control group was not included because it was designed as a within-subject pre/post study, with each participant serving as their own baseline. Future work could investigate this question using an untutored control-group design.

\subsection{Concrete Harm Scenarios}
The FTC's GM/OnStar case illustrates two concrete harm scenarios associated with manipulative vehicle-enrollment interfaces. First, the FTC alleged that a misleading enrollment design (\emph{Bundling Consent}) left many consumers unaware that they had enrolled in the Smart Driver feature and that GM collected and disclosed their precise geolocation and driving-behavior data to consumer reporting agencies, including Verisk and LexisNexis, without informed consent, resulting in loss of privacy, increased insurance premiums, and loss of insurance coverage~\cite{ftc2026gmonstarorder,ftc2026gmcomplaint}. Second, the OnStar enrollment interface warned that declining the terms would deactivate Automatic Crash Response, Emergency Services, and Vehicle Diagnostics (\emph{Forced Consent}), making the decision to limit data sharing contingent on giving up safety-related vehicle functionality~\cite{ftc2026gmcomplaint}. These consequences illustrate why the dark patterns identified in our AAOS audit warrant attention beyond interface usability: similar manipulative consent structures in deployed vehicle systems can have privacy, financial, and safety-related consequences.


\section{Related Work}
\label{sec:related}

\subsection{Dark Patterns Across Digital Ecosystems}

Dark patterns have been documented across a range of digital domains, with their form shaped by platform constraints, interaction modality, and business incentives. In home IoT, an audit of 57 smart-home devices found that dark patterns are concentrated in always-on devices and amplified by multimodal interaction, vendor ecosystems, and lifecycle issues such as device transfer or removal~\cite{kowalczyk2023understanding}. The authors explicitly identify cars as a missing domain where similar audit methods are needed. Cross-platform audits show that dark-pattern exposure also varies by interface modality. Gunawan \textit{et al.} examined 105 services across mobile apps, mobile web, and desktop web, finding that mobile interfaces often contain more, and different, dark patterns than their desktop counterparts~\cite{gunawan2021comparative}. This suggests that single-modality audits can miss harms, especially for rights-relevant controls such as privacy settings, consent, deletion, and cancellation.

Domain-specific studies further show that manipulation is shaped by business models. In mobile games, dark patterns cluster around monetization mechanics such as free-to-play design, advertisements, and in-app purchases~\cite{niknejad2024level}. In e-commerce, large-scale audits of shopping websites find that dark patterns are common, often covert or information-hiding, and sometimes commoditized through third-party vendors and plugins~\cite{mathur2019dark}. Together, these studies show that dark patterns are pervasive, scalable, and shaped by the technical and economic structure of each domain. Automotive infotainment systems combine several of these risk factors: IoT-like embeddedness, platform lock-in, multimodal interaction, and safety-critical functionality. Yet dark patterns in vehicle interfaces remain underexplored. This motivates our focus on AAOS as a production-relevant automotive infotainment platform.

\subsection{Automotive Privacy and Interface-Mediated Consent}

Prior automotive privacy research has examined user perceptions of vehicle sensing and the opacity of connected-car privacy disclosures, but has not studied dark patterns at the in-vehicle UI level. Bloom \textit{et al.} surveyed 302 urban residents about autonomous-vehicle sensing scenarios and found that participants were comfortable with primary sensing uses but less comfortable with secondary uses such as recognition, identification, and tracking~\cite{bloom2017self}. Their study shows that users care about the scope and purpose of vehicle data collection, but it focuses on sensing perceptions rather than the interface mechanisms through which users grant or refuse consent.

Kemp's ``Driving Blind'' report examines connected-car privacy policies across 15 leading vehicle brands in Australia~\cite{kemp2024driving}. The report documents fragmented notice structures, vague definitions of personal information, unclear retention and sharing practices, and limited visibility into third-party or data-broker relationships. This work highlights the limits of policy-text transparency and calls for stronger consent and disclosure requirements. These studies establish that automotive data collection raises privacy concerns, both in public perception and in policy disclosure. However, they do not examine whether the design of in-vehicle interfaces itself steers users toward consent or makes privacy-protective choices harder. Our work addresses this gap by auditing AAOS interfaces for dark patterns and evaluating how users recognize and respond to those patterns across automotive and e-commerce contexts.

\section{Conclusion}
\label{sec:conclusion}

In-vehicle dark patterns deserve attention as a distinct problem. The 17 patterns documented in AAOS are familiar from web and mobile contexts, but they appear in a setting where consent flows are tied to vehicle functionality and sensitive data collection. Across our studies, participants initially extended a benefit of the doubt to automotive interfaces, yet a brief tutorial substantially reduced this rating advantage over e-commerce. Even after instruction, participants struggled to identify specific dark-pattern types in automotive interfaces. Overall, our findings show that both implementation choices and users' ability to recognize manipulation shape the risks posed by dark patterns in vehicle infotainment systems.

\section*{Ethical Considerations}
Phase~1 analyzes publicly available AAOS emulators and involves no human subjects. Phases~2 and~3 were IRB-reviewed and deemed exempt. Participants consented, were compensated through Prolific, and provided no personally identifiable information. We mitigate potential reputational or misuse risks by limiting our claims to the evaluated interfaces and contextualizing OEM-specific observations.

\section*{ACKNOWLEDGMENTS}
We thank our anonymous reviewers and shepherd for providing us with valuable feedback that helped improve our paper. The work reported in this paper was supported in part by NSF under Grant CNS-2541570 and partially by the Clemson University Creative Inquiry program.

\bibliographystyle{IEEEtran}
\bibliography{ref}

\begin{thebibliography}{10}
\providecommand{\url}[1]{#1}
\csname url@samestyle\endcsname
\providecommand{\newblock}{\relax}
\providecommand{\bibinfo}[2]{#2}
\providecommand{\BIBentrySTDinterwordspacing}{\spaceskip=0pt\relax}
\providecommand{\BIBentryALTinterwordstretchfactor}{4}
\providecommand{\BIBentryALTinterwordspacing}{\spaceskip=\fontdimen2\font plus
\BIBentryALTinterwordstretchfactor\fontdimen3\font minus
  \fontdimen4\font\relax}
\providecommand{\BIBforeignlanguage}[2]{{%
\expandafter\ifx\csname l@#1\endcsname\relax
\typeout{** WARNING: IEEEtran.bst: No hyphenation pattern has been}%
\typeout{** loaded for the language `#1'. Using the pattern for}%
\typeout{** the default language instead.}%
\else
\language=\csname l@#1\endcsname
\fi
#2}}
\providecommand{\BIBdecl}{\relax}
\BIBdecl

\bibitem{mozilla2023carsprivacy}
{Mozilla Foundation}, ``It’s official: Cars are the worst product category we
  have ever reviewed for privacy,''
  \url{https://www.mozillafoundation.org/en/privacynotincluded/articles/its-official-cars-are-the-worst-product-category-we-have-ever-reviewed-for-privacy/},
  Sep. 2023, accessed: 2025-07-04.

\bibitem{mozilla2023carsprivacyconcerns}
{Mozilla Foundation and Privacy Not Included}, ``After researching cars and
  privacy, here’s what keeps us up at night,''
  \url{https://www.mozillafoundation.org/en/privacynotincluded/articles/after-researching-cars-and-privacy-heres-what-keeps-us-up-at-night/},
  Sep.~6 2023, accessed: 2025-07-04.

\bibitem{androidautomotive_elevating2025}
{Android Automotive Solutions by Bamboo Apps}, ``Elevating in‑car experiences
  with android automotive apps,''
  \url{https://androidautomotive.dev/blog/android-automotive-apps}, May 2025,
  accessed: 2025-07-18.

\bibitem{trijardcaddy2024enhancing}
\BIBentryALTinterwordspacing
V.~Trijard-Caddy, ``Enhancing automotive interface design: Insights from
  touchscreen usability and driver interaction studies,'' in \emph{Usability
  and User Experience}, ser. AHFE Open Access, T.~Ahram and C.~Falc{\~a}o,
  Eds., vol. 156.\hskip 1em plus 0.5em minus 0.4em\relax USA: AHFE
  International, 2024, aHFE 2024 International Conference. [Online]. Available:
  \url{https://doi.org/10.54941/ahfe1005448}
\BIBentrySTDinterwordspacing

\bibitem{gray2018dark}
C.~M. Gray, Y.~Kou, B.~Battles, J.~Hoggatt, and A.~L. Toombs, ``The dark
  (patterns) side of ux design,'' in \emph{Proceedings of the 2018 CHI
  conference on human factors in computing systems}, 2018, pp. 1--14.

\bibitem{brignull2023deceptive}
\BIBentryALTinterwordspacing
H.~Brignull, M.~Leiser, C.~Santos, and K.~Doshi, ``{Deceptive Patterns -- User
  Interfaces Designed to Trick You},'' 2023, accessed: 2026-08-08. [Online].
  Available: \url{https://www.deceptive.design/}
\BIBentrySTDinterwordspacing

\bibitem{harrison2024privacy}
A.~Dowthwaite, D.~Cook, and A.~L. Cox, ``Privacy preferences in automotive data
  collection,'' \emph{Transportation research interdisciplinary perspectives},
  vol.~24, p. 101022, 2024.

\bibitem{gozubuyuk2024analyzing}
B.~G{\"o}z{\"u}b{\"u}y{\"u}k, B.~Tang, K.~G. Shin, and M.~D. Pes{\'e},
  ``Analyzing privacy implications of data collection in android automotive
  os,'' \emph{arXiv preprint arXiv:2409.15561}, 2024.

\bibitem{pese2023first}
M.~D. Pese, ``A first look at android automotive privacy,'' SAE Technical
  Paper, Tech. Rep., 2023.

\bibitem{luxoft2025gasaaos}
{Luxoft}, ``To gas or not to gas: Optimizing aaos for automotive ivi systems,''
  \url{https://www.luxoft.com/blog/gas-or-not-gas-optimizing-aaos-for-automotive-ivi-systems},
  2025, accessed: 2025-07-06.

\bibitem{wikipedia_androidautomotive}
{Wikipedia contributors}, ``Android automotive,''
  \url{https://en.wikipedia.org/wiki/Android_Automotive}, 2025, accessed:
  2025-07-11.

\bibitem{honda2025carsapps}
{Honda Motor Co., Ltd.}, ``Cars \& apps,''
  \url{https://global.honda/en/cars-apps/}, 2025, accessed: 2025-07-07.

\bibitem{volvo2025emulator}
{Volvo Cars}, ``In-car apps: Android emulator for xc40,''
  \url{https://developer.volvocars.com/in-car-apps/android-emulator-xc40/},
  2025, accessed: 2025-07-07.

\bibitem{gm2025invehicleapps}
{General Motors}, ``In‑vehicle apps — gm developer,''
  \url{https://developer.gm.com/in-vehicle-apps}, 2025, accessed: 2025-07-07.

\bibitem{polestar2025developer}
{Polestar}, ``Developer — polestar global,''
  \url{https://www.polestar.com/global/developer/}, 2025, accessed: 2025-07-07.

\bibitem{widdicks2020backfiring}
K.~Widdicks, D.~Pargman, and S.~Bjork, ``Backfiring and favouring: how design
  processes in hci lead to anti-patterns and repentant designers,'' in
  \emph{Proceedings of the 11th Nordic Conference on Human-Computer
  Interaction: Shaping Experiences, Shaping Society}, 2020, pp. 1--12.

\bibitem{cota_darkpatterns}
\BIBentryALTinterwordspacing
Z.~Cota, ``{Dark Patterns: Unfriendly UX Practices},'' 2019 World Interaction
  Design Day Lightning Talk, UX@UA, University of Arizona, 2019, accessed:
  2026-08-05. [Online]. Available:
  \url{https://uxua.arizona.edu/sites/default/files/documents/dark-patterns-unfriendly-ux-practices.pdf}
\BIBentrySTDinterwordspacing

\bibitem{fpf_darkpatterns}
\BIBentryALTinterwordspacing
{Future of Privacy Forum}, ``{``Dark Patterns:'' Manipulative UX Design and the
  Role of Regulation},'' Webinar and event webpage, Mar. 2021, accessed:
  2026-08-05. [Online]. Available:
  \url{https://fpf.org/fpf-event/dark-patterns-manipulative-ux-design-and-the-role-of-regulation/}
\BIBentrySTDinterwordspacing

\bibitem{prolific}
Prolific, ``Prolific - quickly find research participants you'll love,''
  \url{https://www.prolific.com/}, 2026, accessed: 2026-04-22.

\bibitem{qualtrics_website}
\BIBentryALTinterwordspacing
{Qualtrics}, ``Qualtrics,'' \url{https://www.qualtrics.com/}, 2026, accessed:
  2026-02-05. [Online]. Available: \url{https://www.qualtrics.com/}
\BIBentrySTDinterwordspacing

\bibitem{landis1977measurement}
J.~R. Landis and G.~G. Koch, ``The measurement of observer agreement for
  categorical data,'' \emph{biometrics}, pp. 159--174, 1977.

\bibitem{amazonprime_website}
{Amazon}, ``{Amazon Prime},'' \url{https://www.amazon.com/prime}, n.d.,
  accessed: 2026-08-05.

\bibitem{kay_website}
{KAY Jewelers}, ``{KAY Jewelers},'' \url{https://www.kay.com/}, n.d., accessed:
  2026-08-05.

\bibitem{ftc2026gmonstarorder}
\BIBentryALTinterwordspacing
{Federal Trade Commission}, ``{FTC Finalizes Order Settling Allegations that GM
  and OnStar Collected and Sold Geolocation Data Without Consumers' Informed
  Consent},'' Press release, Jan. 2026, accessed: 2026-08-05. [Online].
  Available:
  \url{https://www.ftc.gov/news-events/news/press-releases/2026/01/ftc-finalizes-order-settling-allegations-gm-onstar-collected-sold-geolocation-data-without-consumers}
\BIBentrySTDinterwordspacing

\bibitem{ftc2026gmcomplaint}
\BIBentryALTinterwordspacing
{U.S. Federal Trade Commission}, ``{Complaint, In the Matter of General Motors
  LLC, General Motors Holdings LLC, and OnStar, LLC},'' FTC Matter No. 2423052,
  Docket No. C-4828, Jan. 2026, final complaint, accessed Aug. 5, 2026.
  [Online]. Available:
  \url{https://www.ftc.gov/system/files/ftc_gov/pdf/2423052c4828gmlexisfinalcomplaint.pdf}
\BIBentrySTDinterwordspacing

\bibitem{kowalczyk2023understanding}
M.~Kowalczyk, J.~T. Gunawan, D.~Choffnes, D.~J. Dubois, W.~Hartzog, and
  C.~Wilson, ``Understanding dark patterns in home iot devices,'' in
  \emph{Proceedings of the 2023 CHI Conference on Human Factors in Computing
  Systems}, 2023, pp. 1--27.

\bibitem{gunawan2021comparative}
J.~Gunawan, A.~Pradeep, D.~Choffnes, W.~Hartzog, and C.~Wilson, ``A comparative
  study of dark patterns across web and mobile modalities,'' \emph{Proceedings
  of the ACM on Human-Computer Interaction}, vol.~5, no. CSCW2, pp. 1--29,
  2021.

\bibitem{niknejad2024level}
S.~Niknejad, T.~Mildner, N.~Zargham, S.~Putze, and R.~Malaka, ``Level up or
  game over: Exploring how dark patterns shape mobile games,'' in
  \emph{Proceedings of the International Conference on Mobile and Ubiquitous
  Multimedia}, 2024, pp. 148--156.

\bibitem{mathur2019dark}
A.~Mathur, G.~Acar, M.~J. Friedman, E.~Lucherini, J.~Mayer, M.~Chetty, and
  A.~Narayanan, ``Dark patterns at scale: Findings from a crawl of 11k shopping
  websites,'' \emph{Proceedings of the ACM on human-computer interaction},
  vol.~3, no. CSCW, pp. 1--32, 2019.

\bibitem{bloom2017self}
C.~Bloom, J.~Tan, J.~Ramjohn, and L.~Bauer, ``Self-driving cars and data
  collection: Privacy perceptions of networked autonomous vehicles,'' in
  \emph{Thirteenth symposium on usable privacy and security (soups 2017)},
  2017, pp. 357--375.

\bibitem{kemp2024driving}
K.~Kemp, ``Driving blind: The unexamined privacy risks of connected cars,''
  \emph{Available at SSRN}, 2024.

\bibitem{underarmour_website}
{Under Armour}, ``{Under Armour},'' \url{https://www.underarmour.com/en-us/},
  n.d., accessed: 2026-08-05.

\bibitem{cnn_website}
{CNN}, ``{CNN},'' \url{https://www.cnn.com/}, n.d., accessed: 2026-08-05.

\bibitem{frontier_website}
{Frontier Airlines}, ``{Frontier Airlines},''
  \url{https://www.flyfrontier.com/}, n.d., accessed: 2026-08-05.

\bibitem{ticketmaster_website}
\BIBentryALTinterwordspacing
{Ticketmaster}, ``{Ticketmaster},'' n.d., accessed: 2026-08-05. [Online].
  Available: \url{https://www.ticketmaster.com/}
\BIBentrySTDinterwordspacing

\end{thebibliography}

\clearpage

\appendix
\renewcommand{\thesection}{\Alph{section}}
\setcounter{section}{0}

\refstepcounter{section}
\section*{Appendix A: Emulators}
\label{app:emulators}

\begin{figure}[H]
\centering

\begin{subfigure}[t]{0.48\columnwidth}
    \centering
    \includegraphics[width=0.9\linewidth,height=7cm,keepaspectratio]{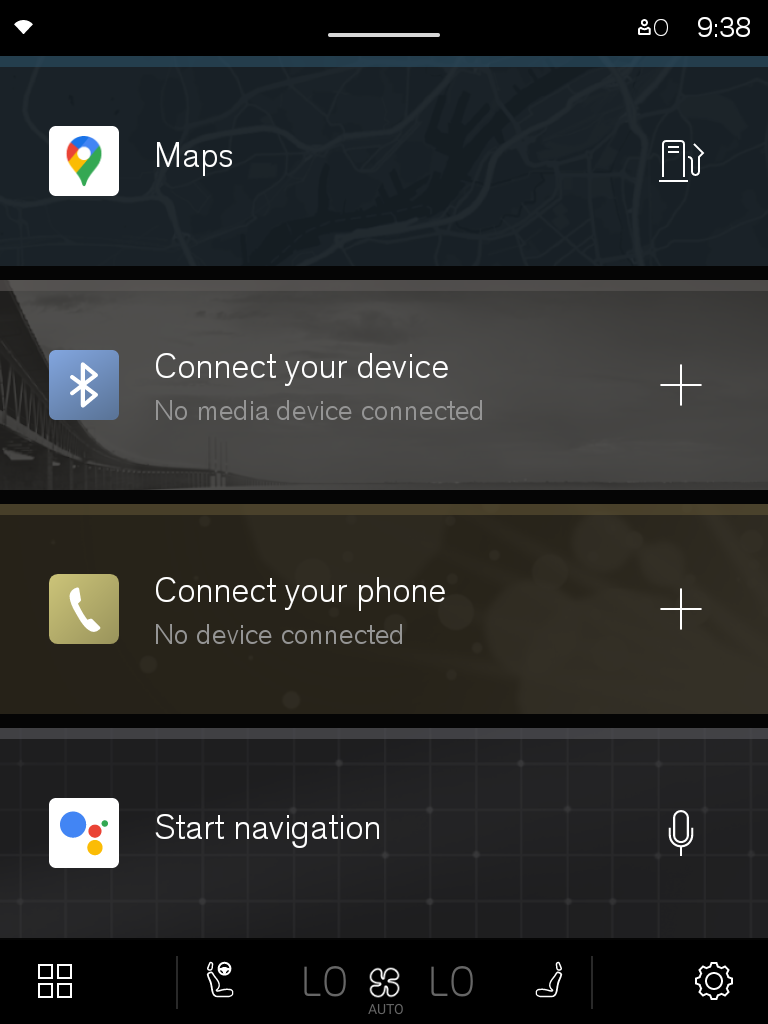}
    \caption{Volvo HOME}
\end{subfigure}
\hfill
\begin{subfigure}[t]{0.48\columnwidth}
    \centering
    \includegraphics[width=0.9\linewidth,height=7cm,keepaspectratio]{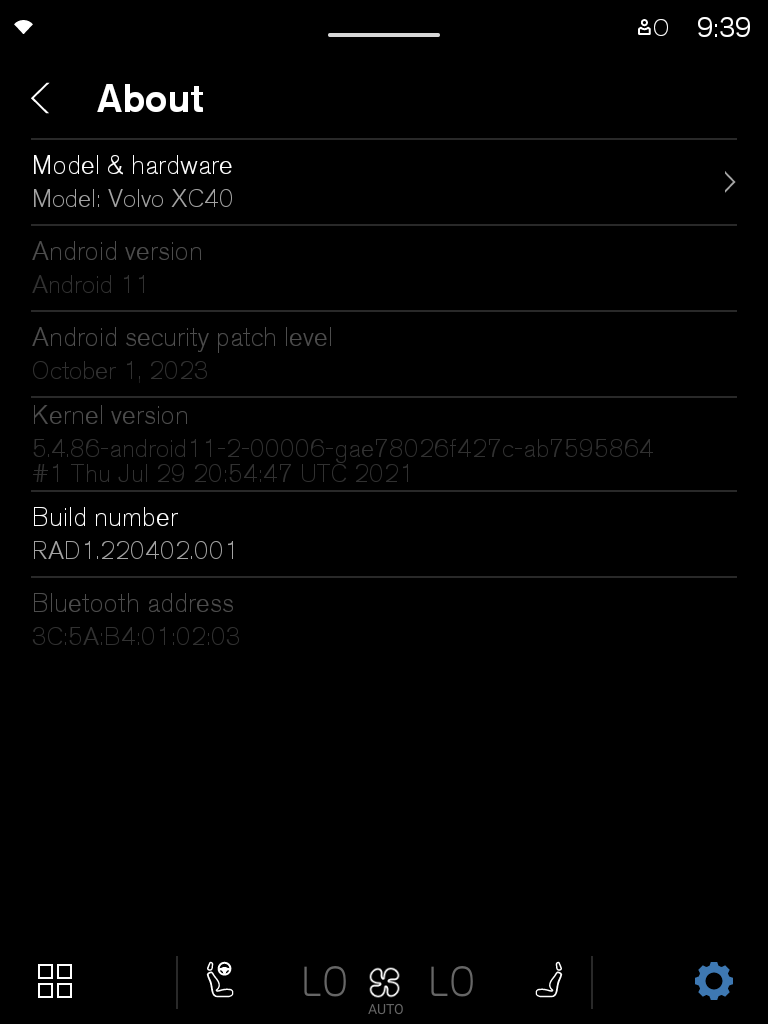}
    \caption{Volvo ABOUT}
\end{subfigure}

\begin{subfigure}[t]{0.48\columnwidth}
    \centering
    \includegraphics[width=0.9\linewidth,height=7cm,keepaspectratio]{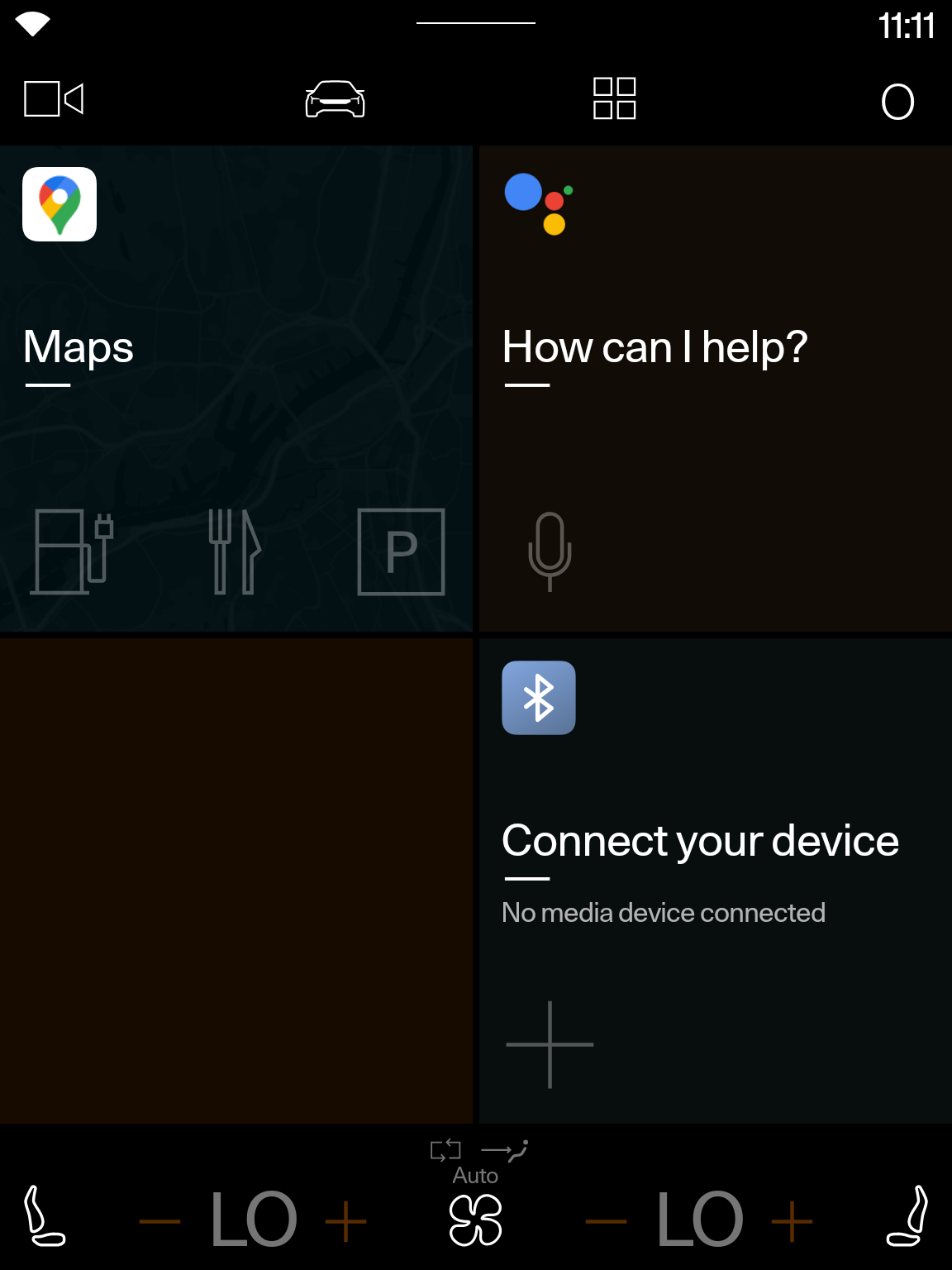}
    \caption{Polestar HOME}
\end{subfigure}
\hfill
\begin{subfigure}[t]{0.48\columnwidth}
    \centering
    \includegraphics[width=0.9\linewidth,height=7cm,keepaspectratio]{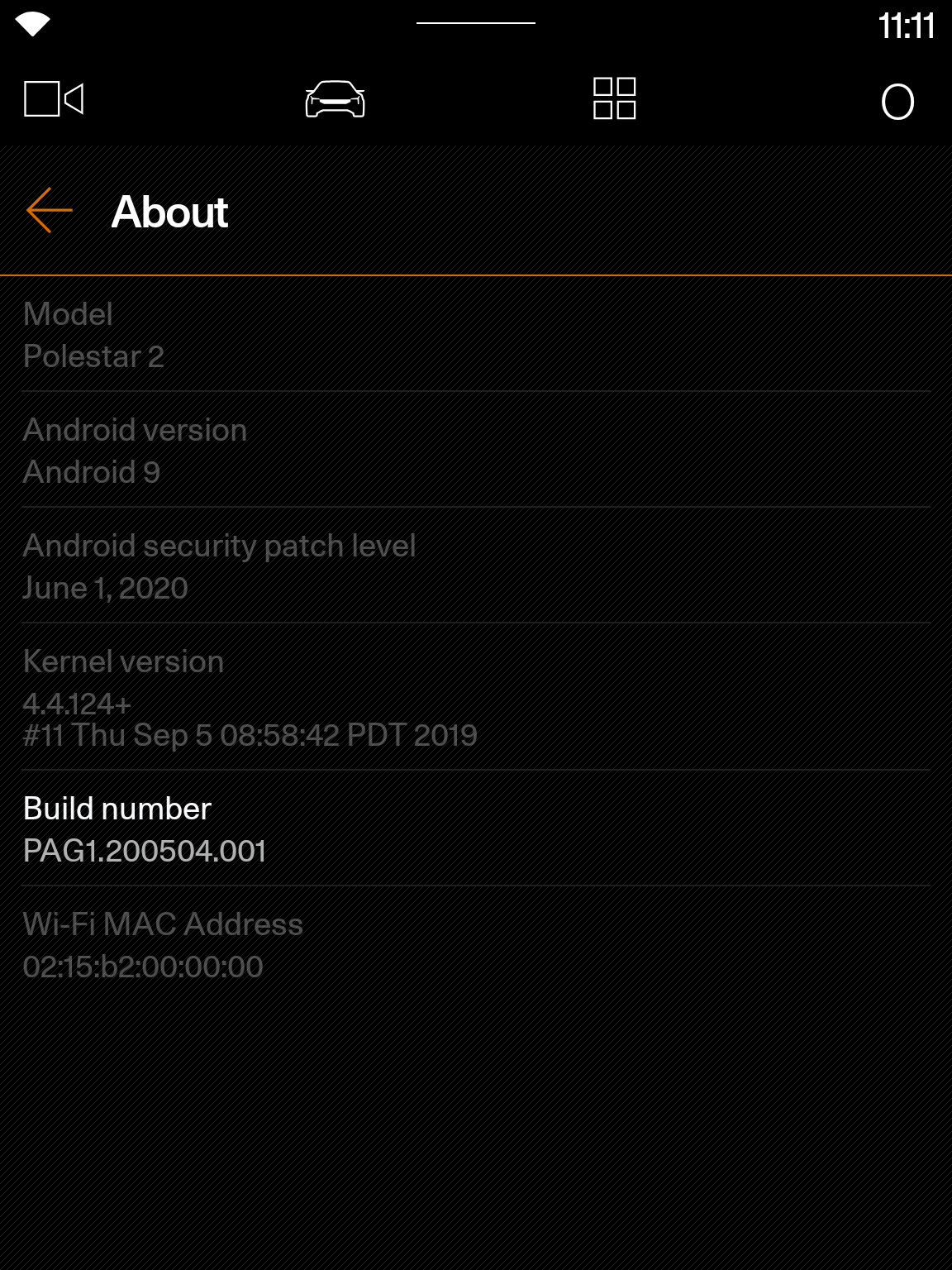}
    \caption{Polestar ABOUT}
\end{subfigure}
\caption{Emulator screenshots (HOME vs ABOUT) for the Volvo and Polestar AAOS builds. Corresponding GMC and Honda emulator screenshots are shown in Figure~\ref{fig:emulator-real-validation}.} 
\label{fig:emulator}
\end{figure}

Table~\ref{tab:emulator-builds} reports the build numbers of the four AAOS emulator builds evaluated in Phase~1. Figure~\ref{fig:emulator} shows the Volvo and Polestar emulator outputs; the corresponding GMC and Honda emulator screenshots are shown in Figures~\ref{fig:emulator-real-validation}. For each emulator shown in Figure~\ref{fig:emulator}, the left image captures the home screen (``HOME''), and the right image captures the \textit{About} screen available in Settings (``ABOUT'').

\begin{table}[H]
\centering
\scriptsize
\setlength{\tabcolsep}{4pt}
\renewcommand{\arraystretch}{1.05}
\caption{Build numbers of the AAOS emulator builds evaluated in Phase~1.}
\label{tab:emulator-builds}

\begin{tabularx}{\columnwidth}{@{} l >{\raggedright\arraybackslash}X @{}}
\toprule
\rowcolor{white}
\textbf{OEM} & \textbf{Build Number} \\
\midrule
\rowcolor{rowgray}
GMC &
\texttt{full\_gminfo37\_car\_emu-user 10 W1S2IY164.2.2-Q HM22BP-543 152 release-keys} \\

\rowcolor{white}
Honda &
\texttt{ACB1216R0.240521.1} \\

\rowcolor{rowgray}
Volvo &
\texttt{RAD1.220402.001} \\

\rowcolor{white}
Polestar &
\texttt{PAG1.200504.001} \\
\bottomrule
\end{tabularx}
\end{table}

\refstepcounter{section}
\section*{Appendix B: Emulator and Real-Vehicle Interface Validation}
\label{app:emulator-validation}

We compared the audited AAOS emulator interfaces with the corresponding production infotainment systems in a 2022 GMC Yukon and a 2023 Honda Accord Hybrid. Figure~\ref{fig:emulator-real-validation} provides side-by-side emulator and real-vehicle images for four corresponding interface screens from each vehicle (Home, Location, Settings, and About), which are visual examples. We performed a full comparison of the audited emulator and corresponding in-vehicle interfaces. Across the examined UIs, we found no fundamental differences, with the visible differences shown in the figure.

\refstepcounter{section}
\section*{Appendix C: Annotator Coverage by Emulator and Pattern}
\label{app:annotator-coverage}

Table~\ref{tab:annotator-emulator-pattern} reports the emulator or emulators examined and the pattern types identified by each annotator. Each row represents one annotator--emulator audit. A value of \(1\) indicates that at least one instance of the corresponding pattern was identified; multiple observations of the same pattern within an audit are collapsed to \(1\). A dash (--) indicates that the annotator did not identify a pattern that was observed by at least one annotator for that OEM, whereas \emph{N/O} indicates that no occurrence of the pattern was observed for that OEM in the audit. The final column gives the number of distinct pattern types identified in that audit.


\begin{table}[H]
\centering

\caption{Annotator--emulator--pattern coverage.}
\label{tab:annotator-emulator-pattern}

{\footnotesize
\setlength{\tabcolsep}{1.2pt}
\renewcommand{\arraystretch}{0.82}

\newcommand{\missdp}{\textcolor{black}{--}}
\newcommand{\nodp}{\textcolor{black}{\emph{N/O}}}

\begin{adjustbox}{max width=0.94\columnwidth}
\begin{tabular}{@{}ll*{17}{c}c@{}}
\toprule
\rowcolor{white}
\textbf{Ann.} &
\textbf{OEM} &
\textbf{Pre} &
\textbf{FC} &
\textbf{PZ} &
\textbf{CS} &
\textbf{PC} &
\textbf{BC} &
\textbf{RM} &
\textbf{Ob} &
\textbf{HC} &
\textbf{IoC} &
\textbf{Sn} &
\textbf{Mis} &
\textbf{IoCh} &
\textbf{II} &
\textbf{FP} &
\textbf{IB} &
\textbf{UN} &
\textbf{Total} \\
\midrule

\rowcolor{rowgray}
A1 & Honda
& \missdp & \missdp & \missdp & \missdp & \missdp & \missdp & \missdp
& \missdp & \missdp & \missdp & 1 & \missdp & \missdp & \missdp
& \nodp & \nodp & \nodp & 1 \\

\rowcolor{white}
A2 & Polestar
& \nodp & \missdp & \missdp & \nodp & \nodp & \nodp & \nodp
& \nodp & \nodp & \nodp & \nodp & \nodp & \nodp & \nodp
& \nodp & \nodp & \nodp & 0 \\

\rowcolor{rowgray}
A3 & Volvo
& \missdp & \missdp & \missdp & \missdp & \missdp & \missdp & \missdp
& \nodp & \missdp & \missdp & \nodp & \nodp & \nodp & \nodp
& \missdp & \missdp & \nodp & 0 \\

\rowcolor{white}
A4 & Honda
& \missdp & 1 & \missdp & \missdp & \missdp & \missdp & \missdp
& \missdp & \missdp & \missdp & \missdp & \missdp & \missdp & \missdp
& \nodp & \nodp & \nodp & 1 \\

\rowcolor{rowgray}
A5 & Volvo
& \missdp & \missdp & \missdp & \missdp & \missdp & \missdp & \missdp
& \nodp & \missdp & \missdp & \nodp & \nodp & \nodp & \nodp
& \missdp & \missdp & \nodp & 0 \\

\rowcolor{white}
A6 & Honda
& \missdp & 1 & \missdp & \missdp & \missdp & \missdp & \missdp
& 1 & \missdp & \missdp & \missdp & \missdp & \missdp & \missdp
& \nodp & \nodp & \nodp & 2 \\

\rowcolor{rowgray}
A7 & Volvo
& \missdp & \missdp & \missdp & \missdp & \missdp & \missdp & \missdp
& \nodp & \missdp & \missdp & \nodp & \nodp & \nodp & \nodp
& 1 & 1 & \nodp & 2 \\

\rowcolor{white}
A8 & Honda
& \missdp & 1 & 1 & \missdp & \missdp & \missdp & \missdp
& 1 & \missdp & \missdp & \missdp & 1 & \missdp & \missdp
& \nodp & \nodp & \nodp & 4 \\

\rowcolor{rowgray}
A9 & GMC
& \missdp & \missdp & \missdp & \nodp & \missdp & 1 & 1
& \nodp & \missdp & \missdp & \nodp & \nodp & \nodp & \nodp
& \nodp & \nodp & \missdp & 2 \\

\rowcolor{white}
A9 & Honda
& \missdp & \missdp & \missdp & \missdp & \missdp & 1 & 1
& \missdp & \missdp & \missdp & \missdp & \missdp & \missdp & \missdp
& \nodp & \nodp & \nodp & 2 \\

\rowcolor{rowgray}
A9 & Volvo
& \missdp & \missdp & \missdp & \missdp & \missdp & 1 & 1
& \nodp & \missdp & \missdp & \nodp & \nodp & \nodp & \nodp
& \missdp & \missdp & \nodp & 2 \\

\rowcolor{white}
A10 & GMC
& \missdp & 1 & \missdp & \nodp & \missdp & \missdp & 1
& \nodp & \missdp & \missdp & \nodp & \nodp & \nodp & \nodp
& \nodp & \nodp & 1 & 3 \\

\rowcolor{rowgray}
A11 & GMC
& 1 & \missdp & \missdp & \nodp & 1 & \missdp & \missdp
& \nodp & 1 & \missdp & \nodp & \nodp & \nodp & \nodp
& \nodp & \nodp & \missdp & 3 \\

\rowcolor{white}
A11 & Honda
& 1 & \missdp & \missdp & \missdp & 1 & \missdp & \missdp
& \missdp & 1 & \missdp & \missdp & \missdp & \missdp & \missdp
& \nodp & \nodp & \nodp & 3 \\

\rowcolor{rowgray}
A11 & Volvo
& 1 & \missdp & \missdp & \missdp & 1 & \missdp & \missdp
& \nodp & 1 & \missdp & \nodp & \nodp & \nodp & \nodp
& \missdp & \missdp & \nodp & 3 \\

\rowcolor{white}
A12 & Honda
& 1 & 1 & \missdp & \missdp & 1 & \missdp & \missdp
& \missdp & \missdp & \missdp & \missdp & \missdp & 1 & \missdp
& \nodp & \nodp & \nodp & 4 \\

\rowcolor{rowgray}
A13 & Honda
& \missdp & \missdp & \missdp & \missdp & \missdp & \missdp & \missdp
& \missdp & \missdp & \missdp & \missdp & \missdp & \missdp & \missdp
& \nodp & \nodp & \nodp & 0 \\

\rowcolor{white}
A13 & Volvo
& \missdp & \missdp & \missdp & \missdp & \missdp & \missdp & \missdp
& \nodp & \missdp & \missdp & \nodp & \nodp & \nodp & \nodp
& \missdp & \missdp & \nodp & 0 \\

\rowcolor{rowgray}
A14 & GMC
& 1 & \missdp & \missdp & \nodp & \missdp & \missdp & \missdp
& \nodp & \missdp & 1 & \nodp & \nodp & \nodp & \nodp
& \nodp & \nodp & \missdp & 2 \\

\rowcolor{white}
A14 & Honda
& 1 & \missdp & \missdp & \missdp & \missdp & \missdp & \missdp
& \missdp & \missdp & 1 & \missdp & \missdp & \missdp & \missdp
& \nodp & \nodp & \nodp & 2 \\

\rowcolor{rowgray}
A14 & Volvo
& 1 & \missdp & \missdp & \missdp & \missdp & \missdp & \missdp
& \nodp & \missdp & 1 & \nodp & \nodp & \nodp & \nodp
& \missdp & \missdp & \nodp & 2 \\

\rowcolor{white}
A15 & Honda
& \missdp & 1 & \missdp & \missdp & \missdp & \missdp & \missdp
& \missdp & \missdp & \missdp & \missdp & \missdp & \missdp & \missdp
& \nodp & \nodp & \nodp & 1 \\

\rowcolor{rowgray}
A16 & Honda
& 1 & 1 & \missdp & 1 & \missdp & \missdp & \missdp
& \missdp & \missdp & \missdp & \missdp & \missdp & \missdp & \missdp
& \nodp & \nodp & \nodp & 3 \\

\rowcolor{white}
A16 & Volvo
& 1 & 1 & \missdp & 1 & \missdp & \missdp & \missdp
& \nodp & \missdp & \missdp & \nodp & \nodp & \nodp & \nodp
& \missdp & \missdp & \nodp & 3 \\

\rowcolor{rowgray}
A17 & GMC
& \missdp & \missdp & 1 & \nodp & \missdp & \missdp & \missdp
& \nodp & \missdp & \missdp & \nodp & \nodp & \nodp & \nodp
& \nodp & \nodp & \missdp & 1 \\

\rowcolor{white}
A17 & Honda
& \missdp & \missdp & 1 & \missdp & \missdp & \missdp & \missdp
& \missdp & \missdp & \missdp & \missdp & \missdp & \missdp & \missdp
& \nodp & \nodp & \nodp & 1 \\

\rowcolor{rowgray}
A17 & Volvo
& \missdp & \missdp & 1 & \missdp & \missdp & \missdp & \missdp
& \nodp & \missdp & \missdp & \nodp & \nodp & \nodp & \nodp
& \missdp & \missdp & \nodp & 1 \\

\rowcolor{white}
A18 & Honda
& 1 & \missdp & \missdp & \missdp & \missdp & \missdp & \missdp
& \missdp & \missdp & \missdp & \missdp & \missdp & \missdp & \missdp
& \nodp & \nodp & \nodp & 1 \\

\rowcolor{rowgray}
A19 & Honda
& \missdp & 1 & \missdp & \missdp & \missdp & \missdp & \missdp
& \missdp & \missdp & \missdp & \missdp & \missdp & \missdp & 1
& \nodp & \nodp & \nodp & 2 \\

\rowcolor{white}
A20 & Volvo
& 1 & \missdp & \missdp & 1 & \missdp & \missdp & \missdp
& \nodp & \missdp & \missdp & \nodp & \nodp & \nodp & \nodp
& \missdp & \missdp & \nodp & 2 \\

\rowcolor{rowgray}
A21 & Honda
& \missdp & \missdp & \missdp & \missdp & \missdp & \missdp & \missdp
& \missdp & \missdp & \missdp & \missdp & \missdp & \missdp & \missdp
& \nodp & \nodp & \nodp & 0 \\

\rowcolor{white}
A22 & Honda
& \missdp & 1 & 1 & \missdp & \missdp & \missdp & \missdp
& \missdp & \missdp & \missdp & \missdp & \missdp & \missdp & \missdp
& \nodp & \nodp & \nodp & 2 \\

\rowcolor{rowgray}
A22 & Polestar
& \nodp & 1 & 1 & \nodp & \nodp & \nodp & \nodp
& \nodp & \nodp & \nodp & \nodp & \nodp & \nodp & \nodp
& \nodp & \nodp & \nodp & 2 \\

\rowcolor{white}
A23 & Honda
& \missdp & \missdp & \missdp & 1 & \missdp & 1 & \missdp
& 1 & \missdp & \missdp & \missdp & \missdp & \missdp & \missdp
& \nodp & \nodp & \nodp & 3 \\

\rowcolor{rowgray}
A24 & Honda
& \missdp & 1 & \missdp & 1 & \missdp & \missdp & \missdp
& \missdp & \missdp & \missdp & \missdp & \missdp & \missdp & \missdp
& \nodp & \nodp & \nodp & 2 \\

\rowcolor{white}
A25 & Volvo
& 1 & \missdp & \missdp & \missdp & \missdp & \missdp & \missdp
& \nodp & \missdp & \missdp & \nodp & \nodp & \nodp & \nodp
& \missdp & \missdp & \nodp & 1 \\

\rowcolor{rowgray}
A26 & GMC
& 1 & \missdp & \missdp & \nodp & 1 & \missdp & \missdp
& \nodp & \missdp & \missdp & \nodp & \nodp & \nodp & \nodp
& \nodp & \nodp & \missdp & 2 \\

\bottomrule
\end{tabular}
\end{adjustbox}
}

\vspace{0.25em}

{\scriptsize
\begin{minipage}{\columnwidth}
\textit{Abbreviations:}
Pre = Preselection;
FC = Forced Consent;
PZ = Privacy Zuckering;
CS = Confirmshaming;
PC = Permission Creep;
BC = Bundling Consent;
RM = Roach Motel;
Ob = Obstruction;
HC = Hidden Costs;
IoC = Illusion of Control;
Sn = Sneaking;
Mis = Misdirection;
IoCh = Illusion of Choice;
II = Interface Interference;
FP = False Promises;
IB = Identical Buttons; and
UN = Upgrade Nagging.
A1--A26 denote annotator identifiers.

\textcolor{black}{A value of \(1\) indicates that the annotator identified the
pattern; ``--'' indicates that the annotator did not identify a pattern that
was observed by at least one annotator for that OEM; and \emph{N/O} indicates
that no occurrence of the pattern was identified by any annotator for that OEM in the audit.}
\end{minipage}
}
\end{table}

\refstepcounter{section}
\section*{Appendix D: Annotator Discovery Convergence}
\label{app:annotator-discovery-convergence}

We define the discovery convergence rate for pattern--OEM combinations in which the pattern was observed at least once as \(n/N\), where \(n\) is the number of annotators who identified the pattern and \(N\) is the number who audited that OEM. Figure~\ref{fig:annotator-discovery-convergence} visualizes the corresponding percentages for these combinations. Combinations with no observed occurrence are shown as \emph{N/O} (not observed), which does not establish absence from the emulator because the audit was exploratory. Each annotator contributed at most one identification to a given pattern--OEM combination, regardless of the number of corresponding observations submitted.

Preselection had convergence rates of \(27.8\%\), \(45.5\%\), and \(50.0\%\) in Honda, Volvo, and GMC, respectively. Forced Consent was identified by nine of the 18 Honda annotators (\(50.0\%\)), compared with \(9.1\%\) in Volvo and \(16.7\%\) in GMC. Privacy Zuckering was identified by \(9.1\%\) to \(16.7\%\) of annotators across these three OEMs, while Sneaking was identified by one Honda annotator (\(5.6\%\)). GMC and Polestar results should be interpreted cautiously because only six and two annotators audited those emulators, respectively. All candidate observations were jointly reviewed by the two authors and the automotive-privacy specialist, and any disagreements were reconciled.

\begin{figure}[H]
\centering

\begin{tikzpicture}[x=1.05cm,y=0.42cm]

\newcommand{\hmcell}[4]{%
    \filldraw[fill=blue!#4,draw=white,line width=0.3pt]
        (#1,#2) rectangle ++(1,-1);
    \node[font=\tiny] at ({#1+0.5},{#2-0.5}) {#3};
}

\newcommand{\hmno}[2]{%
    \filldraw[fill=gray!22,draw=white,line width=0.3pt]
        (#1,#2) rectangle ++(1,-1);
    \node[font=\tiny] at ({#1+0.5},{#2-0.5}) {\textit{N/O}};
}

\node[font=\scriptsize\bfseries,align=center] at (0.5,0.65)
    {Honda\\{\tiny \(N=18\)}};

\node[font=\scriptsize\bfseries,align=center] at (1.5,0.65)
    {Volvo\\{\tiny \(N=11\)}};

\node[font=\scriptsize\bfseries,align=center] at (2.5,0.65)
    {GMC\\{\tiny \(N=6\)}};

\node[font=\scriptsize\bfseries,align=center] at (3.5,0.65)
    {Polestar\\{\tiny \(N=2\)}};
\node[font=\tiny\bfseries,anchor=east] at (-0.12,-0.5)
    {Preselection};

\node[font=\tiny\bfseries,anchor=east] at (-0.12,-1.5)
    {Forced Consent};

\node[font=\tiny\bfseries,anchor=east] at (-0.12,-2.5)
    {Privacy Zuckering};

\node[font=\tiny\bfseries,anchor=east] at (-0.12,-3.5)
    {Confirmshaming};

\node[font=\tiny\bfseries,anchor=east] at (-0.12,-4.5)
    {Permission Creep};

\node[font=\tiny\bfseries,anchor=east] at (-0.12,-5.5)
    {Bundling Consent};

\node[font=\tiny\bfseries,anchor=east] at (-0.12,-6.5)
    {Roach Motel};

\node[font=\tiny\bfseries,anchor=east] at (-0.12,-7.5)
    {Obstruction};

\node[font=\tiny\bfseries,anchor=east] at (-0.12,-8.5)
    {Hidden Costs};

\node[font=\tiny\bfseries,anchor=east] at (-0.12,-9.5)
    {Illusion of Control};

\node[font=\tiny\bfseries,anchor=east] at (-0.12,-10.5)
    {Sneaking};

\node[font=\tiny\bfseries,anchor=east] at (-0.12,-11.5)
    {Misdirection};

\node[font=\tiny\bfseries,anchor=east] at (-0.12,-12.5)
    {Illusion of Choice};

\node[font=\tiny\bfseries,anchor=east] at (-0.12,-13.5)
    {Interface Interference};

\node[font=\tiny\bfseries,anchor=east] at (-0.12,-14.5)
    {False Promises};

\node[font=\tiny\bfseries,anchor=east] at (-0.12,-15.5)
    {Identical Buttons};

\node[font=\tiny\bfseries,anchor=east] at (-0.12,-16.5)
    {Upgrade Nagging};
    

\hmcell{0}{0}{27.8\%}{32}
\hmcell{1}{0}{45.5\%}{50}
\hmcell{2}{0}{50.0\%}{55}
\hmno{3}{0}

\hmcell{0}{-1}{50.0\%}{55}
\hmcell{1}{-1}{9.1\%}{12}
\hmcell{2}{-1}{16.7\%}{20}
\hmcell{3}{-1}{50.0\%}{55}

\hmcell{0}{-2}{16.7\%}{20}
\hmcell{1}{-2}{9.1\%}{12}
\hmcell{2}{-2}{16.7\%}{20}
\hmcell{3}{-2}{50.0\%}{55}

\hmcell{0}{-3}{16.7\%}{20}
\hmcell{1}{-3}{18.2\%}{22}
\hmno{2}{-3}
\hmno{3}{-3}

\hmcell{0}{-4}{11.1\%}{15}
\hmcell{1}{-4}{9.1\%}{12}
\hmcell{2}{-4}{33.3\%}{38}
\hmno{3}{-4}

\hmcell{0}{-5}{11.1\%}{15}
\hmcell{1}{-5}{9.1\%}{12}
\hmcell{2}{-5}{16.7\%}{20}
\hmno{3}{-5}

\hmcell{0}{-6}{5.6\%}{8}
\hmcell{1}{-6}{9.1\%}{12}
\hmcell{2}{-6}{33.3\%}{38}
\hmno{3}{-6}

\hmcell{0}{-7}{16.7\%}{20}
\hmno{1}{-7}
\hmno{2}{-7}
\hmno{3}{-7}

\hmcell{0}{-8}{5.6\%}{8}
\hmcell{1}{-8}{9.1\%}{12}
\hmcell{2}{-8}{16.7\%}{20}
\hmno{3}{-8}

\hmcell{0}{-9}{5.6\%}{8}
\hmcell{1}{-9}{9.1\%}{12}
\hmcell{2}{-9}{16.7\%}{20}
\hmno{3}{-9}

\hmcell{0}{-10}{5.6\%}{8}
\hmno{1}{-10}
\hmno{2}{-10}
\hmno{3}{-10}

\hmcell{0}{-11}{5.6\%}{8}
\hmno{1}{-11}
\hmno{2}{-11}
\hmno{3}{-11}

\hmcell{0}{-12}{5.6\%}{8}
\hmno{1}{-12}
\hmno{2}{-12}
\hmno{3}{-12}

\hmcell{0}{-13}{5.6\%}{8}
\hmno{1}{-13}
\hmno{2}{-13}
\hmno{3}{-13}

\hmno{0}{-14}
\hmcell{1}{-14}{9.1\%}{12}
\hmno{2}{-14}
\hmno{3}{-14}

\hmno{0}{-15}
\hmcell{1}{-15}{9.1\%}{12}
\hmno{2}{-15}
\hmno{3}{-15}

\hmno{0}{-16}
\hmno{1}{-16}
\hmcell{2}{-16}{16.7\%}{20}
\hmno{3}{-16}

\node[font=\tiny,anchor=east] at (0.55,-18.0)
    {Discovery convergence};

\shade[left color=blue!5,right color=blue!55]
    (0.70,-17.75) rectangle (3.50,-18.15);

\node[font=\tiny,anchor=north] at (0.70,-18.18)
    {0\%};

\node[font=\tiny,anchor=north] at (3.50,-18.18)
    {50\%};

\end{tikzpicture}

\caption{\textcolor{black}{Annotator discovery convergence by pattern and OEM. Darker cells indicate greater discovery convergence; \emph{N/O} denotes no observed occurrence in the audit.}}
\label{fig:annotator-discovery-convergence}

\end{figure}

\refstepcounter{section}
\section*{Appendix E: Pilot Study}
\label{app:pilot}

\textbf{Pilot Study for Phase~2.}
Our initial Phase~2 design presented all OEM implementations of a dark pattern side-by-side on a single page. Each page showed three videos arranged in a grid (one per OEM: GMC, Honda, Volvo), each independently playable and replayable, with the eleven attitude scales presented below the grid. With five dark patterns, each participant completed five such pages.

We piloted this design with five participants. Pilot participants reported the side-by-side comparison task as cognitively demanding, and the page layout encouraged rating OEM variants against one another rather than on absolute terms, threatening the independence of the eleven construct ratings. On the basis of these pilot observations, we revised Phase~2 to the mixed design reported in the main paper: OEM as a between-subjects factor per dark pattern, with each participant rating exactly one randomly assigned OEM implementation per pattern. The revised design costs more participants ($N{=}100$ rather than the smaller sample required by the side-by-side version) but eliminates the comparative-anchoring and working-memory issues observed in the pilot.

\textbf{Pilot Study for Phase~3.}
Phase~3's two-round (pre/post-instruction) design amplifies fatigue risk: every clip is rated twice, so each additional stimulus doubles in total rating cost. Because post-instruction ratings are the primary source of our cross-domain inferences, the stimulus count had to balance dark-pattern coverage against sustained response quality in the post-instruction block. We piloted three candidate clip counts (10, 14, and 18 clips, split evenly between automotive and e-commerce) with small convenience samples ($n = 5$ to $8$ per pilot). At 10 clips, pilots showed adequate response quality but insufficient coverage of the Phase~1 pattern catalog, leaving several core pattern categories unrepresented in at least one domain. At 18 clips, pilots showed signs of fatigue in the post-instruction block, with reduced rating variance and longer session times. At 14 clips, pilots achieved adequate coverage (seven clips per domain, permitting each core pattern category to appear at least once per domain) while keeping the post-instruction block within a range where response quality remained stable. We fixed the clip count at 14 before main data collection began.


\refstepcounter{section}
\section*{Appendix F: Survey Demographics}
\label{app:demographics}

\autoref{tab:demographics} presents the demographics of our participants across both empirical studies.

\begin{table}[t]
\centering
\scriptsize
\caption{\textcolor{black}{Survey participant demographics for Phase~2
         ($N{=}100$) and Phase~3 ($N{=}49$). Dashes
         (``--'') indicate items not collected in that
         phase.}}
\label{tab:demographics}
\begin{tabular}{lcccc}
\toprule
 & \multicolumn{2}{c}{\textbf{Phase 2}} &
   \multicolumn{2}{c}{\textbf{Phase 3}} \\
\cmidrule(lr){2-3}\cmidrule(lr){4-5}
\textbf{Demographic Data} & \textbf{N} & \textbf{\%}
                          & \textbf{N} & \textbf{\%} \\
\midrule
\textit{Age Group} \\
\quad 18--25               & 14 & 14.00 &  4 &  8.16 \\
\quad 26--35               & 31 & 31.00 & 17 & 34.69 \\
\quad 36--45               & 26 & 26.00 &  9 & 18.37 \\
\quad 46 and above         & 29 & 29.00 & 19 & 38.78 \\
\midrule
\textit{Race / Ethnicity} \\
\quad White                 & 57 & 57.00 & 38 & 77.55 \\
\quad Asian                 & 13 & 13.00 &  5 & 10.20 \\
\quad Hispanic / Latino     &  6 &  6.00 &  4 &  8.16 \\
\quad Black / African Am.   & 24 & 24.00 &  2 &  4.08 \\
\midrule
\textit{Gender} \\
\quad Female                & 51 & 51.00 & 24 & 48.98 \\
\quad Male                  & 49 & 49.00 & 25 & 51.02 \\
\midrule
\textit{AAOS Awareness} \\
\quad Heard of it, unclear  & 30 & 30.00 & 22 & 44.90 \\
\quad Knows what it is      & 35 & 35.00 & 16 & 32.65 \\
\quad Never heard of it     & 35 & 35.00 & 11 & 22.45 \\
\midrule
\textit{AAOS Experience} \\
\quad Never used            & 83 & 83.00 & 25 & 51.02 \\
\quad Seen used by others   & -- &  --   & 10 & 20.41 \\
\quad Currently uses        & 17 & 17.00 &  5 & 10.20 \\
\quad Used briefly          & -- &  --   &  5 & 10.20 \\
\quad Used several times    & -- &  --   &  4 &  8.16 \\
\bottomrule
\end{tabular}
\end{table}

\refstepcounter{section}
\section*{Appendix G: Detailed RQ2.3 Contingency Analyses}
\label{app:rq2-3-contingency}

We tested associations between the seven qualitative themes and the two post-survey behavioral-intention measures. Each theme was coded as present or absent and crossed with the four response categories for each measure, producing \(2 \times 4\) contingency tables. Because every table contained at least one expected cell count below five, significance was assessed using Fisher--Freeman--Halton tests with \(200{,}000\) Monte Carlo resamples. Bonferroni correction was applied separately across the seven themes for each behavioral-intention measure (\(\alpha=0.05/7=0.0071\)). Table~\ref{tab:rq2-3-contingency} reports theme-present/theme-absent counts, descriptive Pearson \(\chi^2\) statistics, raw and Bonferroni-corrected \(p\)-values, and Cramér's \(V\). Because several response categories contain few participants, subgroup-level associations should be interpreted cautiously.

\begin{table}[H]
\centering
\footnotesize
\setlength{\tabcolsep}{4pt}
\renewcommand{\arraystretch}{0.95}

\caption{Associations between qualitative themes and behavioral-intention measures. Response-category entries are theme-present/theme-absent counts; \(p_{\mathrm{adj}}\) denotes Bonferroni-corrected \(p\)-values.}
\label{tab:rq2-3-contingency}

\textbf{(a) Consent-screen reading intention}

\vspace{2pt}

\resizebox{\columnwidth}{!}{%
\begin{tabular}{lrrrrrrr}
\toprule

\rowcolor{white}
\textbf{Measure} &
\textbf{T\&C} &
\textbf{C\&F} &
\textbf{C\&P} &
\textbf{F/NC} &
\textbf{TD} &
\textbf{TU} &
\textbf{AR} \\
\midrule

\rowcolor{rowgray}
\shortstack[l]{Read More Carefully\\(\(n=80\))}
& 58/22 & 53/27 & 23/57 & 2/78 & 57/23 & 17/63 & 13/67 \\

\rowcolor{white}
\shortstack[l]{No Difference\\(\(n=7\))}
& 3/4 & 1/6 & 1/6 & 3/4 & 0/7 & 6/1 & 0/7 \\

\rowcolor{rowgray}
\shortstack[l]{No Real Choice Anyway\\(\(n=7\))}
& 3/4 & 5/2 & 1/6 & 0/7 & 2/5 & 5/2 & 0/7 \\

\rowcolor{white}
\shortstack[l]{Already Reading Carefully\\(\(n=4\))}
& 2/2 & 3/1 & 1/3 & 0/4 & 1/3 & 3/1 & 0/4 \\

\midrule

\rowcolor{rowgray}
\(\chi^2(3)\)
& 5.332 & 7.970 & 1.284 & 22.319 & 19.794 & 22.059 & 3.372 \\

\rowcolor{white}
\(p_{\mathrm{raw}}\)
& 0.109974 & 0.046205 & 0.870176 & 0.007245
& 0.000055 & 0.000025 & 0.687297 \\

\rowcolor{rowgray}
\(p_{\mathrm{adj}}\)
& 0.769821 & 0.323433 & 1.000000 & 0.050715
& \textbf{0.000385} & \textbf{0.000175} & 1.000000 \\

\rowcolor{white}
\(V\)
& 0.233 & 0.285 & 0.114 & 0.477 & 0.449 & 0.474 & 0.186 \\

\bottomrule
\end{tabular}%
}

\vspace{8pt}

\textbf{(b) Privacy-toggle intention}

\vspace{2pt}

\resizebox{\columnwidth}{!}{%
\begin{tabular}{lrrrrrrr}
\toprule

\rowcolor{white}
\textbf{Measure} &
\textbf{T\&C} &
\textbf{C\&F} &
\textbf{C\&P} &
\textbf{F/NC} &
\textbf{TD} &
\textbf{TU} &
\textbf{AR} \\
\midrule

\rowcolor{rowgray}
\shortstack[l]{Through Each Toggle Carefully\\(\(n=76\))}
& 54/22 & 51/25 & 25/51 & 0/76 & 50/26 & 22/54 & 10/66 \\

\rowcolor{white}
\shortstack[l]{Turn Everything On\\(\(n=2\))}
& 2/0 & 1/1 & 0/2 & 0/2 & 1/1 & 0/2 & 1/1 \\

\rowcolor{rowgray}
\shortstack[l]{Turn Everything Off\\(\(n=12\))}
& 6/6 & 9/3 & 1/11 & 1/11 & 8/4 & 3/9 & 2/10 \\

\rowcolor{white}
\shortstack[l]{Accept Defaults\\(\(n=8\))}
& 4/4 & 1/7 & 0/8 & 4/4 & 1/7 & 6/2 & 0/8 \\

\midrule

\rowcolor{rowgray}
\(\chi^2(3)\)
& 4.181 & 10.216 & 7.229 & 37.760 & 8.923 & 8.380 & 3.691 \\

\rowcolor{white}
\(p_{\mathrm{raw}}\)
& 0.268974 & 0.010495 & 0.071890 & 0.000045
& 0.019750 & 0.047145 & 0.278124 \\

\rowcolor{rowgray}
\(p_{\mathrm{adj}}\)
& 1.000000 & 0.073465 & 0.503227 & \textbf{0.000315}
& 0.138249 & 0.330013 & 1.000000 \\

\rowcolor{white}
\(V\)
& 0.207 & 0.323 & 0.272 & 0.621 & 0.302 & 0.292 & 0.194 \\

\bottomrule
\end{tabular}%
}

\vspace{2pt}

\begin{minipage}{\columnwidth}
\scriptsize
\textit{Abbreviations:}
T\&C = Transparency \& Clarity;
C\&F = Control \& Fair Design;
C\&P = Contextual \& Purposeful Consent;
F/NC = Fatalism / No Change;
TD = Trust Decreased;
TU = Trust Unchanged; and
AR = Awareness Raised.
\end{minipage}

\end{table}

\refstepcounter{section}
\section*{Appendix H: Phase~3 E-Commerce Stimuli}
\label{app:ecomm-stimuli}

Table~\ref{tab:ecomm-stimuli} summarizes the source, access date, and dark-pattern type for each Phase~3 e-commerce stimulus. The seven e-commerce stimuli were selected to represent dark-pattern types also present among the seven automotive stimuli.

\begin{table}[H]
\centering
\scriptsize
\setlength{\tabcolsep}{2pt}
\renewcommand{\arraystretch}{1.10}

\caption{Phase~3 e-commerce stimuli.}
\label{tab:ecomm-stimuli}

\resizebox{\columnwidth}{!}{%
\begin{tabular}{@{}
>{\centering\arraybackslash}m{0.28\columnwidth}
>{\centering\arraybackslash}m{0.25\columnwidth}
>{\centering\arraybackslash}m{0.42\columnwidth}
@{}}
\toprule

\rowcolor{white}
\textbf{\shortstack{Source\\(Access Date)}} &
\textbf{Pattern} &
\textbf{Pattern Description} \\
\midrule

\rowcolor{rowgray}
\shortstack{Under Armour~\cite{underarmour_website}\\
(September 21, 2025)}
&
Bundling Consent
&
Multiple permissions are combined into one decision. \\

\rowcolor{white}
\shortstack{CNN~\cite{cnn_website}\\
(December 10, 2025)}
&
Forced Consent
&
Acceptance is required to continue. \\

\rowcolor{rowgray}
\shortstack{Frontier Airlines~\cite{frontier_website}\\
(September 21, 2025)}
&
Privacy Zuckering
&
The data-sharing control is obscured. \\

\rowcolor{white}
\shortstack{KAY Jewelers~\cite{kay_website}\\
(September 21, 2025)}
&
Confirmshaming
&
The refusal option is discouraged. \\

\rowcolor{rowgray}
\shortstack{Amazon Prime~\cite{amazonprime_website}\\
(December 14, 2025)}
&
Roach Motel
&
Joining is easier than leaving. \\

\rowcolor{white}
\shortstack{Frontier Airlines~\cite{frontier_website}\\
(September 21, 2025)}
&
Illusion of Control
&
The controls provide no meaningful choice. \\

\rowcolor{rowgray}
\shortstack{Ticketmaster~\cite{ticketmaster_website}\\
(September 21, 2025)}
&
Hidden Costs
&
Additional costs appear late in the interaction. \\

\bottomrule
\end{tabular}%
}
\end{table}


\begin{figure*}[p]
\centering
\captionsetup[subfigure]{labelformat=empty}
\setlength{\tabcolsep}{5pt}
\renewcommand{\arraystretch}{1.0}

\begin{tabular}{cc}

\begin{subfigure}[t]{0.47\textwidth}
    \centering
    \includegraphics[width=0.7\linewidth,height=2.25cm]
    {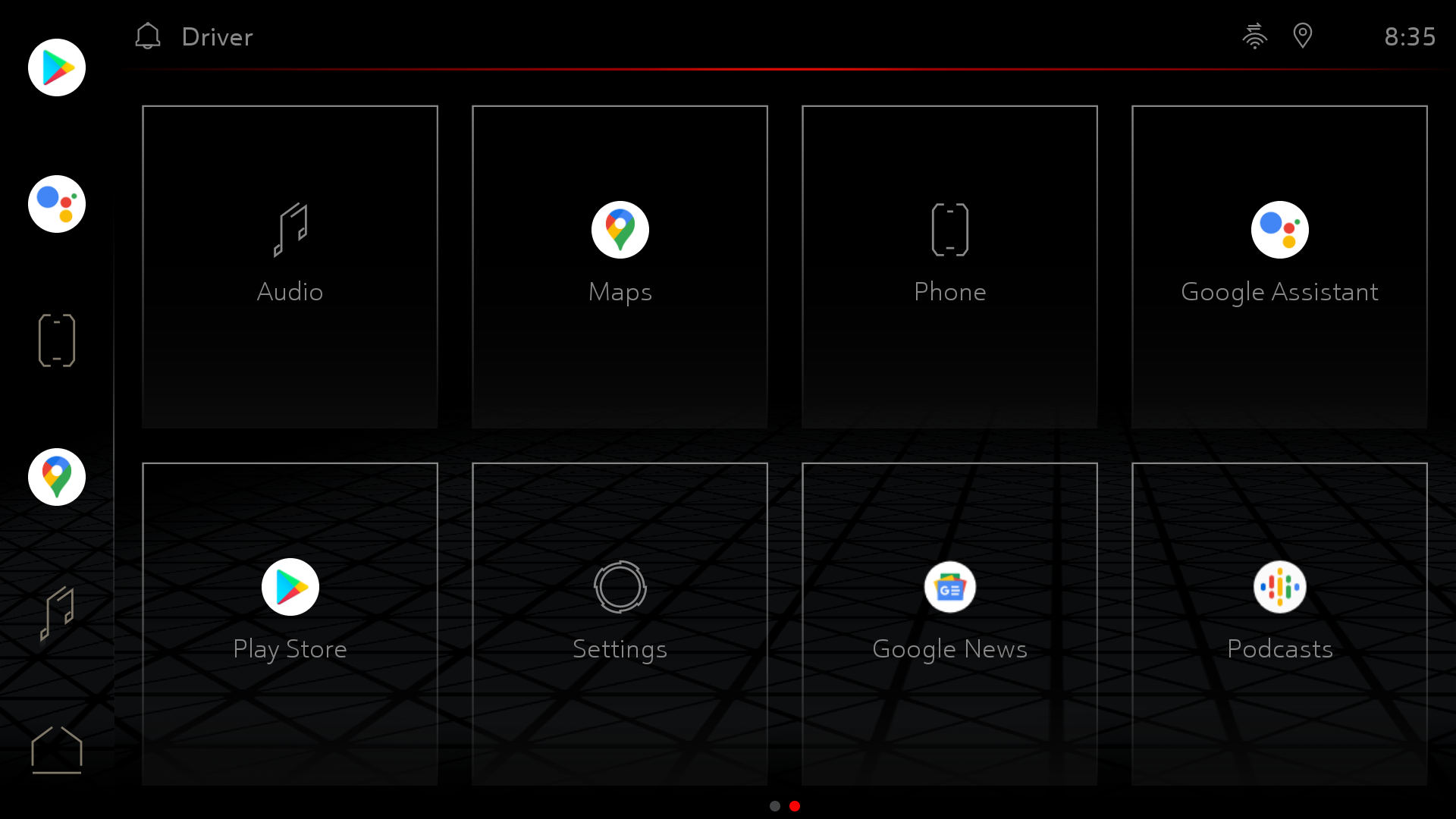}
    \caption{GMC emulator Home.}
    \label{fig:gmc-emulator-home}
\end{subfigure}
&
\begin{subfigure}[t]{0.47\textwidth}
    \centering
    \includegraphics[width=0.7\linewidth,height=2.25cm]
    {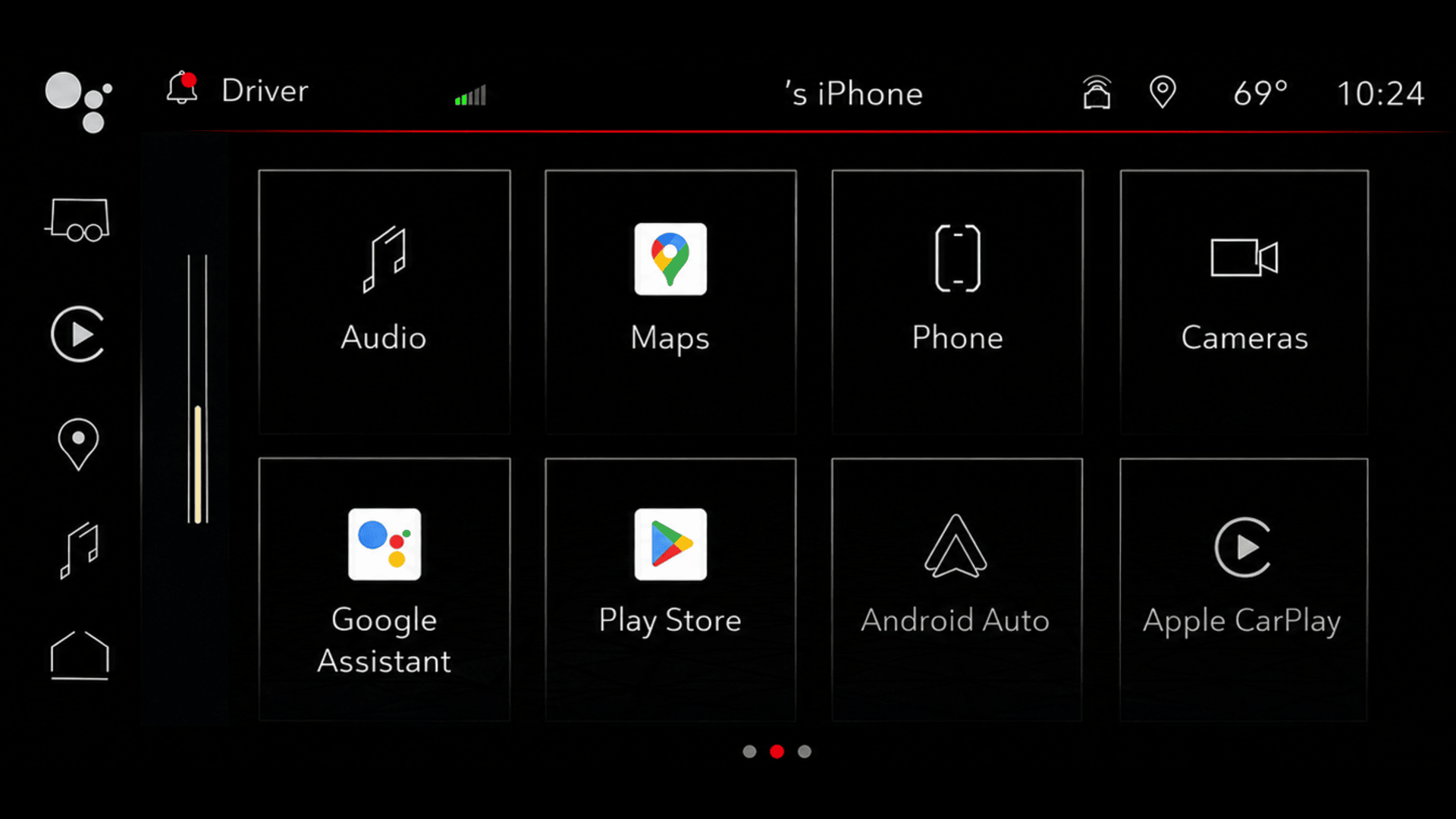}
    \caption{GMC real-vehicle Home.}
    \label{fig:gmc-real-home}
\end{subfigure}
\\[2pt]

\begin{subfigure}[t]{0.47\textwidth}
    \centering
    \includegraphics[width=0.7\linewidth,height=2.25cm]
    {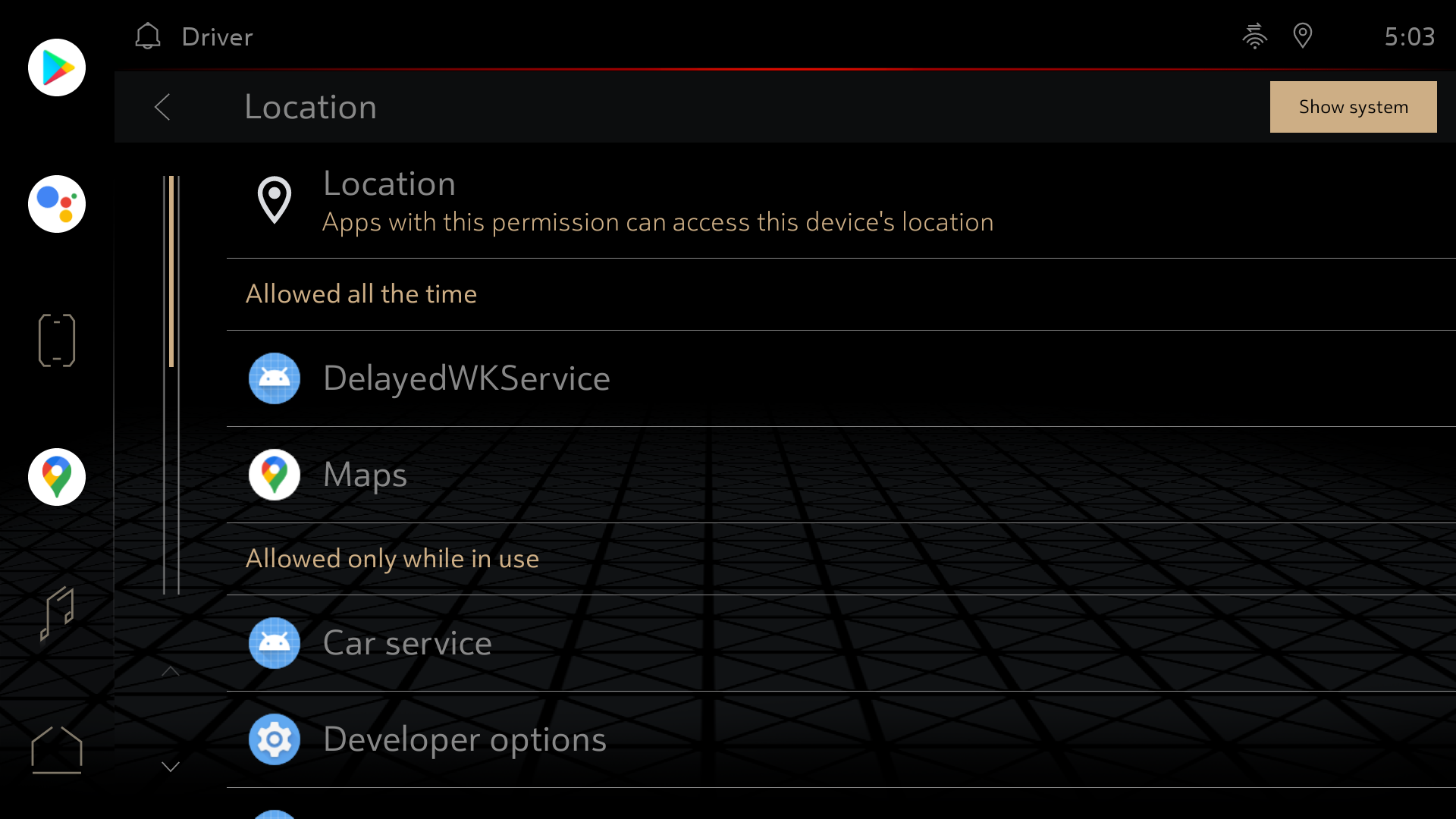}
    \caption{GMC emulator Location.}
    \label{fig:gmc-emulator-location}
\end{subfigure}
&
\begin{subfigure}[t]{0.47\textwidth}
    \centering
    \includegraphics[width=0.7\linewidth,height=2.25cm]
    {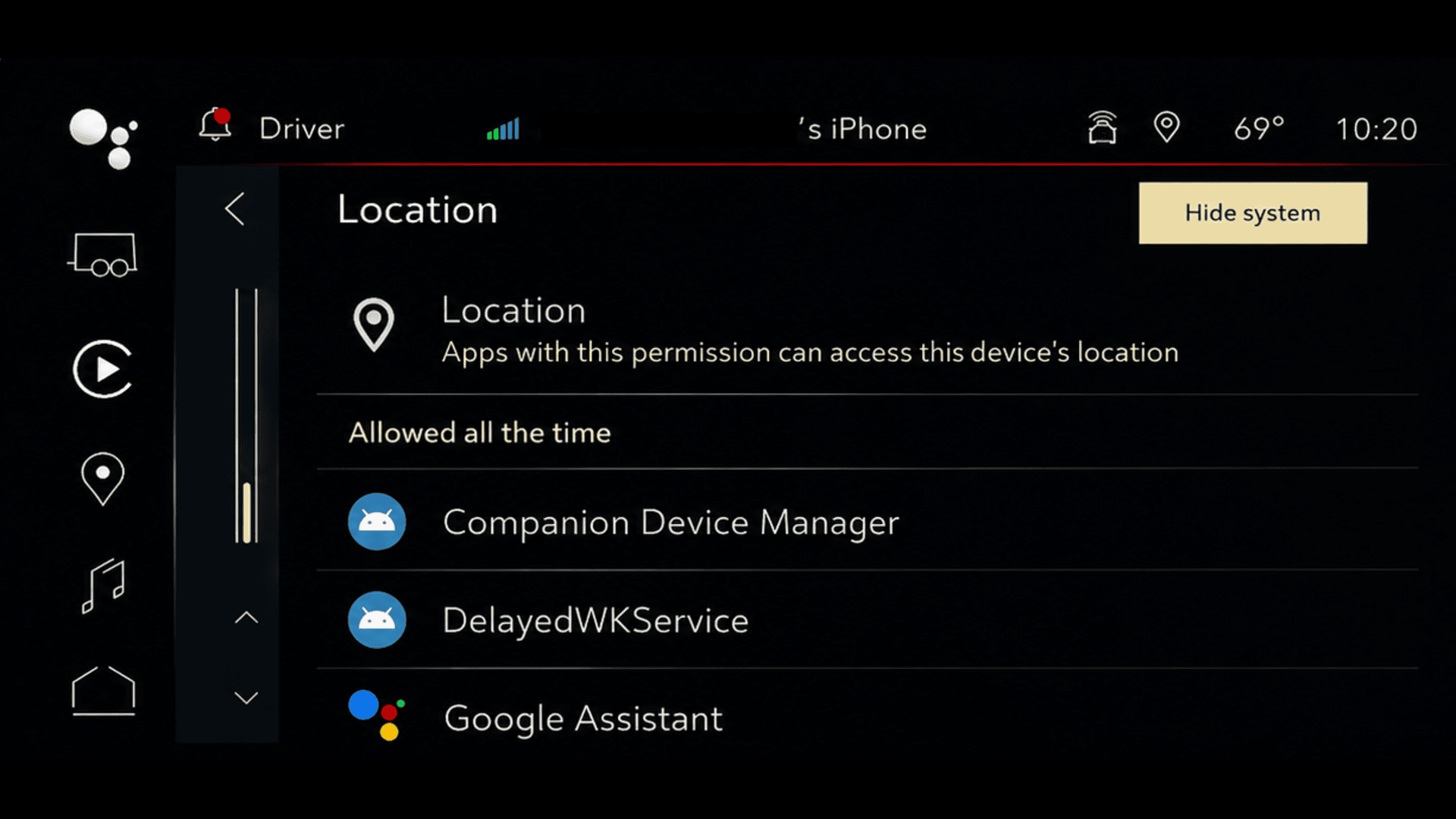}
    \caption{GMC real-vehicle Location.}
    \label{fig:gmc-real-location}
\end{subfigure}
\\[2pt]

\begin{subfigure}[t]{0.47\textwidth}
    \centering
    \includegraphics[width=0.7\linewidth,height=2.25cm]
    {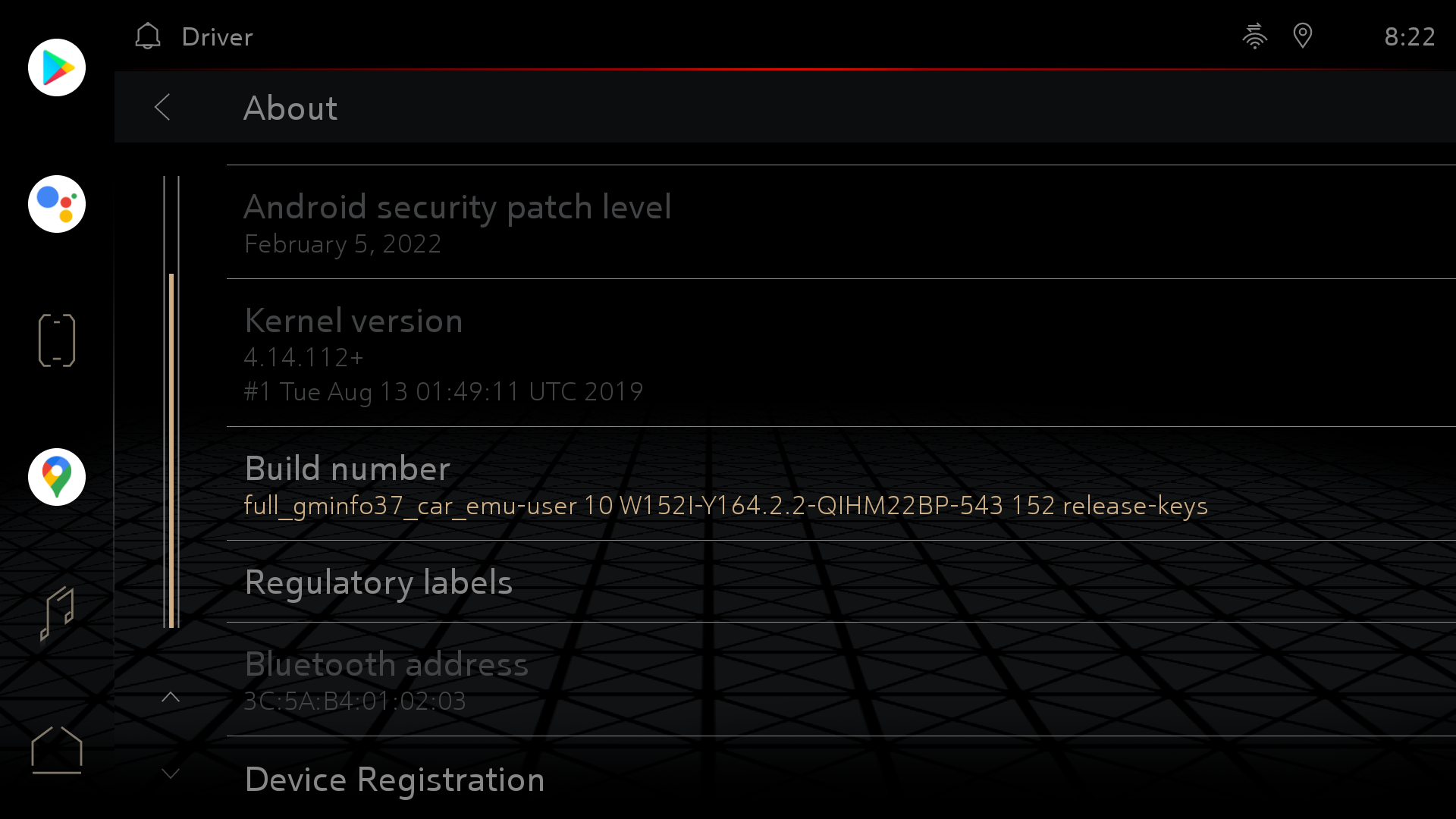}
    \caption{GMC emulator About.}
    \label{fig:gmc-emulator-about}
\end{subfigure}
&
\begin{subfigure}[t]{0.47\textwidth}
    \centering
    \includegraphics[width=0.7\linewidth,height=2.25cm]
    {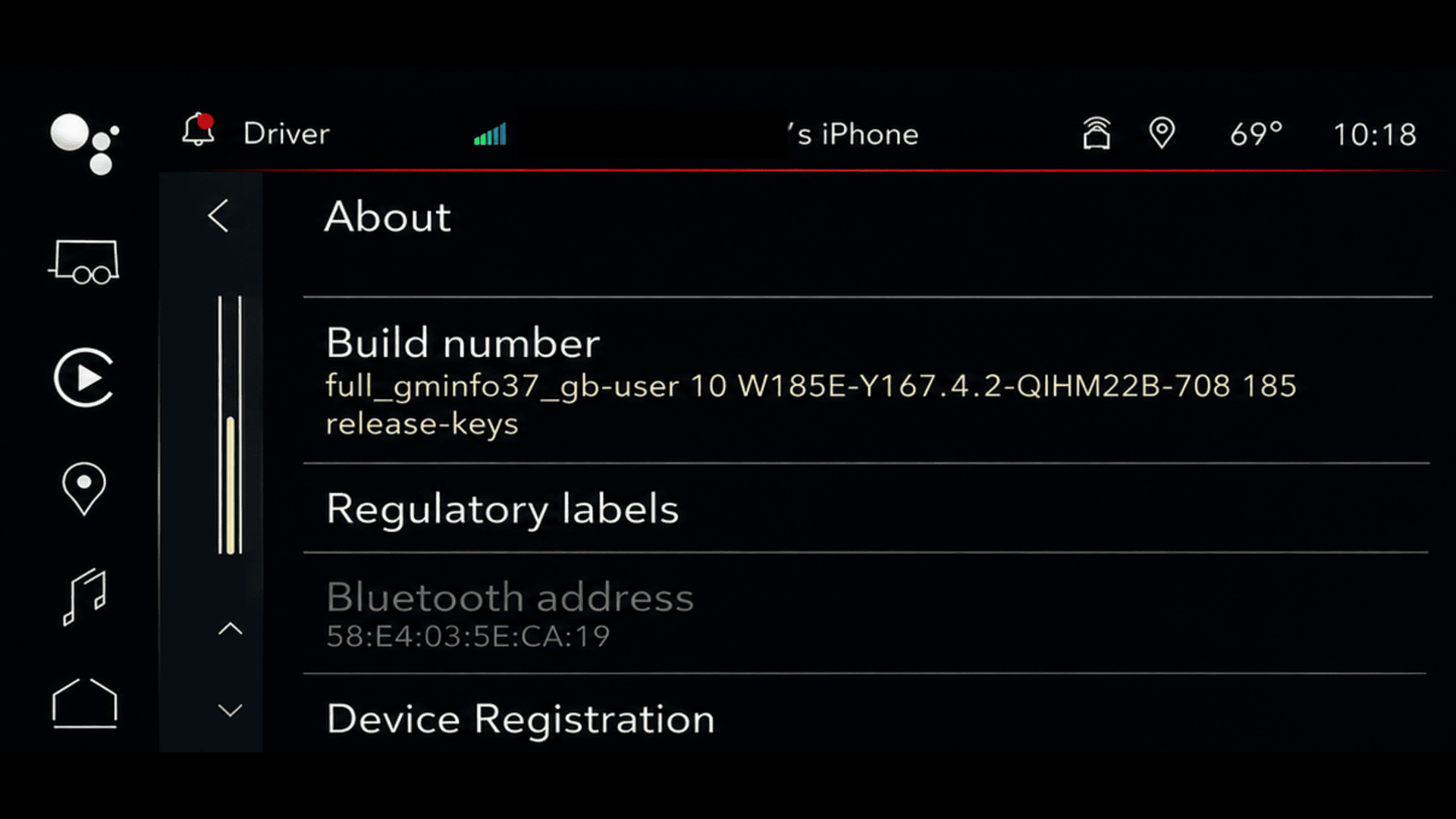}
    \caption{GMC real-vehicle About.}
    \label{fig:gmc-real-about}
\end{subfigure}
\\[2pt]

\begin{subfigure}[t]{0.47\textwidth}
    \centering
    \includegraphics[width=0.7\linewidth,height=2.25cm]
    {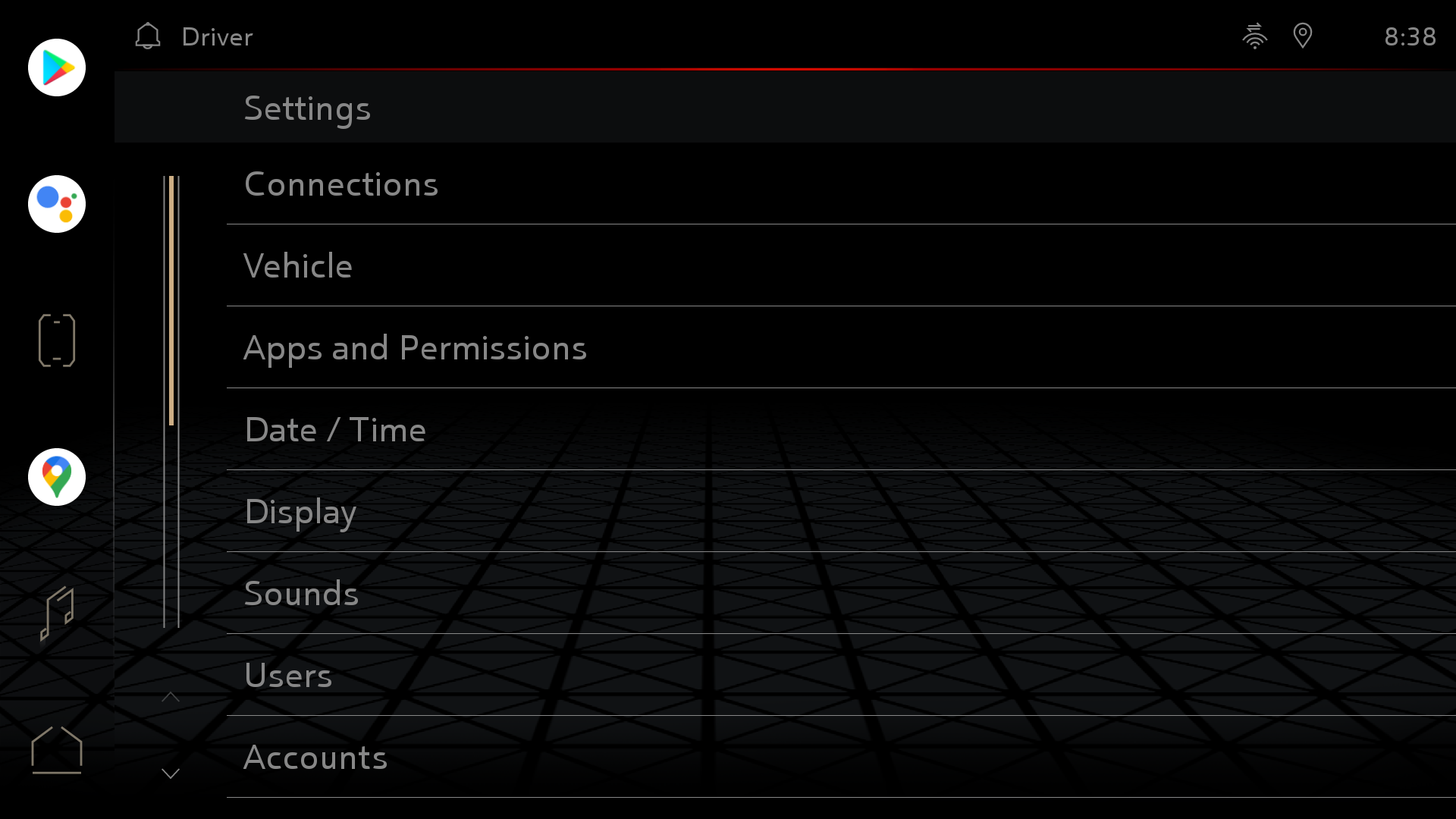}
    \caption{GMC emulator Settings.}
    \label{fig:gmc-emulator-settings}
\end{subfigure}
&
\begin{subfigure}[t]{0.47\textwidth}
    \centering
    \includegraphics[width=0.7\linewidth,height=2.25cm]
    {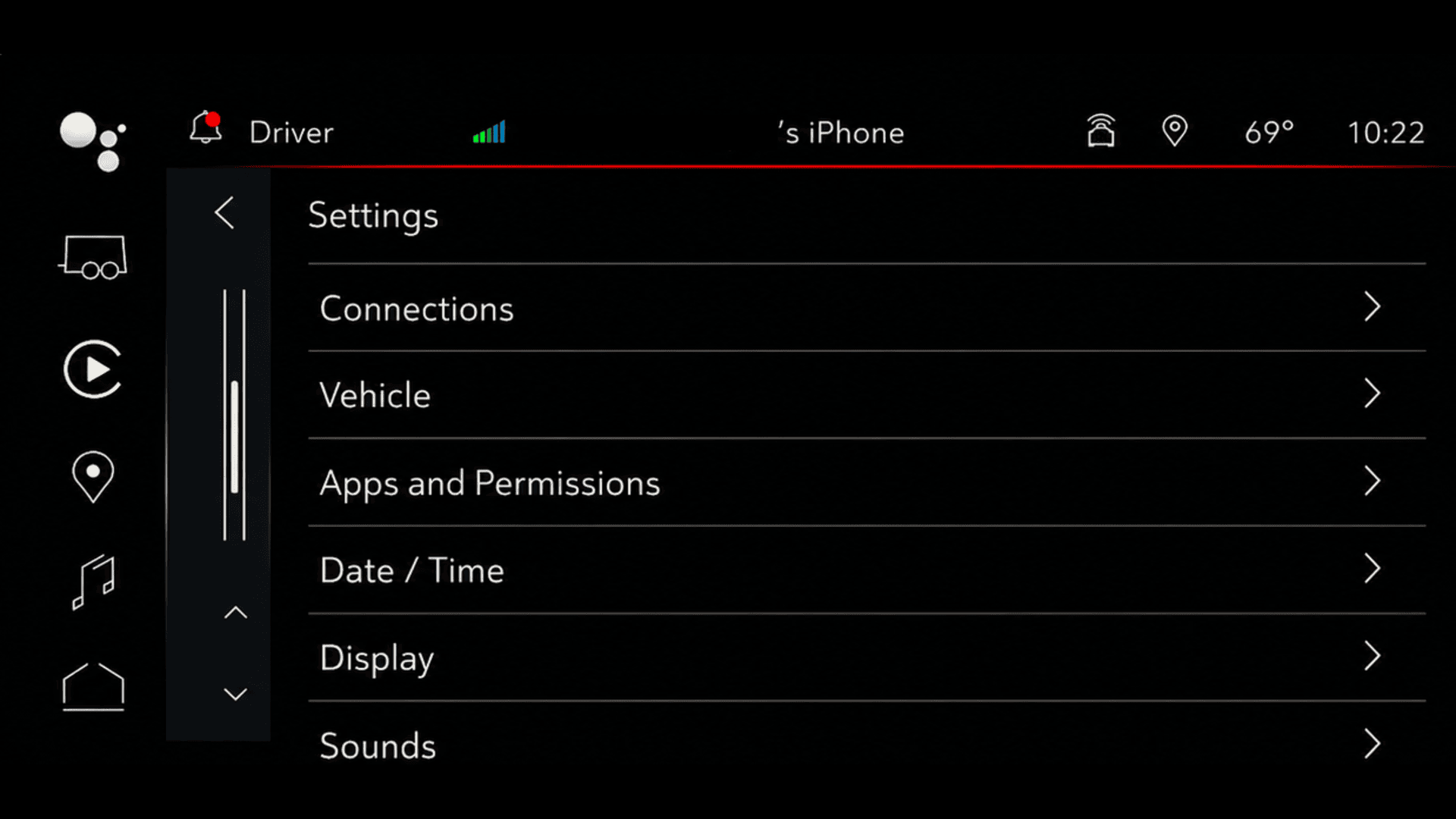}
    \caption{GMC real-vehicle Settings.}
    \label{fig:gmc-real-settings}
\end{subfigure}
\\[4pt]

\midrule

\begin{subfigure}[t]{0.47\textwidth}
    \centering
    \includegraphics[width=0.7\linewidth,height=2.25cm]
    {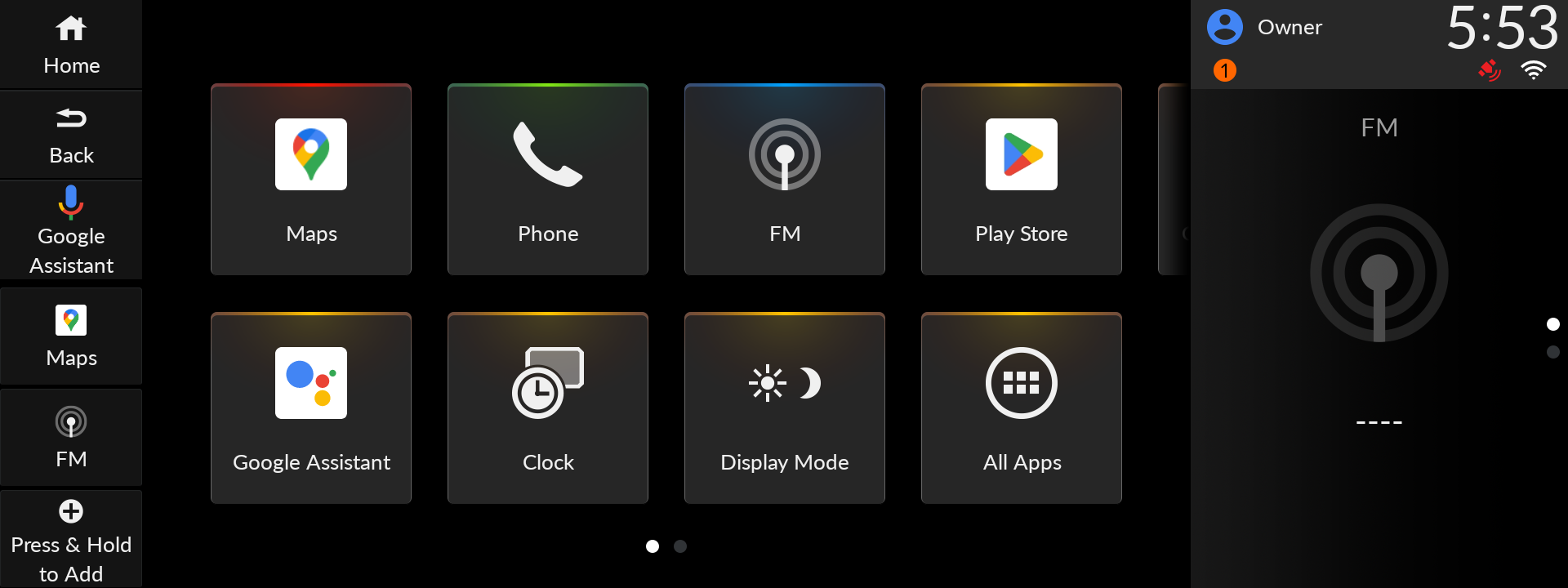}
    \caption{Honda emulator Home.}
    \label{fig:honda-emulator-home}
\end{subfigure}
&
\begin{subfigure}[t]{0.47\textwidth}
    \centering
    \includegraphics[width=0.7\linewidth,height=2.25cm]
    {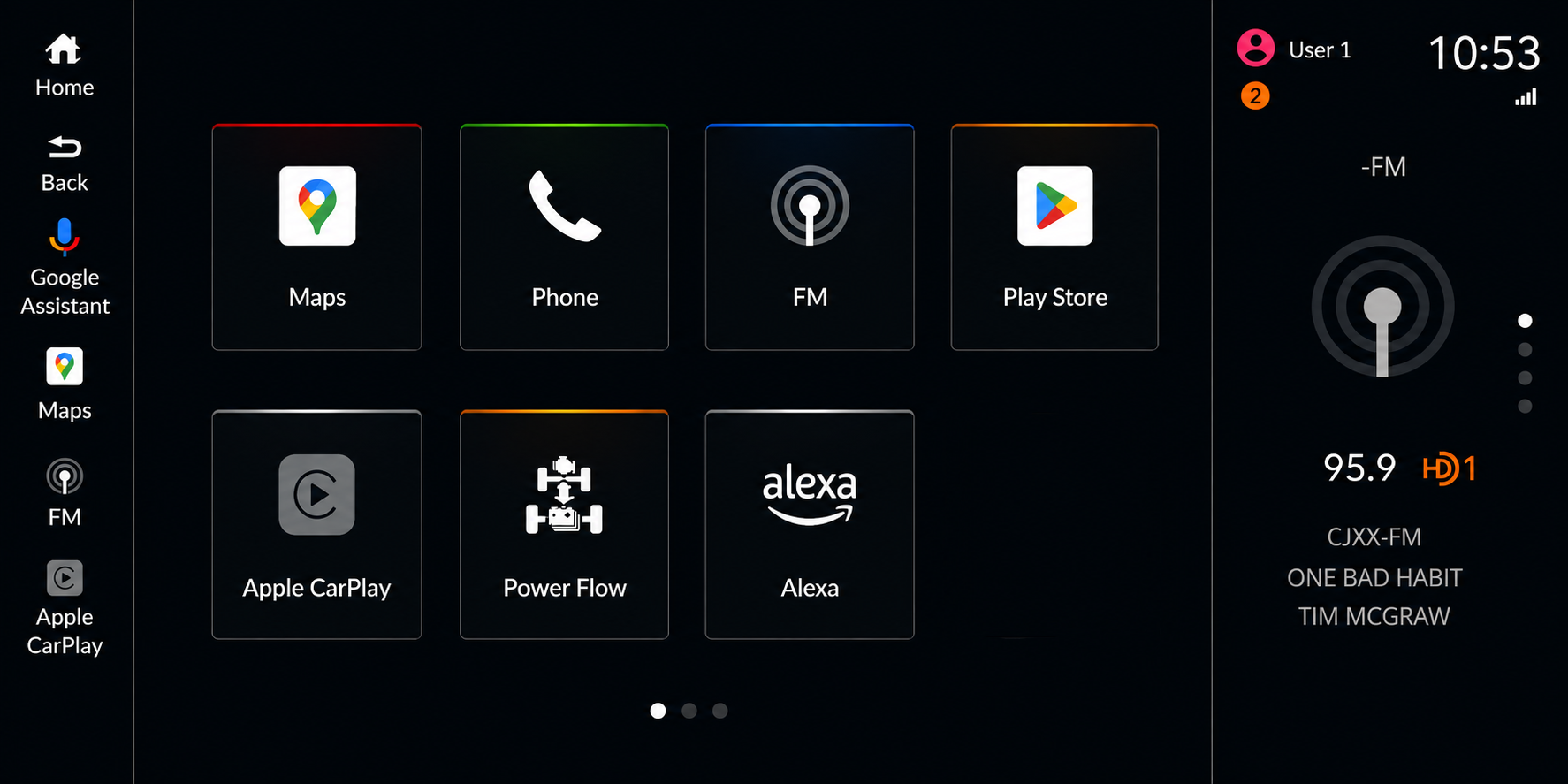}
    \caption{Honda real-vehicle Home.}
    \label{fig:honda-real-home}
\end{subfigure}
\\[2pt]

\begin{subfigure}[t]{0.47\textwidth}
    \centering
    \includegraphics[width=0.7\linewidth,height=2.25cm]
    {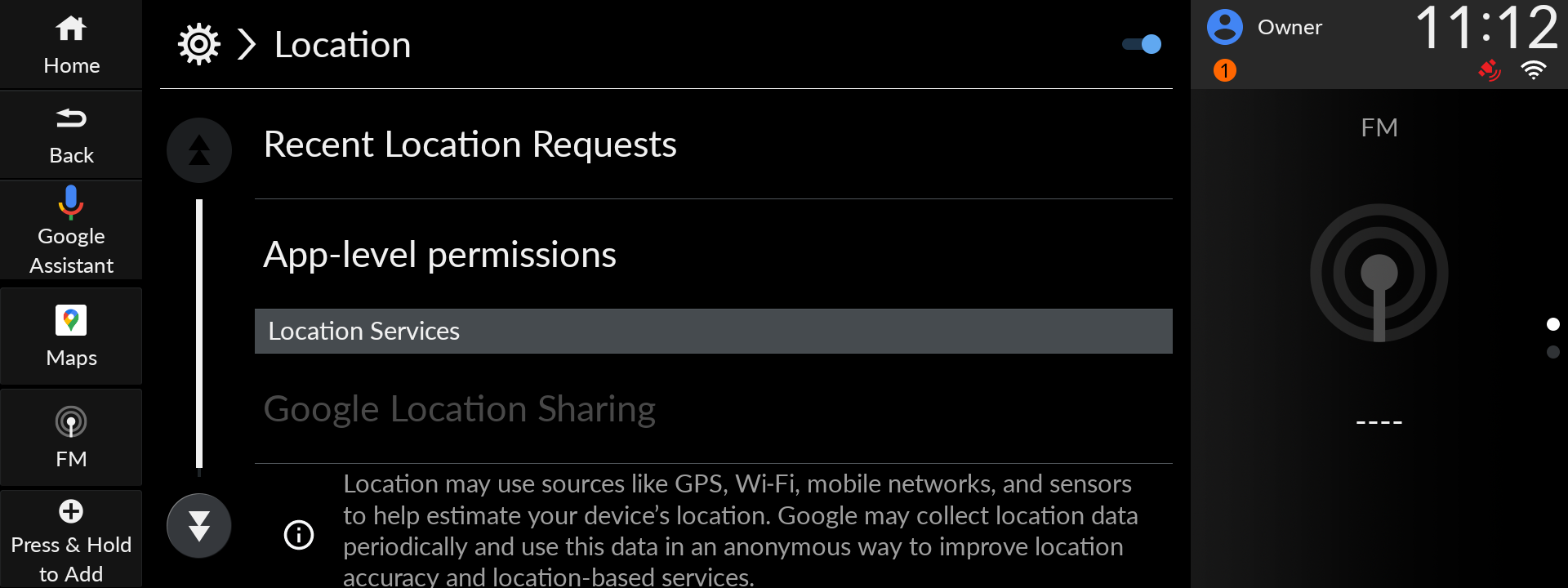}
    \caption{Honda emulator Location.}
    \label{fig:honda-emulator-location}
\end{subfigure}
&
\begin{subfigure}[t]{0.47\textwidth}
    \centering
    \includegraphics[width=0.7\linewidth,height=2.25cm]
    {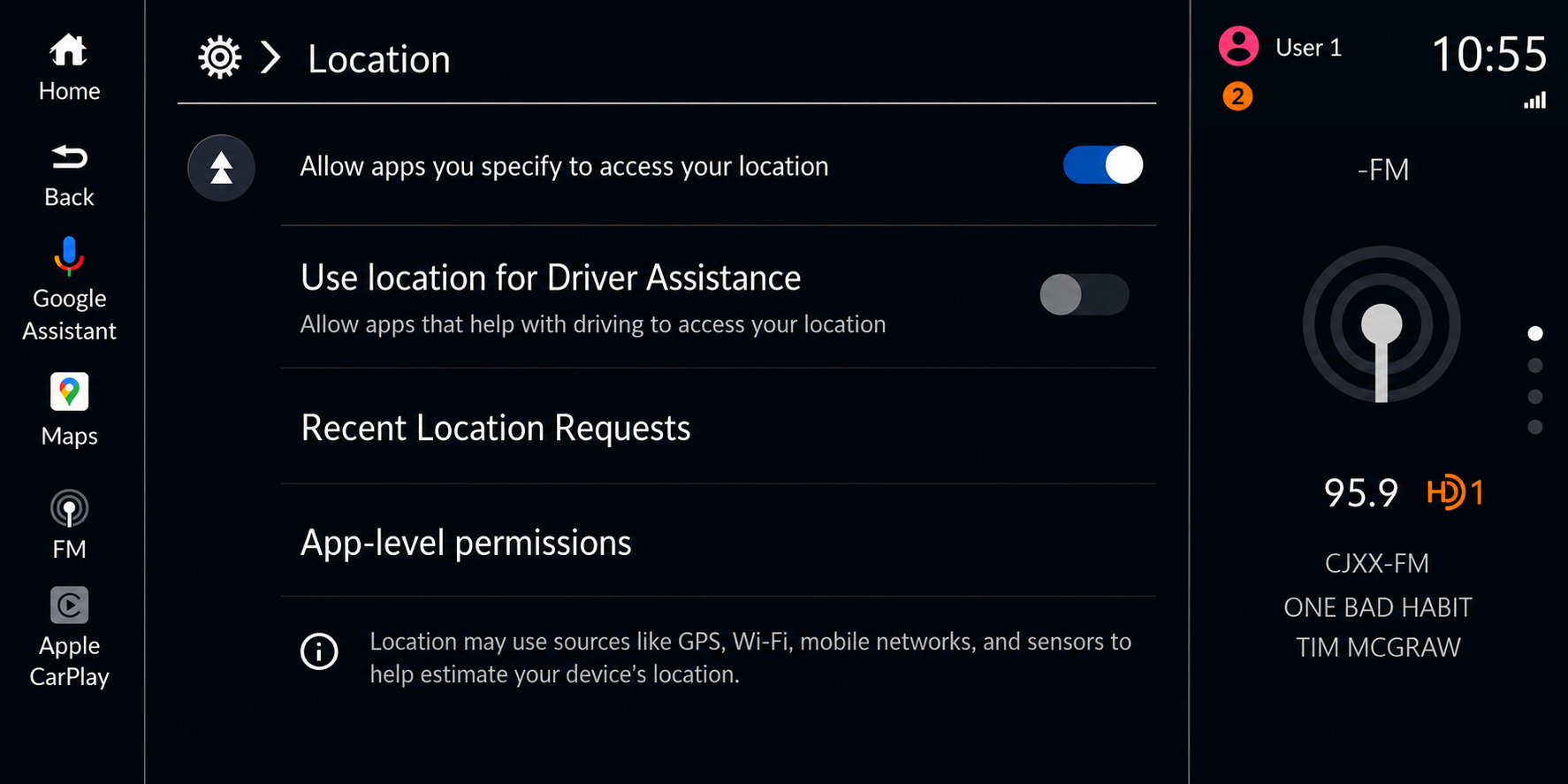}
    \caption{Honda real-vehicle Location.}
    \label{fig:honda-real-location}
\end{subfigure}
\\[2pt]

\begin{subfigure}[t]{0.47\textwidth}
    \centering
    \includegraphics[width=0.7\linewidth,height=2.25cm]
    {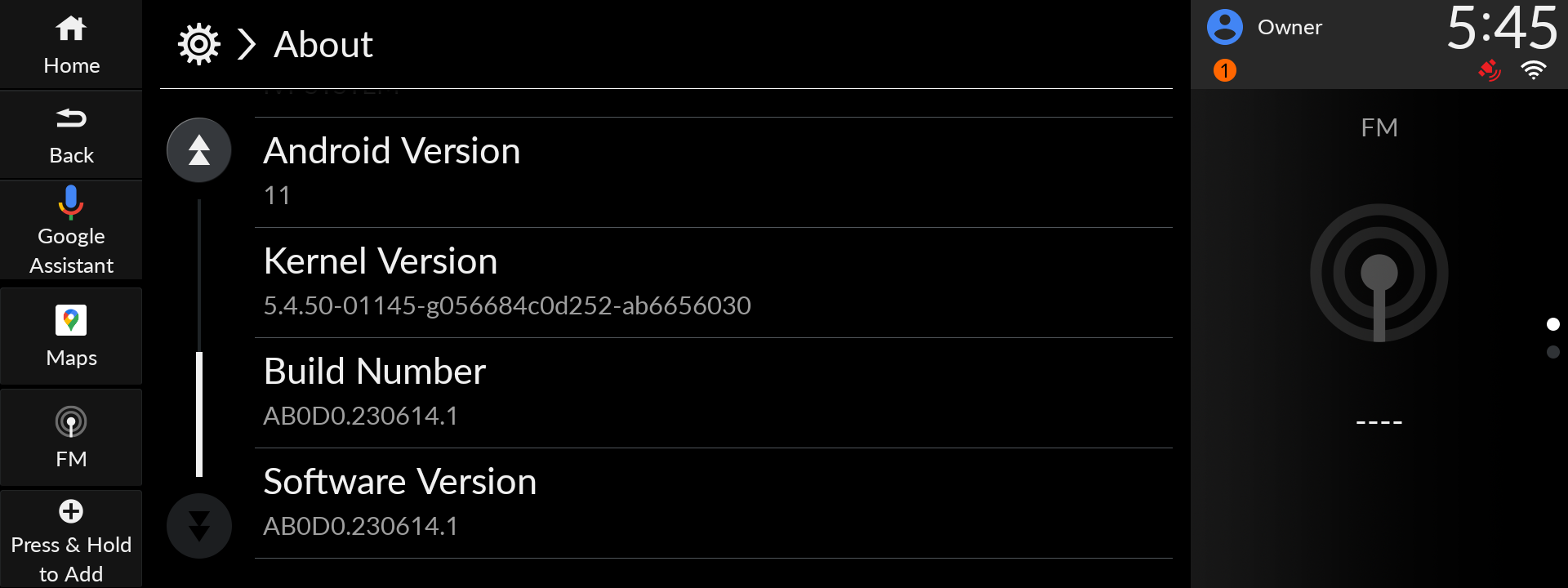}
    \caption{Honda emulator About.}
    \label{fig:honda-emulator-about}
\end{subfigure}
&
\begin{subfigure}[t]{0.47\textwidth}
    \centering
    \includegraphics[width=0.7\linewidth,height=2.25cm]
    {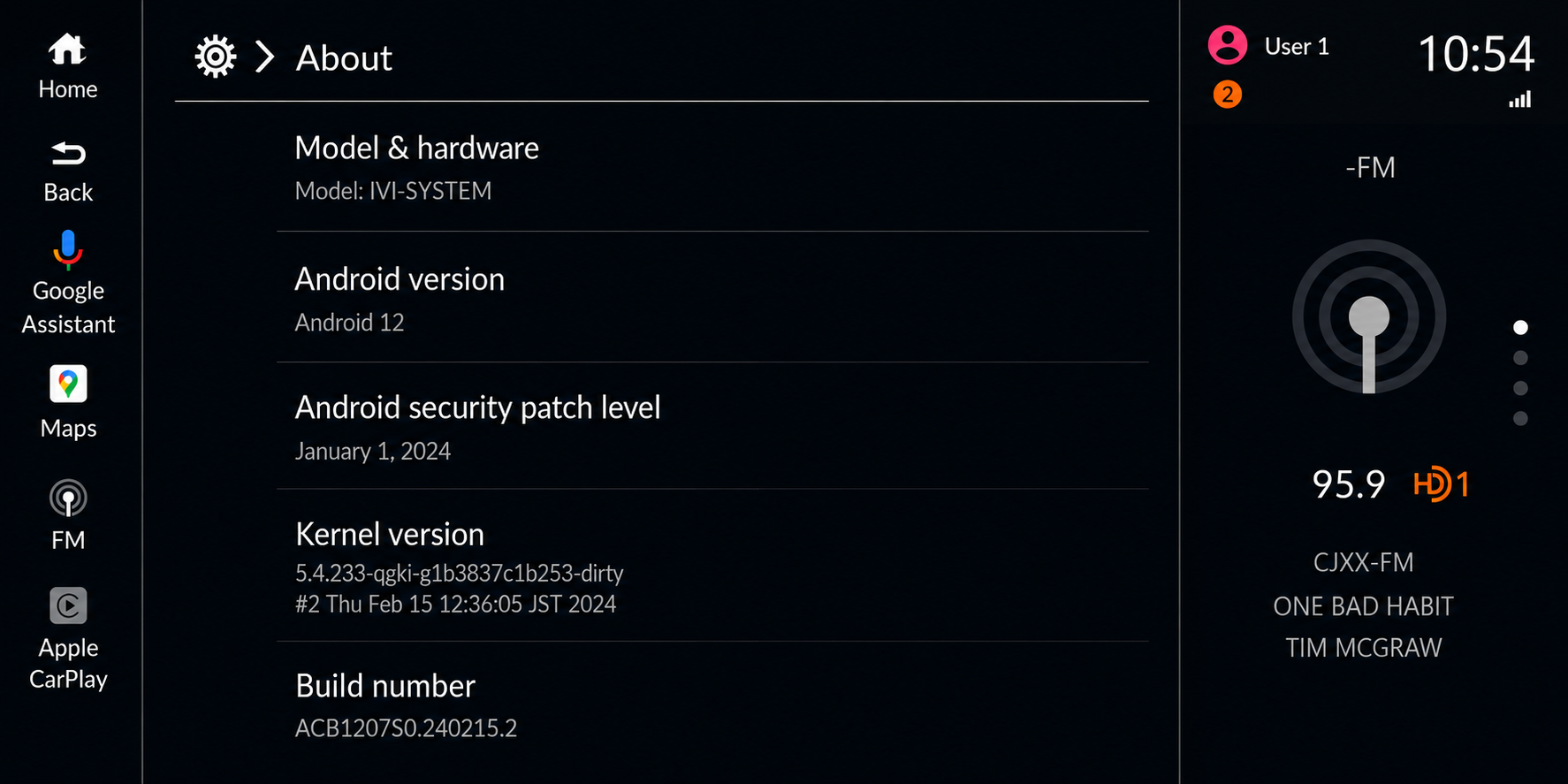}
    \caption{Honda real-vehicle About.}
    \label{fig:honda-real-about}
\end{subfigure}
\\[2pt]

\begin{subfigure}[t]{0.47\textwidth}
    \centering
    \includegraphics[width=0.7\linewidth,height=2.25cm]
    {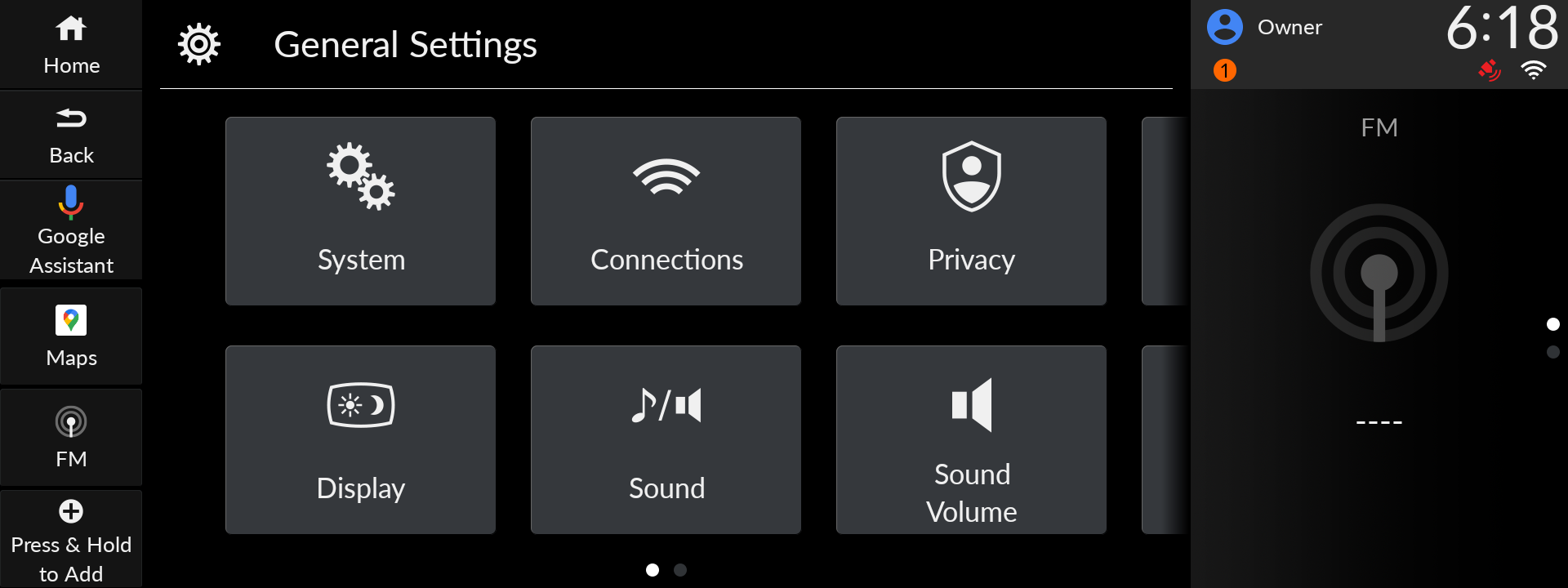}
    \caption{Honda emulator Settings.}
    \label{fig:honda-emulator-gs}
\end{subfigure}
&
\begin{subfigure}[t]{0.47\textwidth}
    \centering
    \includegraphics[width=0.7\linewidth,height=2.25cm]
    {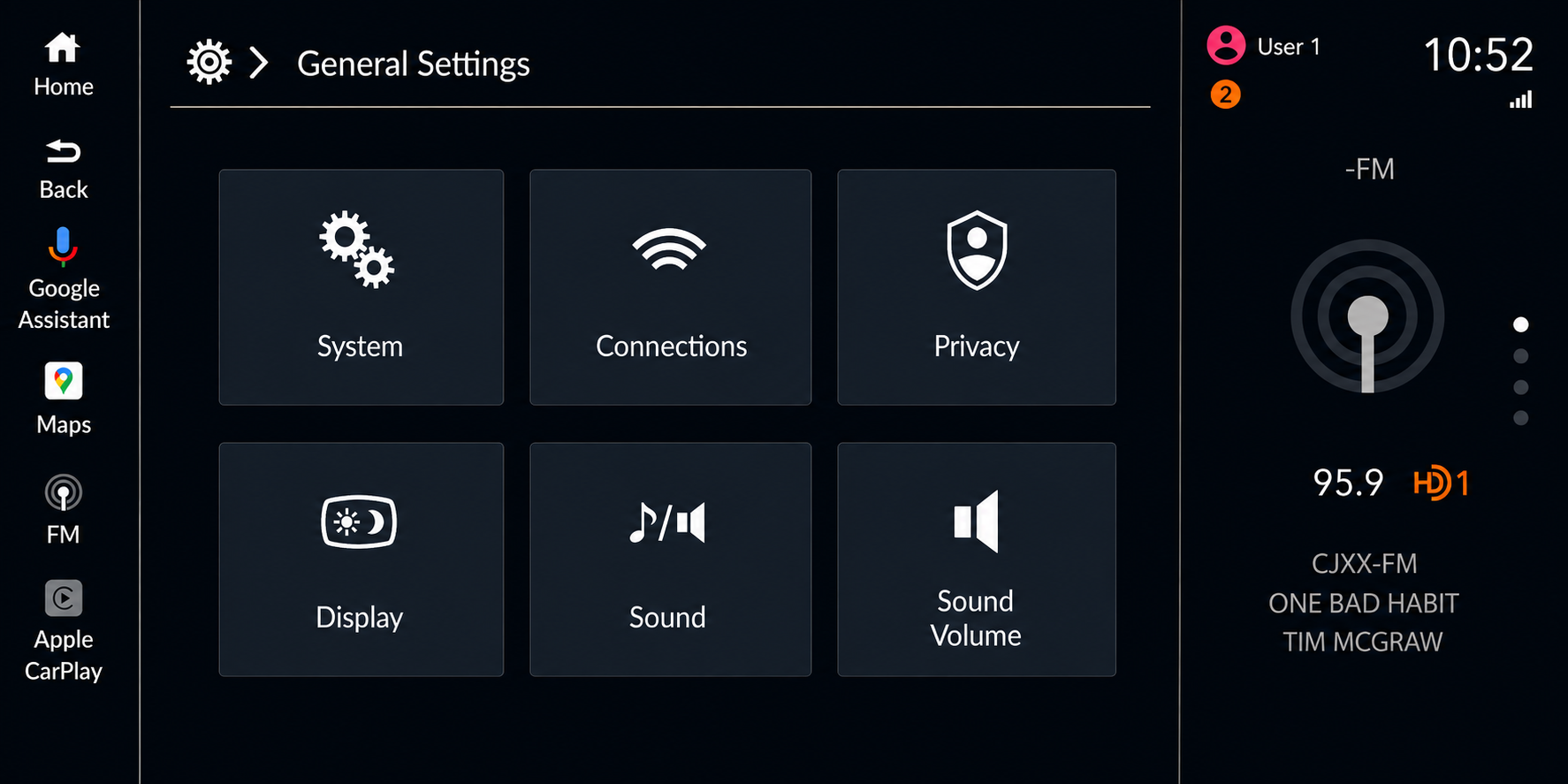}
    \caption{Honda real-vehicle Settings.}
    \label{fig:honda-real-gs}
\end{subfigure}

\end{tabular}
\caption{Side-by-side comparison of corresponding GMC and Honda emulator and real-vehicle interfaces. The displayed screens reflect the broader consistency observed in our full comparison, with no fundamental differences between them.}
\label{fig:emulator-real-validation}
\end{figure*}

\end{document}